\documentclass[prx,amsmath,amssymb, notitlepage, twocolumn,
nofootinbib,
superscriptaddress,
longbibliography
]{revtex4-1}

\usepackage{amsmath}
\usepackage{tabularx,graphicx}
\usepackage{epstopdf}
\usepackage{makecell}
\usepackage{graphicx}
\usepackage{latexsym}
\usepackage{amssymb}
\usepackage{amsmath}
\usepackage{color, colortbl}
\usepackage{psfrag}
\usepackage{bbm}
\usepackage{bm}
\usepackage{titlesec}
\usepackage{dsfont}
\usepackage{feynmp}
\usepackage{slashed}
\usepackage{multirow}
\usepackage{url}

\newcommand{\mc}[1]{\mathcal{#1}}

\newcommand{\beq}{\begin{eqnarray}}
	\newcommand{\eeq}{\end{eqnarray}}

\newcommand{\ra}{\rangle}

\newcommand{\bsp}{\begin{aligned}}
	\newcommand{\esp}{\end{aligned}}

\newcommand{\ie}{{i.e., }}
\newcommand{\eg}{{e.g., }}

\usepackage{xcolor}
\definecolor{darkblue}{rgb}{0.,0.,0.4}
\definecolor{darkred}{rgb}{0.5,0.,0.}
\definecolor{BlueViolet}{RGB}{138,43,226}
\definecolor{SkyBlue}{RGB}{30,144,255}
\definecolor{DarkGreen}{RGB}{0,100,0}

\usepackage[colorlinks=true,linkcolor=darkblue,citecolor=blue,urlcolor=darkred]{hyperref}
\usepackage[normalem]{ulem}

\usepackage{mathrsfs}
\usepackage[title]{appendix}

\usepackage{etoolbox}
\patchcmd{\appendices}{\quad}{: }{}{}
\usepackage{comment}

\newcommand{\Z}{\mathbb{Z}}

\def\U{\mathrm{U}(1)}
\usepackage{pst-all}
\usepackage{leftidx}
\usepackage[off]{auto-pst-pdf} 
\usepackage{booktabs}
\usepackage{yfonts}
\makeatletter

\usepackage[caption=false]{subfig}

\begin{document}

\title{
Classification of symmetry-enriched topological quantum spin liquids 
}

\author{Weicheng Ye}
	
\affiliation{Perimeter Institute for Theoretical Physics, Waterloo, Ontario, Canada N2L 2Y5}

\author{Liujun Zou}
\affiliation{Perimeter Institute for Theoretical Physics, Waterloo, Ontario, Canada N2L 2Y5}

\begin{abstract}

We present a systematic framework to classify symmetry-enriched topological quantum spin liquids in two spatial dimensions. This framework can deal with all topological quantum spin liquids, which may be either Abelian or non-Abelian, chiral or non-chiral. It can systematically treat a general symmetry, which may include both lattice symmetry and internal symmetry, may contain anti-unitary symmetry, and may permute anyons. The framework applies to all types of lattices, and can systematically distinguish different lattice systems with the same symmetry group using their quantum anomalies, which are sometimes known as Lieb-Schultz-Mattis anomalies. We apply this framework to classify $U(1)_{2N}$ chiral states and non-Abelian Ising$^{(\nu)}$ states enriched by a $p6\times SO(3)$ or $p4\times SO(3)$ symmetry, and $\Z_N$ topological orders and $U(1)_{2N}\times U(1)_{-2N}$ topological orders enriched by a $p6m\times SO(3)\times\Z_2^T$, $p4m\times SO(3)\times\Z_2^T$, $p6m\times\Z_2^T$ or $p4m\times\Z_2^T$ symmetry, where $p6$, $p4$, $p6m$ and $p4m$ are lattice symmetries, while $SO(3)$ and $\Z_2^T$ are spin rotation and time reversal symmetries, respectively. In particular, we identify symmetry-enriched topological quantum spin liquids that are not easily captured by the usual parton-mean-field approach, including examples with the familiar $\Z_2$ topological order.

\end{abstract}

\maketitle

\tableofcontents

\section{Introduction} \label{sec: intro}

Topological quantum spin liquids, which are more formally referred to as bosonic topological orders, are exotic gapped quantum phases of matter with long-range entanglement beyond the phenomenon of spontaneous symmetry breaking.{\footnote{All symmetries discussed in this paper are 0-form invertible symmetries, unless otherwise stated.}} In two space dimensions, they can host anyons, \ie quasi-particle excitations that are neither bosons nor fermions \cite{wen2004quantum}. Besides being fundamentally interesting, they are also potential platforms for fault-tolerant quantum computation \cite{Kitaev2003,Nayak2007}.

Roughly speaking, in the absence of symmetries, the universal properties of a topological quantum spin liquid are encoded in the properties of its anyons, such as their fusion rules and statistics. We refer to these properties as the topological properties of a topological quantum spin liquid. In the presence of symmetries, there can be interesting interplay between these topological properties and the symmetries. In particular, even with fixed topological properties, there can be sharply distinct topological quantum spin liquids that cannot evolve into each other without breaking the symmetries or encountering a quantum phase transition. These are known as different symmetry-enriched topological quantum spin liquids.

The goal of this paper is to classify symmetry-enriched topological quantum spin liquids in two space dimensions. That is, given the topological and symmetry properties, we would like to understand which symmetry-enriched topological quantum spin liquids are possible. This problem is first of fundamental conceptual interest, as understanding different types of quantum matter is one of the central goals of condensed matter physics. Moreover, although topological quantum spin liquids have been identified in certain lattice models and small-sized quantum simulators, its realization and detection in macroscopic quantum materials remains elusive \cite{Broholm2020}. The knowledge of which symmetry-enriched topological quantum spin liquids are possible is helpful for identifying the correct observable signatures of them, thus paving the way to the realization and detection of these interesting quantum phases in the future.

In the previous condensed matter literature, there are two widely used approaches to classify symmetry-enriched topological quantum spin liquids, and most of the other approaches can be viewed as variations or generalizations of them. The first approach is based on parton mean fields and projective symmetry groups, which starts by representing the microscopic degrees of freedom (such as spins) via certain fractional degrees of freedom, and then studies the possible projective representations of the symmetries carried by these fractional degrees of freedom \cite{Wen2001}. The advantages of this approach include its simplicity and its intimate connection to the microscopic degrees of freedom of the system. It has been applied to classify various symmetry-enriched quantum spin liquids in many different lattice systems, and it treats internal symmetries and lattice symmetries on equal footing.

However, this approach is not perfect. One of the main disadvantages of this approach is that it is not easily applicable to general topological quantum spin liquids. For example, it is inconvenient to apply this approach to classify symmetry-enriched $\Z_N$ topological orders with $N>2$. This is because such topological orders are most naturally described by partons coupled to a dynamical $\Z_N$ gauge field. When $N>2$, the partons themselves are interacting, so a non-interacting parton mean field description is inadequate. Also, it is often challenging to use this approach to study topological quantum spin liquids where anyons are permuted in a complicated manner by symmetries \cite{barkeshli2014}. Moreover, even for quantum spin liquids that are often assumed to be captured by parton mean fields, some of their symmetry enrichment patterns may not be captured by any parton mean field. Ref. \cite{Ye2021a} presented such phenomenon for the gapless Dirac spin liquid, and in Sec. \ref{sec: Zn} we find that this phenomenon can occur even for the familiar $\Z_2$ topological quantum spin liquid. Another disadvantage of this approach is that the projective representations carried by the fractional degrees of freedom may not be the same as the symmetry properties of the physical anyons, which sometimes require one more nontrivial step to obtain. For examples, see Refs. \cite{Cincio2015, Zou2018}.

The second approach is based on the categorical theoretic description of the topological quantum spin liquids, and studies how the category theory (i.e. a powerful mathematical theory) corresponding to a topological quantum spin liquid can be consistently extended to include a symmetry group \cite{barkeshli2014, Tarantino2015}. The advantage of this approach is that it is fully general and can be applied to situations where the parton-based approach is inconvenient, since it is believed that the topological properties of all topological quantum spin liquids can be described by category theory. Also, it yields the symmetry properties of the anyons directly, without relying on some artificial fractional degrees of freedom as in the first approach. Overall, this approach is very systematic and powerful in studying topological quantum spin liquids with only internal symmetries.

However, this approach also has its disadvantages. Unlike the first approach, it has a relatively weaker connection with the microscopic properties of the physical system, and it is particularly inconvenient when applied to systems with lattice symmetries, which are often physically important. Specifically, within this approach all symmetry properties of the physical system are assumed to be captured by its symmetry group. However, such a description of the symmetries is inadequate for many purposes. For example, spin-1/2 systems defined on a triangular lattice, kagome lattice and honeycomb lattice can all have the same symmetry group, say, the $SO(3)$ spin rotational symmetry and the $p6m$ lattice symmetry. But they are physically distinct, and any symmetry-enriched topological quantum spin liquid that can emerge in one of these three systems cannot emerge in the other two \cite{Po2017, Ye2021a}. A physically relevant classification of symmetry-enriched topological quantum spin liquids should take into account this distinction, which reflects some symmetry properties beyond the symmetry group. In the above example, these properties can be viewed as the lattice types.

In the present paper, we develop a framework to classify symmetry-enriched topological quantum spin liquids, which combines the advantages of the above two approaches while avoiding their disadvantages. Specifically, we will use the language of category theory to directly describe the topological properties of a topological quantum spin liquid and the symmetry properties of the anyons, and we will also keep track of the robust microscopic symmetry-related information of the physical system. Roughly speaking, this information includes:

\begin{itemize}

\item The symmetry group, say, the $SO(3)$ spin rotational symmetry and $p6m$ lattice symmetry.

\item How the microscopic degrees of freedom transform under the symmetry, say, whether the microscopic degrees of freedom are spin-1/2's or spin-1's.

\item The lattice type, say, whether the lattice is a triangular, kagome or honeycomb lattice.

\end{itemize}

All three pieces of information above are taken into account in the aforementioned first approach, but the second approach usually only considers the first piece (see exceptions in Refs. \cite{Cheng2015, Qi2015b, Qi2016, Ning2019, Manjunath2020, Manjunath2020a, Song2022} studying certain specific symmetry-enriched quantum spin liquids, but the methods therein have not been generalized to the generic case before). The reason why the other two pieces of information are physically relevant is because, as long as the symmetries are preserved, they are robust properties of the system that cannot change under perturbations \cite{Po2017}.
Moreover, just like the symmetry group, these two pieces of information are often relatively easy to determine experimentally, and they are the basic aspects of any theoretical model. We will make these concepts more precise in Sec. \ref{sec: anomaly of lattice}.

Ultimately, our framework takes as input 1) the topological properties of a topological quantum spin liquid and 2) a collection of the above symmetry-related properties of a microscopic system, and outputs which symmetry-enriched topological quantum spin liquids are possible. Each of these symmetry-enriched topological quantum spin liquids is characterized by the symmetry properties of its anyons. We will apply this framework to classify various symmetry-enriched topological quantum spin liquids. We will focus on examples with symmetry group $G=G_s\times G_{int}$, where $G_s$ is certain lattice symmetry, and $G_{int}$ is an on-site internal symmetry.

This framework relies on the state-of-the-art in the studies of the quantum anomalies of lattice systems and topological quantum spin liquids. In particular, in Ref. \cite{Ye2021a} we and our coauthors obtained the precise characterizations of the full set of the quantum anomalies of a large class of lattice systems, which exactly encode the aforementioned robust microscopic symmetry-related information of the lattice system. These quantum anomalies are sometimes referred to as Lieb-Schultz-Mattis anomalies. Moreover, in Ref. \cite{Ye2022} we developed a systematic framework to calculate the quantum anomaly of a generic symmetry-enriched topological quantum spin liquid.

We remark that our philosophy of encoding the robust microscopic symmetry-related information of a physical system into its quantum anomaly and using anomaly-matching to classify quantum many-body states is very general. Besides classifying symmetry-enriched topological quantum spin liquids, it can also be used to classify other quantum states. In fact, we and our coauthors have applied it to classify certain symmetry-enriched gapless quantum spin liquids in Ref. \cite{Ye2021a}. Our anomaly-based philosophy can be viewed as a generalization of the theories of symmetry indicators or topological quantum chemistry, which classify band structures \cite{Kruthoff2016, Po2017a, Bradlyn2017, Watanabe2017, Po2020}. In fact, the robust microscopic symmetry-related information of a physical system we take is identical to what those theories take, but those theories only apply to weakly correlated systems that can be described by band theory, while ours applies to generic strongly correlated systems.

\section{Outline and summary} \label{sec: outline and summary}

The outline of the rest of the paper and summary of the main results are as follows.

\begin{itemize}
    
    \item In Sec. \ref{sec: characterization}, we briefly review how to describe a symmetry-enriched topological quantum spin liquid and its quantum anomaly. The description is based on the category theory, but no previous knowledge of category theory is required.

    \item In Sec. \ref{sec: anomaly of lattice}, we discuss the symmetry properties of a lattice system and how to encode them into the quantum anomaly of a lattice system.

    \item In Sec. \ref{sec: framework}, we present our general framework to classify symmetry-enriched topological quantum spin liquid, which may be Abelian or non-Abelian, and chiral or non-chiral. This framework applies to topological quantum spin liquids with any symmetry, which may contain both lattice symmetry and internal symmetry, may contain anti-unitary symmetry and may permute anyons.

    \item In Sec. \ref{sec: U(1) level 2N}, we apply the framework in Sec. \ref{sec: framework} to classify the $U(1)_{2N}$ topological quantum spin liquid enriched by a $p6\times SO(3)$ or $p4\times SO(3)$ symmetry, where $p6$ and $p4$ are lattice symmetries and $SO(3)$ is the spin rotational symmetry. The results are summarized in Table \ref{tab:RealizationNumber U(1) level 2N}.

    \item In Sec. \ref{sec: Ising}, we apply our framework in Sec. \ref{sec: framework} to classify the non-Abelian Ising$^{(\nu)}$ topological quantum spin liquid enriched by a $p6\times SO(3)$ or $p4\times SO(3)$ symmetry.

    \item In Sec. \ref{sec: Zn}, we apply our framework in Sec. \ref{sec: framework} to classify the $\Z_N$ topological quantum spin liquids enriched by one of these four symmetries: $p6m\times SO(3)\times\Z_2^T$, $p4m\times SO(3)\times\Z_2^T$, $p6m\times \Z_2^T$ and $p4m\times \Z_2^T$, where the $p6m$ and $p4m$ are lattice symmetries, and $SO(3)$ and $\Z_2^T$ are on-site spin rotational symmetry and time reversal symmetry, respectively. The results are summarized in Table \ref{tab:RealizationNumber}. In particular, even for the simple case with $N=2$, we find many states beyond the description based on the usual parton mean field.

    \item In Sec. \ref{sec: generalization of double semion}, we apply our framework in Sec. \ref{sec: framework} to classify the $U(1)_{2N}\times U(1)_{-2N}$ topological quantum spin liquids enriched by one of these four symmetries: $p6m\times SO(3)\times\Z_2^T$, $p4m\times SO(3)\times\Z_2^T$, $p6m\times \Z_2^T$ and $p4m\times \Z_2^T$. The resutls are summarized in Table \ref{tab:RealizationNumber}.

    \item We close our paper in Sec. \ref{sec: discussion}.

    \item Various appendices contain further details. Appendix \ref{app: reflection and time reversal} presents a formal treatment of the connection between the data characterizing a topological order enriched by reflection symmetry and the data characterizing a topological order enriched by time reversal symmetry. Appendix \ref{app:cohomology} reviews the properties of the lattice symmetries involved in this paper. Appendix \ref{app:SFclasses} details the information of all symmetry fractionalization classes of all states studied in this paper. Appendix \ref{app: anomaly indicators} presents the anomaly indicators for various symmetries, and their values in the physically relevant cases considered in this paper. Appendix \ref{app: beyond parton} presents the details of the symmetry fractionalization classes of the symmetry-enriched $\Z_2$ topological orders that are beyond the usual parton mean fields.

\end{itemize}

\section{Universal characterization of symmetry-enriched topological quantum spin liquids} \label{sec: characterization} 

In this section, we review the universal characterization of symmetry-enriched topological quantum spin liquids. This characterization is divided into two parts. We first specify the topological order corresponding to the topological quantum spin liquid, which is reviewed in Sec. \ref{subsec: UMTC review}. After this, assuming the symmetry is an internal symmetry, we will specify the global symmetry of the topological quantum spin liquid and how it acts on the topological order, which is reviewed in Sec. \ref{subsec: global symmetry}. We review the anomaly of a symmetry-enriched topological order in Sec. \ref{subsec: anomaly review}. Finally, we review the crystalline equivalence principle in Sec. \ref{subsec: crystalline equivalence principle}, which allows us to connect a topological order with lattice symmetry to one with internal symmetry.

\subsection{Topological order} \label{subsec: UMTC review}

The characteristic feature of a topological order is the presence of anyons, point-like excitations that can have self-statistics other than bosonic or fermionic statistics, and nontrivial mutual braiding statistics. When multiple anyons are present, there may also be a protected degenerate state space. It is believed that the category theory can universally characterize all topological orders or the anyons therein. In this subsection we briefly review the concepts relevant to this paper. Our review will be minimal and does not assume any knowledge of the category theory itself. For a more comprehensive review, see \eg Refs.~\cite{barkeshli2014, Kitaev2006, Kong2022} for a more physics oriented introduction, or Refs.~\cite{bakalov2001lectures,etingof2016tensor,turaev1994quantum,Selinger2011} for a more mathematical treatment.

Anyons in a topological order are denoted by $a, b, c, \cdots$. A single anyon cannot be converted into a different single anyon via any local process. There is always a trivial anyon in all topological orders, obtained by performing some local operation on the ground state. Roughly speaking, in a many-body system where a topological order emerges at low energies, the quantum state is specified by two pieces of data: the global data characterizing the anyons, which cannot be changed by any local process, and the local data independent of the anyons, which can change by local operations.

For the purpose of this paper, two important properties of an anyon $a$ will be used, its quantum dimension $d_a$ and topological spin $\theta_a$. Suppose $n$ anyons $a$ are created, the dimension of the degenerate state space scales as $d_a^n$ when $n$ is large. So the quantum dimension $d_a$ effectively measures how rich the internal degree of freedom the anyon $a$ carries. If $d_a=1$, then $a$ is said to be an Abelian anyon. Otherwise it is a non-Abelian anyon. The topological spin $\theta_a$ measures the self-statistics of $a$. For bosons and fermions, the topological spin is $\pm 1$. For a generic anyon, the topological spin can take other values.

Two anyons $a$ and $b$ can fuse into other anyons, expressed using the equation $a\times b\cong\sum_cN^c_{ab}c$, where $N^c_{ab}$ are positive integers and $c$ is said to be a fusion outcome of $a$ and $b$. Generically, $a$ and $b$ may have multiple fusion outcomes, and there may also be multiple different ways to get each fusion outcome, which is why the right hand side of the fusion equation involves a summation and $N^c_{ab}$ can be larger than 1.
Physically, fusion means that if we only perform measurements far away from the anyons $a$ and $b$, we will think they look like the fusion outcome $c$. Diagramatically, we can use
$$
\begin{pspicture}(-0.1,-0.2)(1.5,-1.2)
  \small
  \psset{linewidth=0.9pt,linecolor=black,arrowscale=1.5,arrowinset=0.15}
  \psline{-<}(0.7,0)(0.7,-0.35)
  \psline(0.7,0)(0.7,-0.55)
  \psline(0.7,-0.55) (0.25,-1)
  \psline{-<}(0.7,-0.55)(0.35,-0.9)
  \psline(0.7,-0.55) (1.15,-1)	
  \psline{-<}(0.7,-0.55)(1.05,-0.9)
  \rput[tl]{0}(0.4,0){$c$}
  \rput[br]{0}(1.4,-0.95){$b$}
  \rput[bl]{0}(0,-0.95){$a$}
 \scriptsize
  \rput[bl]{0}(0.85,-0.5){$\mu$}
  \end{pspicture}
$$ 
to represent a fusion process, or
$$
\begin{pspicture}[shift=-0.65](-0.1,-0.2)(1.5,1.2)
  \small
  \psset{linewidth=0.9pt,linecolor=black,arrowscale=1.5,arrowinset=0.15}
  \psline{->}(0.7,0)(0.7,0.45)
  \psline(0.7,0)(0.7,0.55)
  \psline(0.7,0.55) (0.25,1)
  \psline{->}(0.7,0.55)(0.3,0.95)
  \psline(0.7,0.55) (1.15,1)	
  \psline{->}(0.7,0.55)(1.1,0.95)
  \rput[bl]{0}(0.4,0){$c$}
  \rput[br]{0}(1.4,0.8){$b$}
  \rput[bl]{0}(0,0.8){$a$}
 \scriptsize
  \rput[bl]{0}(0.85,0.35){$\mu$}
  \end{pspicture}
$$
to represent a splitting process, which can be viewed as the reversed process of fusion. In the above, $\mu\in\{1, 2, \cdots, N^c_{ab}\}$, and the arrows can be viewed as the world lines of the anyons. If the trivial anyon is in the fusion product of two anyons, we say that these two anyons are anti-particles of each other, and we denote the anti-particle of $a$ by $\bar a$.

When multiple anyons are present, we may imagine fusing them in different orders. Similarly, when a given anyon splits into multiple other anyons, it may also do so in different orders. For example, the two sides of the following equation represent two different orders of splitting an anyon $d$ into three anyons, $a$, $b$ and $c$. Generically, the state obtained after these two processes are not identical, but they can be related by the so-called $F$-symbol, which is a unitary matrix acting on the degenerate state space.
$$
 \raisebox{-0.5\height}{\begin{pspicture}[shift=-1.0](0,-0.45)(1.8,1.8)
  \small
  \psset{linewidth=0.9pt,linecolor=black,arrowscale=1.5,arrowinset=0.15}
  \psline(0.2,1.5)(1,0.5)
  \psline(1,0.5)(1,0)
  \psline(1.8,1.5) (1,0.5)
  \psline(0.6,1) (1,1.5)
   \psline{->}(0.6,1)(0.3,1.375)
   \psline{->}(0.6,1)(0.9,1.375)
   \psline{->}(1,0.5)(1.7,1.375)
   \psline{->}(1,0.5)(0.7,0.875)
   \psline{->}(1,0)(1,0.375)
   \rput[bl]{0}(0.05,1.6){$a$}
   \rput[bl]{0}(0.95,1.6){$b$}
   \rput[bl]{0}(1.75,1.6){${c}$}
   \rput[bl]{0}(0.5,0.5){$e$}
   \rput[bl]{0}(0.9,-0.3){$d$}
 \scriptsize
   \rput[bl]{0}(0.3,0.8){$\alpha$}
   \rput[bl]{0}(0.7,0.25){$\beta$}
\end{pspicture}}
= \sum_{f, \mu,\nu}
\left[F_d^{abc}\right]_{(e,\alpha,\beta),(f,\mu,\nu)}
  \raisebox{-0.5\height}{\begin{pspicture}[shift=-1.0](0,-0.45)(1.8,1.8)
  \small
  \psset{linewidth=0.9pt,linecolor=black,arrowscale=1.5,arrowinset=0.15}
  \psline(0.2,1.5)(1,0.5)
  \psline(1,0.5)(1,0)
  \psline(1.8,1.5) (1,0.5)
  \psline(1.4,1) (1,1.5)
   \psline{->}(0.6,1)(0.3,1.375)
   \psline{->}(1.4,1)(1.1,1.375)
   \psline{->}(1,0.5)(1.7,1.375)
   \psline{->}(1,0.5)(1.3,0.875)
   \psline{->}(1,0)(1,0.375)
   \rput[bl]{0}(0.05,1.6){$a$}
   \rput[bl]{0}(0.95,1.6){$b$}
   \rput[bl]{0}(1.75,1.6){${c}$}
   \rput[bl]{0}(1.25,0.45){$f$}
   \rput[bl]{0}(0.9,-0.3){$d$}
 \scriptsize
   \rput[bl]{0}(1.5,0.8){$\mu$}
   \rput[bl]{0}(0.7,0.25){$\nu$}
  \end{pspicture}}
$$
In this equation, $e$ is a fusion out come of $a$ and $b$, where $\alpha\in\{1, \cdots N^e_{ab}\}$, $d$ is a fusion outcome of $c$ and $e$, where $\beta\in\{1, \cdots, N^d_{ec}\}$, $f$ is a fusion outcome of $b$ and $c$, where $\mu\in\{1, \cdots, N^f_{bc}\}$, and $f$ and $a$ also fuse into $d$, where $\nu\in\{1, \cdots, N^d_{af}\}$.

When anyons move around each other, the state may acquire some nontrivial braiding phase factor, and it may even be acted by a unitary matrix in general. The statistics and braiding properties of the anyons are encoded in the $R$-symbol, which is also a unitary matrix acting in  the degenerate state space and is defined via the the following
diagram:
$$
 \raisebox{-0.5\height}{\begin{pspicture}[shift=-0.65](-0.1,-0.2)(1.5,1.2)
  \small
  \psset{linewidth=0.9pt,linecolor=black,arrowscale=1.5,arrowinset=0.15}
  \psline{->}(0.7,0)(0.7,0.43)
  \psline(0.7,0)(0.7,0.5)
 \psarc(0.8,0.6732051){0.2}{120}{240}
 \psarc(0.6,0.6732051){0.2}{-60}{35}
  \psline (0.6134,0.896410)(0.267,1.09641)
  \psline{->}(0.6134,0.896410)(0.35359,1.04641)
  \psline(0.7,0.846410) (1.1330,1.096410)	
  \psline{->}(0.7,0.846410)(1.04641,1.04641)
  \rput[bl]{0}(0.4,0){$c$}
  \rput[br]{0}(1.35,0.85){$b$}
  \rput[bl]{0}(0.05,0.85){$a$}
 \scriptsize
  \rput[bl]{0}(0.82,0.35){$\mu$}
  \end{pspicture}}
=\sum\limits_{\nu }\left[ R_{c}^{ab}\right] _{\mu \nu}
 \raisebox{-0.5\height}{\begin{pspicture}[shift=-0.65](-0.1,-0.2)(1.5,1.2)
  \small
  \psset{linewidth=0.9pt,linecolor=black,arrowscale=1.5,arrowinset=0.15}
  \psline{->}(0.7,0)(0.7,0.45)
  \psline(0.7,0)(0.7,0.55)
  \psline(0.7,0.55) (0.25,1)
  \psline{->}(0.7,0.55)(0.3,0.95)
  \psline(0.7,0.55) (1.15,1)	
  \psline{->}(0.7,0.55)(1.1,0.95)
  \rput[bl]{0}(0.4,0){$c$}
  \rput[br]{0}(1.4,0.8){$b$}
  \rput[bl]{0}(0,0.8){$a$}
 \scriptsize
  \rput[bl]{0}(0.82,0.37){$\nu$}
  \end{pspicture}}
  .
$$
The topological spin can be expressed using the $R$-symbol via
$$
\theta _{a}=\theta _{\overline{a}}
=\sum\limits_{c,\mu } \frac{d_{c}}{d_{a}}\left[ R_{c}^{aa}\right] _{\mu \mu }
= \frac{1}{d_{a}}
 \raisebox{-0.5\height}{\begin{pspicture}[shift=-0.5](-1.3,-0.6)(1.3,0.6)
\small
  \psset{linewidth=0.9pt,linecolor=black,arrowscale=1.5,arrowinset=0.15}
  \psarc[linewidth=0.9pt,linecolor=black] (0.7071,0.0){0.5}{-135}{135}
  \psarc[linewidth=0.9pt,linecolor=black] (-0.7071,0.0){0.5}{45}{315}
  \psline(-0.3536,0.3536)(0.3536,-0.3536)
  \psline[border=2.3pt](-0.3536,-0.3536)(0.3536,0.3536)
  \psline[border=2.3pt]{->}(-0.3536,-0.3536)(0.0,0.0)
  \rput[bl]{0}(-0.2,-0.5){$a$}
  \end{pspicture}}
$$

The $F$- and $R$-symbols satisfy strong constraints and have some ``gauge" freedom \ie two sets of $\{F, R\}$ data related by certain gauge transformations are physically equivalent. We remark that the $F$- and $R$-symbols can all be defined microscopically \cite{Kawagoe2019}.

\subsection{Global symmetry}
\label{subsec: global symmetry}

In the presence of global symmetry, anyons can display interesting phenomena including (1) anyon permutation and (2) symmetry fractionalization. We will review the concepts relevant to this paper below, and more comprehensive reviews can be found in, \eg Refs.~\cite{barkeshli2014,etingof2016tensor,Etingof2009}. In this subsection and the next, we will assume that all symmetries are internal symmetries, \ie they do not move the locations of the degrees of freedom. We will comment on how to deal with the case with lattice symmetry in Sec. \ref{subsec: crystalline equivalence principle}.

Before specifying any microscopic symmetry of interest, it is useful to first discuss the abstract topological symmetry group of a topological order, whose elements can be viewed as invertible maps that take one anyon into another, so that the fusion properties of the topological order are invariant. For unitary (anti-unitary) topological symmetry action, the braiding properties, encoded in the $F$- and $R$-symbols, is preserved (conjugated). Later in the paper we will see many examples of topological symmetries of various topological orders.

Now we consider a microscopic symmetry described by a group $G$. Suppose $R_{\bf g}$ is a symmetry action, where ${\bf g}$ labels an element of $G$. This action can change anyons into other types, for example, it change an anyon $a$ into another anyon $^{\bf g}a$. We use
\beq \label{eq: anyon permutation}
\rho_{\bf g}: a\rightarrow\ ^{\bf g}a
\eeq
to represent the anyon permutation pattern. Mathematically, we may say that $\rho_{\rm g}$ describes a group homomorphism from $G$ to the topological symmetry group of a topological order. From here we see why the notion of topological symmetry is important: It encodes all possible anyon permutation patterns.

However, $\rho_{\rm g}$ by itself is insufficient to fully characterize a symmetry-enriched topological order. Consider creating three anyons $a_1$, $a_2$ and $a_3$ from the ground states, such that $a_1\times a_2\rightarrow \bar a_3$, \ie these three anyons can fuse into the trivial anyon. After separating these anyons far away from each other, there are generically $N^{\bar a_3}_{a_1a_2}$ degenerate such states. What effect does $R_{\bf g}$ have when it acts on a state in this degenerate space, denoted by $|\Psi_{a_1, a_2, a_3}\ra$?

Since the state of a topological order is specified by the two pieces of information, the global one and the local one, the symmetry localization ansatz states that \cite{Essin2012}
\beq \label{eq: symmetry localization}
\begin{aligned}
&R_{\bf g}|\Psi_{a_1, a_2, a_3}\ra\\
&=V_{\bf g}^{(1)}V_{\bf g}^{(2)}V_{\bf g}^{(3)}U_{\bf g}(^{\bf g}a_1, ^{\bf g}a_2; ^{\bf g}\bar{a}_3)|\Psi_{^{\bf g}a_1, ^{\bf g}a_2, ^{\bf g}a_3}\ra
\end{aligned}
\eeq
where $V_{\bf g}^{(i)}$ is a local unitary operation supported only around the anyon $a_i$, for $i=1,2,3$, and $U_{\rm g}(^{\bf g}a_1, ^{\bf g}a_2; ^{\bf g}\bar{a}_3)$ is a unitary matrix with rank $N^{\bar a_3}_{a_1a_2}$, which acts on the degenerate state space and describes the symmetry action on the global part of the information contained in the state. Notice that the state appearing on the right hand side is $|\Psi_{^{\bf g}a_1, ^{\bf g}a_2, ^{\bf g}a_3}\ra$, \ie generically the anyons are permuted by the symmetry.

It can be shown that the local operations $V$ satisfy
\beq
\eta_{a_i}({\bf g}, {\bf h})V_{\bf gh}^{(i)}|\Psi_{a_1, a_2, a_3}\ra=R_{\bf g}V_{\bf h}^{(i)}R_{\bf g}^{-1}V_{\bf g}^{(i)}|\Psi_{a_1,a_2,a_3}\ra
\eeq
for a pair of group elements ${\bf g}$ and ${\bf h}$. Here $\eta_{a_i}({\bf g}, {\bf h})$ are generically nontrivial phase factors, implying that the anyons may carry fractional charge or projective quantum number under the symmetry, \ie there can be symmetry fractionalization.{\footnote{In the context of symmetry fractionalization, it is well known that the notion of fractional charge or projective quantum number is generically not the same as the projective representations more commonly discussed in mathematics, which are classified by $H^2(G, U(1))$.}}

For a given topological order and a symmetry group $G$, it turns out that the data $\{\rho_{\bf g}; U_{\bf g}(a,b;c), \eta_a({\bf g}, {\bf h})\rbrace$ completely characterizes how this topological order is enriched by the symmetry $G$. We will often call $U_{\bf g}(a, b; c)$ the $U$-symbol, and $\eta_a({\bf g}, {\bf h})$ the $\eta$-symbol. This data $\{\rho_{\bf g}; U_{\bf g}(a,b;c), \eta_a({\bf g}, {\bf h})\rbrace$ also satisfies strong constraints and have some ``gauge" freedom, \ie two sets of $\{\rho_{\bf g}; U_{\bf g}(a,b;c), \eta_a({\bf g}, {\bf h})\rbrace$ data related by certain gauge transformations are physically equivalent. These gauge transformations will not be explicitly used in this paper, but they are summarized in, \eg Ref. \cite{Ye2022}. Moreover, even if two sets of data $\{\rho_{\bf g}; U_{\bf g}(a,b;c), \eta_a({\bf g}, {\bf h})\rbrace$ are not related by a gauge transformation, they still correspond to the physical state if they are related to each other by anyon relabeling that preserves the fusion and braiding properties \cite{Essin2012, barkeshli2014}, \ie such relabeling is precisely a unitary element in the topological symmetry group.{\footnote{Namely, for any unitary element ${\bf k}_0$ in the topological symmetry group (which does not have to be a microscopic symmetry), we can relabel each anyon $a$ by $^{{\bf k}_0}a$ (under this relabeling the data $\{F^{abc}_d, R^{ab}_c\}$ is the same as $\{F^{^{{\bf k}_0}a^{{\bf k}_0}b^{{\bf k}_0}c}_{^{{\bf k}_0}d}, R^{^{{\bf k}_0}a^{{\bf k}_0}b}_{^{{\bf k}_0}c}\}$ up to gauge transformations). Then the symmetry-enriched topological order with data $\{\rho_{\bf g}: a\rightarrow\,  ^{\bf g} a; U_{\bf g}(a, b; c), \eta_{a}({\bf g}, {\bf h})\}$ is the same as the one with data $\{\rho'_{\bf g}: a\rightarrow\, ^{{\bf k}_0}\left[^{\bf g}\left(^{\overline{{\bf k}_0}}a\right)\right]; U'_{\bf g}(a, b; c)=U_{\bf g}(^{\overline{{\bf k}_0}}a, ^{\overline{{\bf k}_0}}b; ^{\overline{{\bf k}_0}}c), \eta'_a({\bf g}, {\bf h})=\eta_{^{\overline{{\bf k}_0}}a}({\bf g}, {\bf h})\}$.}}

For a given anyon permutation pattern, one can show that distinct symmetry fractionalization classes form a torsor over $H^2_{\rho}(G, \mathcal{A})$. Namely, different possible symmetry fractionalization classes can be related to each other by elements of $H^2_{\rho}(G, \mathcal{A})$, where $\mc{A}$ is an Abelian group whose group elements correspond to the Abelian anyons in this topological order, and the group multiplication corresponds to the fusion of these Abelian anyons. The subscript $\rho$ represents the permutation action of $G$ on these Abelian anyons. In particular, given an element $[t] \in H^2_{\rho}(G, \mathcal{A})$, we can go from one symmetry fractionalization class with data $\eta_a({\bf g}, {\bf h})$ to another with data $\tilde{\eta}_a({\bf g}, {\bf h})$ given by
\begin{align}\label{eq:torsor}
\tilde{\eta}_a({\bf g}, {\bf h}) = \eta_a({\bf g}, {\bf h})  M_{a, t({\bf g},{\bf h})}
\end{align}
where $t({\bf g},{\bf h}) \in \mathcal{A}$ is a representative 2-cocyle for the cohomology class $[t]$ and $ M_{a, t({\bf g},{\bf h})} = \frac{\theta_{a\times t({\bf g},{\bf h})}}{\theta_a \theta_{t({\bf g},{\bf h})}}$ is the braiding statistics between $a$ and $t({\bf g},{\bf h})$ \cite{Bonderson2008}.

In the case where no anyon is permuted by any symmetry, there is always a canonical notion of a trivial symmetry fractionalization class, where $\eta_a({\bf g},{\bf h}) = 1$ for all anyon $a$ and all ${\bf g}, {\bf h} \in G$. In this case, an element of $H^2(G, \mathcal{A})$ is sufficient to completely characterize the symmetry fractionalization class.

Later in the paper, for a symmetry-enriched topological order we will often just specify $\rho_{\bf g}$, without explicitly specifying the $U$- and $\eta$-symbols. However, we will specify the $U$- and $\eta$-symbols of the topological symmetry group of this topological order, which allows us to determine the $U$- and $\eta$-symbols of the microscopic symmetry as follows.

Denote the microscopic symmetry group by $G$ and the topological symmetry group by $G_0$, then $\rho_{\bf g}$ defines a group homomorphism $\varphi: G\rightarrow G_0$. Denote the $U$- and a set of $\eta$-symbols of $G_0$ by $U^{(0)}_{\bf g_0}(a, b; c)$ and $\eta_a^{(0)}({\bf g}_0, {\bf h}_0)$, for any ${\bf g}_0, {\bf h}_0\in G_0$. These $U$- and $\eta$-symbols are some data intrinsic to the topological order, independent of the microscopic symmetry $G$, just like the $F$- and $R$-symbols. The $U$- and a set of $\eta$-symbols of the microscopic symmetry $G$ can be written as
\beq \label{eq: microscopic U and eta}
\begin{aligned}
&U_{\bf g}(a, b; c)=U_{\varphi(\bf g)}(a, b; c),\\
&\eta_a({\bf g}, {\bf h})=\eta_a(\varphi({\bf g}), \varphi({\bf h}))
\end{aligned}
\eeq
for any ${\bf g}, {\bf h}\in G$. Other symmetry fractionalization classes correponding to other sets of $\eta$-symbols can be related to this one via Eq.~\eqref{eq:torsor}.

\subsection{Anomalies of symmetry-enriched topological orders} \label{subsec: anomaly review}

A symmetry-enriched topological order can have a quantum anomaly. Roughly speaking, the anomaly characterizes the interplay between locality and the symmetry of the system. There are a few different definitions of quantum anomaly that are believed to be equivalent. In general, the anomaly can be viewed as an obstruction to gauging the symmetry, an obstruction to having a symmetric short-range entangled ground state, an obstruction to describing the system using a Hilbert space with a tensor product structure and on-site symmetry actions, or as the boundary manifestation of a higher dimensional bulk.

For a symmetry-enriched topological order, we can characterize its anomaly via anomaly indicators, a set of quantities expressed in terms of the data like $\{F^{abc}_d, R^{ab}_c, \rho_{\bf g}, U_{\bf g}(a, b; c), \eta_a({\bf g}, {\bf h})\}$. For example, consider a $\Z_2\times \Z_2$ symmetry. The anomalies of $(2+1)$-dimensional bosonic systems with a $\Z_2\times \Z_2$ symmetry are classified by $(\Z_2)^2$. Suppose the two generators of $\Z_2\times \Z_2$ are $C_1$ and $C_2$. The two anomaly indicators can be given by $\mc{I}_3(C_1, C_2)$ and $\mc{I}_3(C_2, C_1)$, where

\begin{widetext}
\beq\label{eq:indicator_with_z2z2 main}
\begin{aligned}
\mathcal{I}_3\left(C_1, C_2\right) = \frac{1}{D^2}\sum_{\substack{a,b,x,u\\ \mu\nu\tilde{\mu}\tilde\nu\rho\sigma\alpha\\
^{C_1}a = a\\
a \times b \times ^{C_1}b \rightarrow ^{C_2}a}}
&{d_b}\frac{\theta_x}{\theta_a}
\left(R_u^{b, ^{C_1}b}\right)_{\rho\sigma}  \left(F_{{^{C_2}}a}^{a,b,{^{C_1}}b}\right)^*_{(x,\tilde{\mu},\tilde{\nu})(u,\sigma,\alpha)}
\left(F_{{^{C_2}}a}^{a,{^{C_1}}b,b}\right)_{({^{C_1}}x,\mu,\nu)(u,\rho,\alpha)}\\
&\times U_{C_1}^{-1}(a,b;x)_{\tilde{\mu}\mu} U_{C_1}^{-1}(x,^{C_1}b;^{C_2}a)_{\tilde{\nu}\nu}\times \frac{1}{\eta_b(C_1, C_1)}\frac{\eta_a(C_2, C_1)}{\eta_a(C_1, C_2)}
\end{aligned}
\eeq   
\end{widetext}
and $\mc{I}_3(C_2, C_1)$ is obtained from the above equation by swapping $C_1\leftrightarrow C_2$ \cite{Ye2022}. The reason for the subscript of $\mc{I}_3$ can be seen in Appendix \ref{app: anomaly indicators}. The summation is over all anyon types $a$ and $b$ satisfying $^{C_1}a=a$ and that $^{C_2}a$ is a fusion outcome of $a\times b\times^{C_1}b$, $x$ also denotes anyon types, and the Greek letters index different ways of getting a particular fusion outcome (\eg $\mu=1, 2, \cdots, N^{^{C_2}a}_{a\,^{C_1}b}$). Both $\mc{I}_3(C_1, C_2)$ and $\mc{I}_3(C_2, C_1)$ are actually partition functions of a $3+1$ dimensional $\Z_2\times\Z_2$ symmetry-protected topological phase whose boundary is the relevant symmetry-enriched topological order, as discussed in detail in Ref. \cite{Ye2022}. These two anomaly indicators take values in $\pm 1$, and each set of values of these anomaly indicators specifies an element in the $(\Z_2)^2$ group, which classifies these anomalies.

Later in the paper we will discuss systems with symmetries different from $\Z_2\times\Z_2$, but it turns out that many anomaly indicators for topological orders with those other symmetries can be expressed via $\mc{I}_{1,2}$.

\subsection{Crystalline equivalence principle} \label{subsec: crystalline equivalence principle}

In the above two subsections our focus is topological orders with purely internal symmetries. But as mentioned in the Introduction, one of the main goals of this paper is to classify topological quantum spin liquids enriched by a general symmetry, which may contain both lattice symmetry and internal symmetry. The crystalline equivalence principle \cite{Song2016, Else2018} provides a convenient way to describe a topological order with lattice symmetry (and possibly also internal symmetry) using a topological order with a purely internal symmetry.

More concretely, the crystalline equivalence principle asserts that for each topological phase with a symmetry group $G$, where $G$ may contain both lattice symmetry and internal symmetry, there is a corresponding topological phase with only internal symmetries, where the symmetry group is still $G$, and all orientation reversing symmetries in the original topological phase should be viewed as anti-unitary symmetries in the corresponding topological phase. For example, Appendix \ref{app: reflection and time reversal} explains how to translate the data characterizing a symmetry-enriched topological order with reflection symmetry into the data for a time reversal symmetry-enriched topological order.

Strictly speaking, the above statement only applies to bosonic systems, which is the focus of the present paper. The fermionic version of this statement is still under development (see Refs. \cite{Cheng2018, Deb2021, Zhang2022, Manjunath2022} for recent progress).

\section{Symmetry properties and quantum anomalies of lattice systems}
\label{sec: anomaly of lattice}

\begin{figure}
    \centering
    \includegraphics[width=0.45\textwidth]{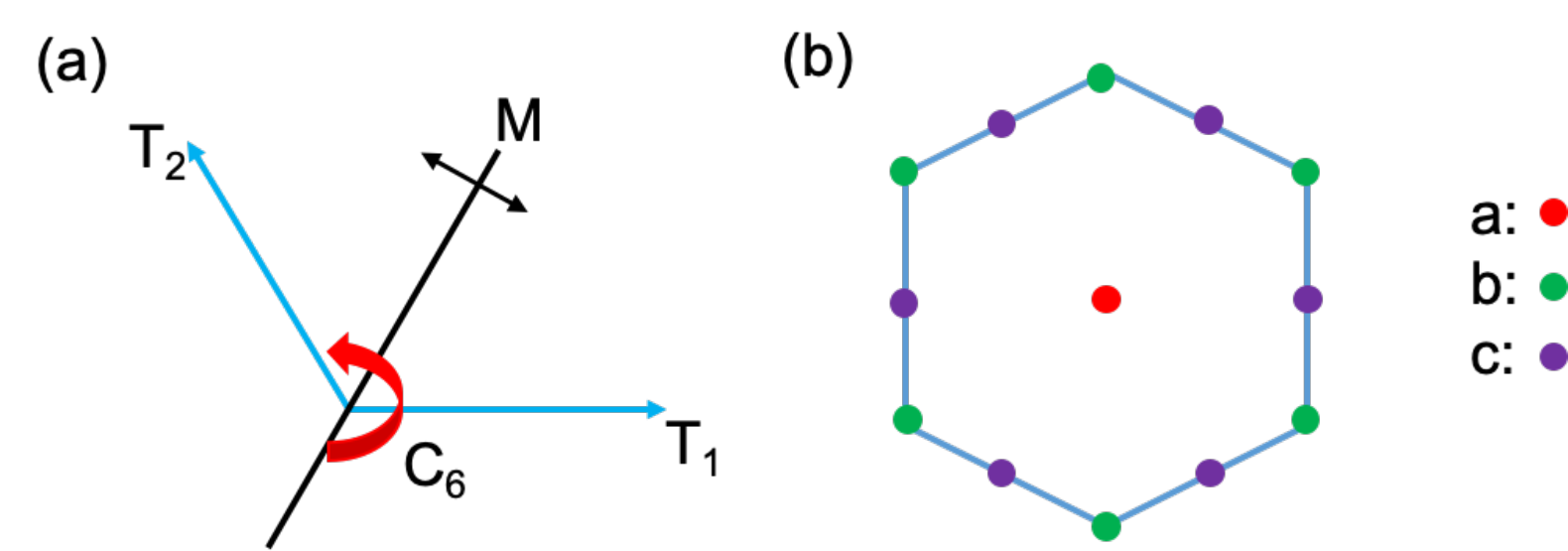}
    \caption{Panel (a) shows the generators of the $p6m$ group, including translations $T_1$ and $T_2$, a 6-fold rotation $C_6$ and a mirror reflection $M$. The two translation vectors have the same length, and their angle is $2\pi/3$. The reflection axis of $M$ bisects these two translation vectors. The $p6$ symmetry is generated by $T_1$, $T_2$ and $C_6$. Namely, $p6$ has no $M$ compared to $p6m$. In panel (b), the hexagon is a translation unit cell of either $p6m$ or $p6$ lattice symmetry. There are three types of high symmetry points, labelled by $a$, $b$ and $c$, and they form the sites of the triangular, honeycomb and kagome lattices, respectively. The $C_6$ rotation center in panel (a) is at a type-$a$ point.}
    \label{fig:p6m}
\end{figure}

\begin{figure}
    \centering
    \includegraphics[width=0.45\textwidth]{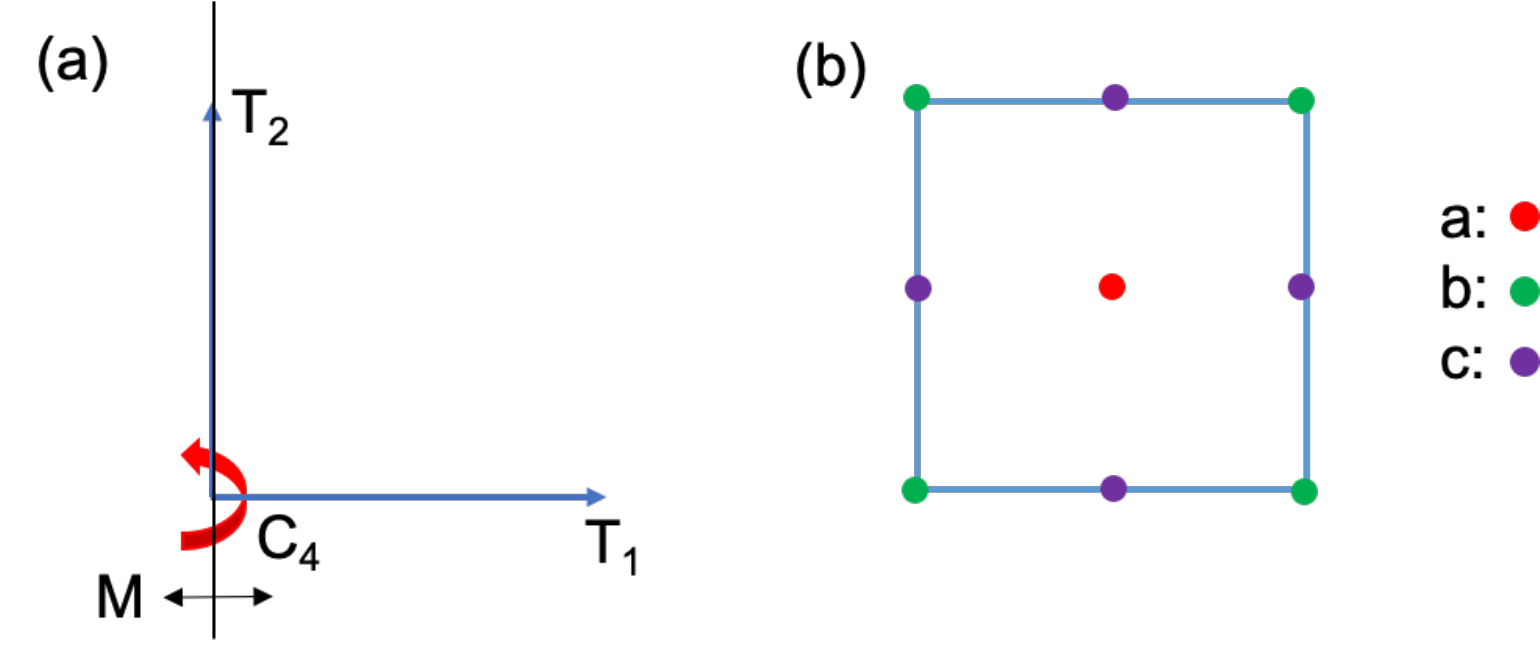}
    \caption{Panel (a) shows the generators of the $p4m$ group, including translations $T_1$ and $T_2$, a 4-fold rotation $C_4$ and a mirror reflection $M$. The two translation vectors have the same length, and their angle is $\pi/2$. The reflection axis of $M$ is parallel to the translation vector of $T_2$. The $p4$ symmetry is generated by $T_1$, $T_2$ and $C_4$. Namely, $p4$ has no $M$ compared with $p4m$. In panel (b), the square is a translation unit cell of either $p4m$ or $p4$ lattice symmetry. There are three high symmetry points, labelled by $a$, $b$ and $c$. Both type-$a$ and type-$b$ points form a square lattice, and type-$c$ points form a checkerboard lattice. The $C_4$ rotation center in panel (a) is taken to be at a type-$a$ point.}
    \label{fig:p4m}
\end{figure}

In the Introduction we mentioned that the three pieces of robust symetry-related information of a lattice system can be encoded in its quantum anomaly. In this section, we make this notion more precise. Although this idea is general, for concreteness, we focus on lattice spin systems in two spatial dimensions with one of these six symmetry groups: $p6\times SO(3)$, $p4\times SO(3)$, $p6m\times\Z_2^T$, $p4m\times\Z_2^T$, $p6m\times SO(3)\times\Z_2^T$, and $p4m\times SO(3)\times\Z_2^T$. Here $p6$, $p4$, $p6m$ and $p4m$ are lattice symmetry groups, whose definitions are explained in Figs. \ref{fig:p6m} and \ref{fig:p4m}. These lattice symmetries are assumed to only move the locations of the microscopic degrees of freedom, without acting on their internal states, \ie there is no spin-orbit coupling. The $SO(3)$ and $\Z_2^T$ are on-site spin rotational symmetry and time reversal symmetry, respectively. These symmetry settings are relevant to many theoretical, experimental and numerical studies, and the examples we consider in the later part of the paper will also be based on these symmetry settings.

Given such a symmetry group, different lattice systems can be organized into the so-called lattice homotopy classes \cite{Po2017, Else2019}. Two lattice systems are in the same lattice homotopy class if and only if they can be deformed into each other by these operations while preserving the lattice symmetry: 1) moving the microscopic degrees of freedom, 2) identifying degrees of freedom with the same type of projective representation under the on-site symmetry, and 3) adding or removing degrees of freedom with linear representation (\ie trivial projective representation) under the on-site symmetry. Lattice systems within the same class share the same robust symmetry-related properties, while those in different classes have distinct symmetry properties and cannot be smoothly connected without breaking the symmetry. So the robust symmetry-related information of a lattice system is the lattice homotopy class it belongs to, while colloquially it is the three pieces of information mentioned in the Introduction.

To make the above discussion less abstract, consider systems with $p6\times SO(3)$ symmetry. From Fig. \ref{fig:p6m}, there are three types of high symmetry points, forming a triangular, honeycomb and kagome lattice, respectively. The on-site symmetry $SO(3)$ has two types of projective representations: half-odd-integer spins and integer spins. According to the above discussion, although spin-1/2 systems defined on triangular, honeycomb and kagome lattices have the same symmetry group, they are in different lattice homotopy classes and have sharply distinct symmetry properties, because they cannot be deformed into each other via the above operations.

Below we enumerate all lattice hotomopy classes in our symmetry settings. To this end, we first specify the types of projective representations under the internal symmetries we consider. As mentioned above, there are two types of projective representations for the $SO(3)$ symmetry. For time reversal symmetry $\Z_2^T$, there are also two types of projective representations: Kramers singlet and Kramers doublet. For symmetry $SO(3)\times\Z_2^T$, there are actually four types of projective representations: integer spin under $SO(3)$ while Kramers singlet under $\Z_2^T$, half-odd-integer spin under $SO(3)$ while Kramers doublet under $\Z_2^T$, half-odd-integer spin under $SO(3)$ while Kramers singlet under $\Z_2^T$, and integer spin under $SO(3)$ while Kramers doublet under $\Z_2^T$. The first two types of projective representations are more common in physical systems and theoretical models than the last two,{\footnote{The last two cases can appear in the following (less common) scenarios. To have half-odd-integer spin under $SO(3)$ while Kramers singlet under $\Z_2^T$, consider a spin-1/2 system with the usual $SO(3)$ spin rotational symmetry, while the $\Z_2^T$ acts simply by complex conjugation. To have integer spin under $SO(3)$ while Kramers doublet under $\Z_2^T$, consider a system made of 1) Kramers doublet degrees of freedom that do not transform under the $SO(3)$ spin rotational symmetry and 2) the usual spin-1 degrees of freedom that carry integer spin under $SO(3)$ while Kramers singlet under $\Z_2^T$.}} so below we will only consider the first two. Therefore, for all three types of internal symmetries we consider, \ie $SO(3)$, $\Z_2^T$ and $SO(3)\times\Z_2^T$, there is a trivial projective representation and a nontrivial one under consideration.

Then for lattice systems with symmetry group being either $p6\times SO(3)$, $p6m\times\Z_2^T$ or $p6m\times SO(3)\times\Z_2^T$, there are 4 different lattice homotopy classes \cite{Po2017}:

\begin{enumerate}

\item Class ``0". A representative configuration: A system with degrees of freedom only carrying the trivial projective representation under the internal symmetry.

\item Class ``a". A representative configuration: A system with degrees of freedom carrying the nontrivial projective representation under the internal symmetry, which locate at the triangular lattice sites (type-$a$ high symmetry points in Fig. \ref{fig:p6m}).

\item Class ``c". A representative configuration: A system with degrees of freedom carrying the nontrivial projective representation under the internal symmetry, which locate at the kagome lattice sites (type-$c$ high symmetry points in Fig. \ref{fig:p6m}).

\item Class ``a+c". A representative configuration: A system with degrees of freedom carrying the nontrivial projective representation under the internal symmetry, which locate at both the triangular and kagome lattice sites (both type-$a$ and type-$c$ high symmetry points in Fig. \ref{fig:p6m}).
\end{enumerate}
Note that a system with degrees of freedom carrying nontrivial projective representation that locate at the honeycomb lattice sites (type-$b$ high symmetry points) is in class 0 \cite{Po2017}.

For lattice systems with symmetry group being either $p4\times SO(3)$, $p4m\times\Z_2^T$ or $p4m\times SO(3)\times\Z_2^T$, there are 8 different lattice homotopy classes. Using labels similar to the above, these are classes ``0", ``a", ``b", ``c", ``a+b", ``a+c", ``b+c" and ``a+b+c", respectively, where each label represents the type of the high symmetry points at which the degrees of freedom carrying the nontrivial projective representation locate (``0" means all degrees of freedom carry the regular representation, \ie the trivial projective representation). Note that type-$a$ and type-$c$ high symmetry points are physically distinct once the $C_4$ rotation center is specified, although they may look identical at the first glance.

To turn the above picture into useful mathematical formulations, Ref. \cite{Ye2021a} shows how to characterize each lattice homotopy class using its quantum anomaly, which in this context is also known as Lieb-Schultz-Mattis anomaly. Different lattice homotopy classes have different anomalies, and the lattice homotopy class 0 has a trivial anomaly. In the context of topological orders, the anomalies can be expressed via the anomaly indicators. For topological quantum spin liquids with $p6\times SO(3)$ symmetry, the anomaly indicators are given in Appendix \ref{app: anomaly indicators}:
\beq \label{eq: anomaly indicator p6 x SO(3)}
\begin{aligned}
\mathsf{I}_1 &= \mc{I}_3(C_2U_\pi, C_2 U'_\pi)\,,\\   
\mathsf{I}_2 &= \mc{I}_3(T_1C_2U_\pi, T_1C_2 U'_\pi)\,.\\    
\end{aligned}
\eeq
where the expression of $\mc{I}_3$ is given by Eq. \eqref{eq:indicator_with_z2z2 main}, $C_2$ is a 2-fold rotation symmetry (\ie $C_2\equiv C_6^3$), while $U_\pi$ and $U'_\pi$ are $\pi$ spin rotations around two orthogonal axes. We can think of $\mathsf{I}_1$ and $\mathsf{I}_2$ as respectively detecting half-odd-integer spins at type-$a$ and type-$c$ high symmetry points, which are respectively the 2-fold rotation centers of the $C_2$ and $T_1C_2$ symmetries. More generally, the values of these anomaly indicators for the 4 lattice homotopy classes enumerated above are shown in Table \ref{tab: anomaly indicators p6 x SO(3)}.
\begin{table}[!htbp]
\centering
\renewcommand{\arraystretch}{1.3}
\begin{tabular}{c|c|c|c|c}
\toprule[2pt]
 & 0 & a & c & a+c \\ \hline
$\mathsf{I}_1$ & 1 & $-1$ & 1 & $-1$\\
$\mathsf{I}_2$ & 1 & $1$ & $-1$ & $-1$ \\
\bottomrule[2pt]
\end{tabular}
\caption{Values of the anomaly indicators for the 4 lattice homotopy classes with symmetry group $p6\times SO(3)$.}
\label{tab: anomaly indicators p6 x SO(3)}
\end{table}

For topological quantum spin liquids with $p4\times SO(3)$ symmetry, the anomaly indicators are
\beq \label{eq: anomaly indicator p4 x SO(3)}
\begin{aligned}
\mathsf{I}_1 &= \mc{I}_3(C_2U_\pi, C_2 U'_\pi)\,,\\    
\mathsf{I}_2 &= \mc{I}_3(T_1T_2C_2U_\pi, T_1T_2C_2 U'_\pi)\,,\\    
\mathsf{I}_3 &= \mc{I}_3(T_1C_2U_\pi, T_1C_2 U'_\pi)\,,\\    
\end{aligned}
\eeq
where $C_2$ is still a 2-fold rotational symmetry (but $C_2=C_4^2$ in this case), while $U_\pi$ and $U'_\pi$ are still $\pi$ spin rotations around two orthogonal axes.
We can think of $\mathsf{I}_1$, $\mathsf{I}_2$ and $\mathsf{I}_3$ as respectively detecting half-odd-integer spins at type-$a$, type-$b$ and type-$c$ high symmetry point, which are respectively the 2-fold rotation centers of the $C_2$, $T_1T_2C_2$ and $T_1C_2$ symmetries. More generally, the values of these anomaly indicators for the 8 lattice homotopy classes enumerated above are shown in Table \ref{tab: anomaly indicators p4 x SO(3)}.
\begin{table}[!htbp]
\centering
\renewcommand{\arraystretch}{1.3}
\begin{tabular}{c|c|c|c|c|c|c|c|c}
\toprule[2pt]
 & 0 & a & b & c & a+b & a+c & b+c & a+b+c \\ \hline
$\mathsf{I}_1$ & 1 & $-1$ & 1 & $1$ & $-1$ & $-1$ & 1 & $-1$\\
$\mathsf{I}_2$ & 1 & $1$ & $-1$ & $1$ & $-1$ & 1 & $-1$ & $-1$\\
$\mathsf{I}_3$ & 1 & $1$ & $1$ & $-1$ & 1 & $-1$ & $-1$ & $-1$\\
\bottomrule[2pt]
\end{tabular}
\caption{Values of the anomaly indicators for the 8 lattice homotopy classes with symmetry group $p4\times SO(3)$.}
\label{tab: anomaly indicators p4 x SO(3)}
\end{table}

The anomaly indicators for the other symmetry groups we consider (\ie $p6m\times SO(3)\times\Z_2^T$, $p4m\times SO(3)\times\Z_2^T$, $p6m\times\Z_2^T$ and $p4m\times\Z_2^T$) and their values in each lattice homotopy classes are more complicated, and they are presented in Appendix \ref{app: anomaly indicators}.

We remark that, strictly speaking, Eqs. \eqref{eq: anomaly indicator p6 x SO(3)} and \eqref{eq: anomaly indicator p4 x SO(3)} are anomaly indicators of topological quantum spin liquids with purely internal $p6\times SO(3)$ or $p4\times SO(3)$ symmetry, but the crystalline equivalence principle discussed in Sec. \ref{sec: framework} still allows us to use them to classify symmetry-enriched topological quantum spin liquids.

Moreover, it is straightforward to generalize the above idea to other types of systems. For instance, to get the anomaly of a system with spin-orbit coupling, whose symmetry is just a subgroup of the symmetry of a system without spin-orbit coupling, one can simply start from the anomaly of a system without spin-orbit coupling and restrict the symmetry to that subgroup.

\section{Framework of classification} \label{sec: framework}

Now we are ready to present our framework to classify symmetry-enriched topological quantum spin liquids.

Our framework is based on the hypothesis of emergibility \cite{Ye2021a, Zou2021}. Namely, suppose the anomaly of the lattice system is $\omega$, then, by tuning the parameters of this lattice system, a quantum many-body state (or its low-energy effective theory) with anomaly $\Omega$ can emerge at low energies if and only if the anomaly-matching condition holds: $\omega=\Omega$.

The ``only if" part of this statement is established and well known \cite{Hooft1980}. The ``if" part is hypothetical, but there is no known counterexample to it and it is supported by multiple nontrivial examples \cite{Kimchi2012, Jian2015, Kim2015, Latimer2020}. So we will assume this hypothesis to be true and use it as our basis of analysis.

With this hypothesis, the framework to classify symmetry-enriched topological quantum spin liquids, or equivalently, to obtain the possible symmetry-enriched topological quantum spin liquids that can emerge in the lattice system of interest, is as follows.

\begin{enumerate}

\item Given the symmetry group, which may contain both lattice symmetry and internal symmetry, we first use the crystalline equivalence principle in Sec. \ref{subsec: crystalline equivalence principle} to translate it into a purely internal symmetry.

\item Based on the above internal symmetry and the topological quantum spin liquid, we use the method in Ref. \cite{barkeshli2014} to obtain the classification of the internal symmetry enriched topological quantum spin liquids.

\item For each of the internal symmetry enriched topological quantum spin liquids, we use the method in Ref. \cite{Ye2022} to obtain its anomaly, $\Omega$.

\item As discussed in Sec. \ref{sec: anomaly of lattice}, the original lattice system has its own quantum anomaly, $\omega$. We check if the anomaly-matching condition, $\omega=\Omega$, holds. If it does (doesn't), then the corresponding symmetry-enriched topological quantum spin liquid can (cannot) emerge in this lattice system, according to the hypothesis of emergibility.

\end{enumerate}

If there is only an internal symmetry but no lattice symmetry, then step 1 in the framework can be ignored. In this case, sometimes one is only interested in anomaly-free states, then $\omega$ in step 4 should be the trivial anomaly. For steps 3 and 4, in our context $\Omega$ ($\omega$) can be represented by the values of the anomaly indicators of the topological quantum spin liquid (lattice system), so checking whether $\omega=\Omega$ becomes checking whether the values of these two sets of anomaly indicators match. For instance, as discussed in Sec. \ref{sec: anomaly of lattice}, the anomaly $\omega$ for lattice systems with $p6\times SO(3)$ symmetry can be specified by the values of $\mathsf{I}_{1,2}$ defined in Eq. \eqref{eq: anomaly indicator p6 x SO(3)}, and these values for different lattice homotopy classes are given in Table \ref{tab: anomaly indicators p6 x SO(3)}. The anomaly $\Omega$ or the corresponding anomaly indicators $\mathsf{I}_{1,2}$ for a $p6\times SO(3)$ symmetric topological quantum spin liquid can be calculated using the results in Ref. \cite{Ye2022}, and we will give detailed analysis below (see Eqs. \eqref{eq: U(1) level 2N p6xSO(3) 1 anomaly calculation} and \eqref{eq: U(1) level 2N p6xSO(3) 2 anomaly calculation} for some of the final results of the calculations).

We reiterate that the above framework can be straightforwardly generalized to classify quantum states other than symmetry-enriched topological quantum spin liquids. For example, it has been used to classify some gapless quantum spin liquids in Ref. \cite{Ye2021a}.

In the following sections, we will apply the above framework to obtain the classification of some representative two dimensional symmetry-enriched topological quantum spin liquids on various lattice systems.

\section{\texorpdfstring{$\U_{2N}$}{} topological orders: Generalized Abelian chiral spin liquids} \label{sec: U(1) level 2N}

Our first class of examples are topological quantum spin liquids with $U(1)_{2N}$ topological orders. These are Abelian chiral  states, where the $N=1$ case is the well-known Kalmeyer-Laughlin state \cite{Kalmeyer1987, Kalmeyer1989}, the $N=2$ case is the $\nu=2$ state in Kitaev's 16-fold way \cite{Kitaev2006}, and a general $U(1)_{2N}$ topological order can be obtained by putting bosons into an interacting bosonic integer quantum Hall state with Hall conductance $2N$ (in natural units) and coupling them to a dynamical $U(1)$ gauge field \cite{Senthil2012, Barkeshli2013, He2015, Zou2018}. We will classify the $U(1)_{2N}$ topological quantum spin liquid enriched by $p6\times SO(3)$ or $p4\times SO(3)$ symmetry. As discussed in Sec. \ref{subsec: crystalline equivalence principle}, these symmetries can be viewed as purely internal symmetries according to the crystalline equivalence principle. Our results are summarized in Table \ref{tab:RealizationNumber U(1) level 2N}.

The topological properties of the $U(1)_{2N}$ topological order can be described by either the Laughlin-$(1/{2N})$ wave function or a Chern-Simons theory with Lagrangian $\mc{L}=-\frac{2N}{4\pi}\epsilon^{\mu\nu\lambda}A_\mu\partial_\nu A_\lambda$, with $A$ a dynamical $U(1)$ gauge field. These states also allow a description using non-interacting parton mean field. Specifically, one can consider $2N$ species of fermionic partons with an $SU(2N)$ gauge structure. When all species are in a Chern band with a unit Chern number, the resulting state is the $U(1)_{2N}$ topological order.{\footnote{The special case with $N=1$ allows another parton mean field description, described by fermionic partons with a $U(1)$ gauge structure. When these fermionic partons are in a Chern band with Chern number 2, the resulting state is the $U(1)_2$ topological order. The special case with $N=2$ also allows another parton mean field description, described by fermionic partons with a $\Z_2$ gauge structure. When these fermionic partons form a $d+id$ superconductor, the resulting state is the $U(1)_4$ topological order \cite{Kitaev2006}.}}

The above descriptions of these topological quantum spin liquids all suffer from some disadvantages. Concretely, the Laughlin wave function is a single specific state, and it cannot describe different symmetry-enriched states. To capture the symmetry actions in the Chern-Simons theory, one needs to invoke the concept of 2-group symmetries \cite{Benini2019}, which are not exact symmetries of the physical system. Also, in the $SU(2N)$ parton mean field description, the projective quantum number of the fermionic partons are not exactly the same as the symmetry fractionalization class of the anyons.

Below we will discuss the topological properties of these states in the language of Sec. \ref{sec: characterization}, which does not suffer from the above disadvantages, since it can describe general symmetry-enriched $U(1)_{2N}$ topological quantum spin liquids directly in terms of the symmetry properties of the anyons.

We label anyons in $\U_{2N}$ by $(a)$, where $a$ is an element in $\{0,\dots,2N-1\}$. These are all Abelian anyons with $d_{(a)}=1$. The fusion rule is given by addition modulo $2N$, \ie
\beq
(a)\otimes (b)=([a+b]_{2N})\,
\eeq
In this paper, we use the notation $[x]_y$ to denote $x$ modulo $y$ for any integer $x$ and positive integer $y$, and $[x]_y$ takes values in $\{0,\dots, y-1\}$.
The $F$-symbols can be written as
\begin{equation} \label{eq: F U(1) level 2N}
F^{(a)(b)(c)}=e^{\frac{i\pi}{2N}a(b+c-[b+c]_{2N})},
\end{equation}
the $R$-symbols are
\begin{equation} \label{eq: R U(1) level 2N}
R^{(a)(b)}=e^{\frac{i\pi}{2N}ab}
\end{equation}
which yield the topological spins:
\beq \label{eq: theta of U(1) level 2N}
\theta_{(a)}=e^{\frac{i\pi}{2N} a^2}
\eeq

The topological symmetry of $\U_{2N}$ is complicated for general $N$ \cite{Delmastro2019}. For $N=1$, there is no nontrivial topological symmetry. For $N\geqslant 2$, there is always a $\Z_2$ topological symmetry generated by the charge conjugation symmetry $C$, such that anyon $(a) \rightarrow ([-a]_{2N})$ under $C$. For this topological symmetry, we can take the $U$-syombols as
\begin{align}\label{eq: U U(1) level 2N}
U_{C}((a),(b); ([a+b]_{2N})) = 
\begin{cases}
(-1)^{a}, & b>0\\
1, & b=0
\end{cases}
\end{align}
and a set of $\eta$-symbols all equal to 1. From this set of $\eta$-symbols, we can obtain all other possible $\eta$-symbols via Eq. \eqref{eq:torsor}. When $2\leqslant N \leqslant 5$, this $\Z_2$ is the full topological symmetry group. When $N\geqslant 6$, there can be other topological symmetries.{\footnote{For example, when $N=6$, the action that takes $(a)\rightarrow ([5a]_{12})$ is a unitary $\Z_2$ topological symmetry. One can easily check that this map preserves the fusion and braiding properties of the $U(1)_{12}$ state. Furthermore, a consistent set of $U$- and $\eta$-symbols can indeed be constructed for this symmetry.}} In the later discussion we consider general $N\geqslant 2$, but limit to the cases where the microscopic symmetry can permute anyons only as charge conjugation (\ie we ignore anyon permutation patterns other than $C$, if any). To read off the results for $N=1$ from those for $N\geqslant 2$, we just need to ignore the cases where the microscopic symmetry permutes anyons.

\subsection{Example: $\Z_2\times SO(3)$}\label{subsec: Z2}

To illustrate our calculation of the anomaly of the $U(1)_{2N}$ topological order with $p6\times SO(3)$ or $p4\times SO(3)$ symmetry, let us first discuss the example where the symmetry is $\Z_2\times SO(3)$ in detail. It turns out that the calculation of the anomaly when the symmetry is $p6\times SO(3)$ or $p4\times SO(3)$ can be reduced to this example, by restricting $p6$ or $p4$ to its various $\Z_2$ subgroups.

The anomalies associated with the $\Z_2\times SO(3)$ symmetry are classified by 
\beq
H^4(\Z_2\times SO(3), \U)\cong \Z_2\,.
\eeq
Hence there is only one type of nontrivial anomaly, which can be detected by the anomaly indicator $\mathsf{I} = \mc{I}_3(C_2 U_\pi, C_2 U'_\pi)$, where $C_2$ is the generator of $\Z_2$, while $U_\pi$ and $U'_\pi$ are elements of $SO(3)$, representing the $\pi$-rotations about two orthogonal axes.

The $SO(3)$ symmetry cannot permute anyons because all elements of $SO(3)$ are continuously connected to the identity element. Hence, only the generator of $\Z_2$, denoted by $C_2$ here, can permute anyons by charge conjugation, and there are two possibilities.

\subsubsection{No anyon permutation} \label{subsubsec: no permutation}

The first possibility is that the action of $C_2$ is trivial and there is no anyon permutation. For the case with $N=1$, this is the only possibility to be considered. Then the symmetry fractionalization classes are classified by 
\beq
\begin{aligned}
& H^2_{(1)}(\Z_2\times SO(3), \Z_{2N})\\ = & H^2(\Z_2, \Z_{2N}) \oplus H^2(SO(3), \Z_{2N}) \\
= & (\Z_2)^2\,.
\end{aligned}
\eeq 
Namely, there are 2 generators that generate 4 different symmetry fractionalization classes. To understand these symmetry fractionalization classes, we can directly write down representative cochains of them. A representative cochain of the first generator, which we denote by $\widetilde{\beta}(x)$ and comes from $H^2(\Z_2, \Z_{2N})$, is:
\beq
\widetilde{\beta}(x)(C_2^i, C_2^j) = \frac{i + j - [i+j]_2}{2} = ij \mod {2N} \,,
\eeq
with $i,j\in\{0,1\}$. The reason for the name of this generator is explained in Appendix \ref{app:SFclasses}. Physically, this generator detects whether the anyon $(1)$ carries a fractional charge under the $\Z_2$ symmetry.

The second generator, which comes from $H^2(SO(3), \Z_{2N})$, detects whether the anyon $(1)$ carries a half-odd-integer spin under the $SO(3)$ symmetry, and we denote it by $N w_2$, for reasons explained in Appendix \ref{app:SFclasses}. To have a representative cochain of $Nw_2$, it is convenient to consider a $\Z_2\times \Z_2$ subgroup of $SO(3)$ generated by $U_\pi$ and $U'_\pi$, and an element in this subgroup can be written as $U_\pi^i {U'}_\pi^j$, with $i, j\in\{0, 1\}$. Then, restricting $SO(3)$ to this $\Z_2\times\Z_2$ subgroup, the representative cochain of $Nw_2$ is
\beq\label{eq:rep_cochain_w2}
\begin{aligned}
&\left(Nw_2\right)(U_\pi^{i_1} {U'}_\pi^{i_2}, U_\pi^{j_1} {U'}_\pi^{j_2})\\
=& N(i_1j_1 + i_2j_2 + i_1j_2)
\mod {2N}\,.
\end{aligned}
\eeq

So the symmetry fractionalization classes can be written as
\beq \label{eq: adding generators}
w = n_1\cdot \widetilde{\beta}(x) + n_2 \cdot Nw_2\,,
\eeq
and labeled as $\{n_1, n_2\}$ with $n_{1, 2}\in\{0, 1\}$. When the $SO(3)$ symmetry is restricted to its $\Z_2\times\Z_2$ subgroup generated by $U_\pi$ and $U'_\pi$, a representative cochain can be taken as
\beq \label{eq: representative cochain Z2 x SO(3) 1}
\begin{aligned}
&w(C_2^{i_1}U_\pi^{i_2}{U'}_\pi^{i_3}, C_2^{j_1}U_\pi^{j_2}{U'}_\pi^{j_3})\\
=&n_1i_1j_1 + n_2N(i_2j_2 + i_3j_3 + i_2j_3) \mod {2N}
\end{aligned}
\eeq

Combining the above equation and Eq. \eqref{eq:torsor}, we get
\beq \label{eq: eta U(1) no permutation}
\begin{aligned}
&\eta_{(a)}(C_2 U_\pi, C_2 U'_\pi) = \eta_{(a)}(C_2 U_\pi, C_2 U_\pi) \\
&=
\exp\left(\frac{i\pi}{N}a(n_1 + Nn_2)\right),\\
&\eta_{(a)}(C_2 U'_\pi, C_2 U_\pi)
= \exp\left(\frac{i\pi}{N}an_1\right)
\end{aligned}
\eeq

Now we plug Eqs. \eqref{eq: F U(1) level 2N}, \eqref{eq: R U(1) level 2N}, \eqref{eq: theta of U(1) level 2N} and \eqref{eq: eta U(1) no permutation} into Eq. \eqref{eq:indicator_with_z2z2 main} (the $U$-symbols therein can all be taken as 1 since there is no anyon permutation), and get the anomaly indicator of the state in symmetry fractionalization class $\{n_1, n_2\}$:
\beq \label{eq: anomaly Z2 x SO(3) no permutation}
\mc{I}_3(C_2 U_\pi, C_2 U'_\pi)=(-1)^{n_1 n_2}\,.
\eeq

\subsubsection{$C_2$ acts as charge conjugation} \label{subsubsec: with permutation}

The second possibility is that $C_2$ acts by charge conjugation. This possibility occurs only if $N\geqslant 2$. Then the symmetry fractionalization classes are classified by 
\beq
H^2_{(2)}(\Z_2\times SO(3), \Z_{2N}) = (\Z_2)^2\,.
\eeq
There are also 2 generators that generate 4 different symmetry fractionalization classes, but these symmetry fractionalization classes are different from those in Sec. \ref{subsubsec: no permutation}. Explicitly, a representative cochain of the first generator, which we denote by $Nx^2$, is:
\beq
(Nx^2)(C_2^i, C_2^j) = Nij \mod {2N} \,,
\eeq
with $i,j\in\{0,1\}$. The second generator also detects whether the anyon $(1)$ carries a half-odd-integer spin under the $SO(3)$ symmetry, and we also denote it by $N w_2$. The representative cochain restricted to the $\Z_2\times\Z_2$ subgroup generated by $U_\pi$ and $U'_\pi$ is still given by Eq.~\eqref{eq:rep_cochain_w2}.

So the symmetry fractionalization classes can be written as
\beq
w = n_1\cdot Nx^2 + n_2 \cdot Nw_2\,,
\eeq
and also labeled as $\{n_1, n_2\}$ with $n_{1, 2}\in\{0, 1\}$. When the $SO(3)$ symmetry is restricted to its $\Z_2\times\Z_2$ subgroup generated by $U_\pi$ and $U'_\pi$, a representative cochain can be taken as
\beq \label{eq: representative cochain Z2 x SO(3) 2}
\begin{aligned}
&w(C_2^{i_1}U_\pi^{i_2}{U'}_\pi^{i_3}, C_2^{j_1}U_\pi^{j_2}{U'}_\pi^{j_3})\\
=&n_1Ni_1j_1 + n_2 N(i_2j_2 + i_3j_3 + i_2j_3) \mod {2N}
\end{aligned}
\eeq

Combining the above equation and Eq. \eqref{eq:torsor}, we get
\beq \label{eq: eta U(1) permutation}
\begin{aligned}
&\eta_{(a)}(C_2 U_\pi, C_2 U'_\pi) = \eta_{(a)}(C_2 U_\pi, C_2 U_\pi) \\
&=
\exp\left(i\pi a(n_1 + n_2)\right),\\
&\eta_{(a)}(C_2 U'_\pi, C_2 U_\pi)
= \exp\left(i\pi an_1\right)
\end{aligned}
\eeq

Now we plug Eqs. \eqref{eq: F U(1) level 2N}, \eqref{eq: R U(1) level 2N}, \eqref{eq: theta of U(1) level 2N}, \eqref{eq: U U(1) level 2N} and \eqref{eq: eta U(1) permutation} into Eq. \eqref{eq:indicator_with_z2z2 main}, and get the anomaly indicator of the state in symmetry fractionalization class $\{n_1, n_2\}$:
\beq \label{eq: anomaly Z2 x SO(3) permutation}
\mc{I}_3(C_2 U_\pi, C_2 U'_\pi)= (-1)^{(n_1 +1)n_2 N}\,.
\eeq
Hence when $N$ is even the anomaly is always absent, and when $N$ is odd, $n_1=0, n_2=1$ gives nonzero anomaly and otherwise the anomaly is absent.

With this warm-up, we are ready to classify $U(1)_{2N}$ topological quantum spin liquids enriched by $p6\times SO(3)$ or $p4\times SO(3)$ symmetry using the framework in Sec. \ref{sec: framework}. The results are summarized in Table \ref{tab:RealizationNumber U(1) level 2N}.

\subsection{$p6\times SO(3)$}

The generator $T_{1,2}$ and $SO(3)$ cannot permute anyons{\footnote{The generator $T_{1,2}$ cannot permute anyons because $C_6 T_2 C_6^{-1}=T_1^{-1}$, implying that $T_1$ and $T_2$ should permute anyons in opposite ways, which, combined with $C_6 T_1 C_6^{-1}=T_1T_2$, implies that neither $T_1$ nor $T_2$ can permute anyons.}}, and only the generator $C_6$ can permute anyons by charge conjugation. Hence, there are two possibilities regarding how $p6\times SO(3)$ can permute anyons:

\begin{enumerate}

\item Trivial $C_6$ action: no anyon permutation. 

In this case, the possible symmetry fractionalization classes are classified by 
\beq
H^2_{(1)}(p6\times SO(3), \Z_{2N}) = \Z_{2N}\oplus\Z_{(2N,6)}\oplus\Z_2\,,
\eeq
whose elements can be written as 
\beq \label{eq: element p6 SO(3) 1}
w = n_1\cdot \widetilde{\mathscr{B}}^{(1)}_{xy} + n_2\cdot  \widetilde{\mathscr{B}}^{(1)}_{c^2} + n_3\cdot Nw_2\,,
\eeq
and labeled by $\{n_1, n_2, n_3\}$, with $n_1\in\{0,\dots,2N-1\}$, $n_2\in\{0,\dots,(2N,6)-1\}$, $n_3\in\{0,1\}$. Here $\widetilde{\mathscr{B}}^{(1)}_{xy}$, $\widetilde{\mathscr{B}}^{(1)}_{c^2}$ and $Nw_2$ are generators of $\Z_{2N}$, $\Z_{(2N,6)}$ and $\Z_2$, respectively (the representative cochains and the reason for the names of these generators are given in Appendix \ref{app:SFclasses}). Physically, we can think of $\widetilde{\mathscr{B}}^{(1)}_{xy}$, $\widetilde{\mathscr{B}}^{(1)}_{c^2}$ and $Nw_2$ as detecting whether the anyon $(1)$ carries projective representation under translation symmetries, $C_6$ and $SO(3)$, respectively.{\footnote{In fact, the generator $A_c^2$ actually detects whether the anyon $(1)$ simutaneously carries projective representation under four different symmetries, generated by $C_6$, $T_1C_6^3$, $T_2C_6^3$ and $T_1T_2C_6^3$, respectively.}} For each symmetry fractionalization class, the $U$- and $\eta$-symbols can be obtained via Eqs. \eqref{eq:torsor}, \eqref{eq: microscopic U and eta} and \eqref{eq: U U(1) level 2N}.

Without considering anomaly matching, $p6\times SO(3)$ symmetric $U(1)_{2N}$ topological quantum spin liquids are classified by $\{n_1, n_2, n_3\}$, if no symmetry permutes anyons. Recall that two symmetry fractionalization classes related to each other by relabeling anyons are physically identical, so $\{n_1, n_2, n_3\}$ and $\{[-n_1]_{2N}, [-n_2]_{(2N, 6)}, n_3\}$ are identified.

As argued in the Introduction and Sec. \ref{sec: anomaly of lattice}, in systems with lattice symmetry it is important to consider anomaly matching for the classification of symmetry-enriched topological quantum spin liquids. The values of the anomaly indicators for different lattice homotopy classes with $p6\times SO(3)$ symmetry are given in Table \ref{tab: anomaly indicators p6 x SO(3)}. The anomaly indicators for the $U(1)_{2N}$ state in the symmetry fractionalization class $\{n_1, n_2, n_3\}$ can be calculated in a way similar to Sec. \ref{subsec: Z2}, which yields \footnote{Another way to calculate these anomaly indicators is as follows. We can restrict the $p6\times SO(3)$ symmetry into two of its $\Z_2\times SO(3)$ subgroups, where the $\Z_2$ in the first subgroup is generated by $C_2\equiv C_6^3$, and in the second it is generated by $T_1C_2$. By comparing the representative cochains of $B_{xy}$, $A_c^2$ and $Nw_2$ in Appendix \ref{app:SFclasses} with Eq. \eqref{eq: representative cochain Z2 x SO(3) 1}, we see that, after the restriction to the first $\Z_2\times SO(3)$ subgroup, the $p6\times SO(3)$ symmetry fractionalization class labeled by $\{n_1, n_2, n_3\}$ becomes the class labeled by $\{n_2, n_3\}$ in Sec. \ref{subsubsec: no permutation}, while after the restriction to the second $\Z_2\times SO(3)$ subgroup, the class labeled by $\{n_1, n_2, n_3\}$ becomes the class labeled by $\{n_1+n_2, n_3\}$ in Sec. \ref{subsubsec: no permutation}. According to Eq. \eqref{eq: anomaly Z2 x SO(3) no permutation}, the anomaly indicators in Eq. \eqref{eq: anomaly indicator p6 x SO(3)} for $\U_{2N}$ with the symmetry fractionalization class labeled by $\{n_1, n_2, n_3\}$ is Eq. \eqref{eq: U(1) level 2N p6xSO(3) 1 anomaly calculation}.}
\beq \label{eq: U(1) level 2N p6xSO(3) 1 anomaly calculation}
\mathsf{I}_1 = (-1)^{n_2 n_3}\,,\quad\mathsf{I}_2 = (-1)^{(n_1+n_2)n_3}\,.
\eeq

Therefore, by matching these anomaly indicators with Table \ref{tab: anomaly indicators p6 x SO(3)}, we arrive at the classification in Table~\ref{tab:Anomaliesp6SO3_1}.
\begin{table}[!htbp]
\centering
\renewcommand{\arraystretch}{1.3}
\begin{tabular}{|c|c|}
\toprule[2pt]
\makecell[c]{Lattice homotopy\\ class} &  \makecell[c]{Symmetry fractionalization\\ class} \\ \hline
0 & \makecell[c]{$\{n_1, n_2, 0\}$,\\ or $\{n_1, n_2, 1\}$ with $n_{1, 2}$ even}\\ \hline
a & $\{n_1, n_2, 1\}$ with $n_{1, 2}$ odd\\ \hline 
c & $\{n_1, n_2, 1\}$ with $n_1$ odd, $n_2$ even\\ \hline 
a+c & $\{n_1, n_2, 1\}$ with $n_1$ even, $n_2$ odd\\ \hline 
\bottomrule[2pt]	
\end{tabular}
\caption{Classification of symmetry-enriched $U(1)_{2N}$ topological quantum spin liquids in lattice systems with $p6\times SO(3)$ symmetry, if no symmetry permutes anyons.}
\label{tab:Anomaliesp6SO3_1}
\end{table}

\item Nontrivial $C_6$ action: $C_6$ as charge conjugation.

In this case, the possible symmetry fractionalization classes are given by 
    \beq
    H^2_{(2)}(p6\times SO(3), \Z_{2N}) = (\Z_2)^3 
    \eeq
whose elements can be written as
\beq \label{eq: element p6 SO(3) 2}
w = n_1\cdot NB_{xy} + n_2 \cdot NA_c^2 + n_3 \cdot Nw_2\,,
\eeq
and labeled by $\{n_1, n_2, n_3\}$ with $n_{1,2,3}\in\{0,1\}$. Here $ NB_{xy}$, $NA_c^2$ and $Nw_2$ are generators of the three $\Z_2$ pieces respectively (the representative cochains and the reason for the names of these generators are given in Appendix \ref{app:SFclasses}). For each symmetry fractionalization class, the $U$- and $\eta$-symbols can be obtained via Eqs. \eqref{eq:torsor}, \eqref{eq: microscopic U and eta} and \eqref{eq: U U(1) level 2N}.

Similar to the previous case, without considering anomaly matching, $p6\times SO(3)$ symmetric $U(1)_{2N}$ topological quantum spin liquids are classified by the above $\{n_1, n_2, n_3\}$ if $C_6$ acts as charge conjugation. Calculating the anomaly indicators for $U(1)_{2N}$ with symmetry fractionalization class labeled by $\{n_1, n_2, n_3\}$ as before, we get
\beq \label{eq: U(1) level 2N p6xSO(3) 2 anomaly calculation}
\mathsf{I}_1 = (-1)^{(n_2+1) n_3 N}\,,\quad\mathsf{I}_2 = (-1)^{(n_1 + n_2+1) n_3 N}\,.
\eeq

Therefore, by matching these anomaly indicators with Table \ref{tab: anomaly indicators p6 x SO(3)}, we arrive at the classification in Table~\ref{tab:Anomaliesp6SO3_2}.
\begin{table}[!htbp]
\centering
\renewcommand{\arraystretch}{1.3}
\begin{tabular}{|c|c|}
\toprule[2pt]
\makecell[c]{Lattice homotopy\\ class} &  \makecell[c]{Symmetry fractionalization\\ class} \\ \hline
0 & \begin{tabular}{c}$\{n_1, n_2, n_3\}$ for $N$ even,\\ $\{n_1, n_2, 0\}$ or $\{0, 1, 1\}$ for $N$ odd\end{tabular}\\ \hline
a & $\{1, 0, 1\}$ for $N$ odd\\ \hline 
c & $\{1, 1, 1\}$ for $N$ odd\\ \hline 
a+c & $\{0, 0, 1\}$ for $N$ odd\\ \hline 
\bottomrule[2pt]	
\end{tabular}
\caption{Classification of symmetry-enriched $U(1)_{2N}$ topological quantum spin liquids in lattice systems with $p6\times SO(3)$ symmetry, if $C_6$ acts as charge conjugation.}
\label{tab:Anomaliesp6SO3_2}
\end{table}   
\end{enumerate}

Summarizing all cases, the total number of different $p6\times SO(3)$ symmetry-enriched $U(1)_{2N}$ topological quantum spin liquids is summarized in Table \ref{tab:RealizationNumber U(1) level 2N}. Note that this classification is complete for $N\leqslant 5$, but incomplete for $N\geqslant 6$, because we have assumed that the only way the symmetry can permute anyons is via charge conjugation, while for $N\geqslant 6$ a symmetry can in principle permute anyons in other manners.

\begin{widetext}

\begin{table}[!htbp]
\centering
\renewcommand{\arraystretch}{1.3}
\begin{tabular}{|c|c|c|c|c|}
\toprule[2pt]
Symmetry group & Lattice homotopy class &  $N=1$ & odd $N>1$ & even $N$ \\ \hline
\multirow{4}{*}{$p6\times SO(3)$} & 0 & 5 & $\left(\frac{5(N(N,3)+1)}{2}\right)+(5)$ & $\left(\frac{5N(N,3)}{2}+3\right)+(8)$ \\ \cline{2-5}
&a & 1 & $\left(\frac{N(N,3)+1}{2}\right)+(1)$ & $\left(\frac{N(N, 3)}{2}\right)+(0)$ \\  \cline{2-5}
&c & 1 & $\left(\frac{N(N,3)+1}{2}\right)+(1)$ & $\left(\frac{N(N, 3)}{2}\right)+(0)$ \\  \cline{2-5}
&a+c & 1 & $\left(\frac{N(N,3)+1}{2}\right)+(1)$ & $\left(\frac{N(N, 3)}{2}+1\right)+(0)$ \\  \hline

\multirow{8}{*}{$p4\times SO(3)$} & 0 & 9 & $\left(\frac{9(N+1)}{2}\right)+(9)+(9)+(9)$ & $(9N+6)+(12)+(20)+(20)$ \\ \cline{2-5}
&a & 1 & $\left(\frac{N+1}{2}\right)+(1)+(1)+(1)$ & $(N)+(0)+(4)+(0)$ \\  \cline{2-5}
&b & 1 & $\left(\frac{N+1}{2}\right)+(1)+(1)+(1)$ & $(N)+(0)+(0)+(4)$ \\  \cline{2-5}
&c & 1 & $\left(\frac{N+1}{2}\right)+(1)+(1)+(1)$ & $(N+2)+(4)+(0)+(0)$ \\  \cline{2-5}
&a+b & 1 & $\left(\frac{N+1}{2}\right)+(1)+(1)+(1)$ & $(N)+(0)+(0)+(0)$ \\  \cline{2-5}
&a+c & 1 & $\left(\frac{N+1}{2}\right)+(1)+(1)+(1)$ & $(N)+(0)+(0)+(0)$ \\  \cline{2-5}
&b+c & 1 & $\left(\frac{N+1}{2}\right)+(1)+(1)+(1)$ & $(N)+(0)+(0)+(0)$ \\  \cline{2-5}
&a+b+c & 1 & $\left(\frac{N+1}{2}\right)+(1)+(1)+(1)$ & $(N)+(0)+(0)+(0)$ \\ \cline{2-5}

\bottomrule[2pt]	
\end{tabular}
\caption{Number of $p6\times SO(3)$ and $p4\times SO(3)$ symmetry-enriched $U(1)_{2N}$ topological quantum spin liquids. For the case with a $p6\times SO(3)$ symmetry, each of the last two columns is written as a sum of two terms, representing the number of states where no anyon is permuted by symmetries and where $C_6$ acts as charge conjugation, respectively. For the case with $p4\times SO(3)$ symmetry, each of the last two columns is written as a sum of four terms, representing the number of states where no anyon is permuted by symmetries, where $C_4$ acts as charge conjugation while $T_{1,2}$ do not, where $T_{1, 2}$ act as charge conjugation while $C_4$ does not, and where $C_4$ and $T_{1,2}$ all act as charge conjugation, respectively. The details of the symmetry fractionalization class of each state can be found in Tables \ref{tab:Anomaliesp6SO3_1}, \ref{tab:Anomaliesp6SO3_2}, \ref{tab:Anomaliesp4SO3_1}, \ref{tab:Anomaliesp4SO3_2}, \ref{tab:Anomaliesp4SO3_3} and \ref{tab:Anomaliesp4SO3_4}.}
\label{tab:RealizationNumber U(1) level 2N}
\end{table}

\end{widetext}

\subsection{$p4\times SO(3)$}

$SO(3)$ cannot permute anyons, and both the generators $T_{1,2}$ and the generator $C_4$ can permute anyons by charge conjugation. Hence, there are four possibilities regarding how $p4\times SO(3)$ can permute anyons: 

\begin{enumerate}
\item Trivial $T_{1,2}$ and $C_4$ action. 

In this case, the possible symmetry fractionalization classes are given by 
\beq
H^2_{(1)}(p4\times SO(3), \Z_{2N}) = \Z_{2N}\oplus\Z_{(2N,4)}\oplus(\Z_2)^2\,.
\eeq
whose elements can be labeled as
\beq
\begin{aligned}
&w = n_1\cdot \widetilde{\mathscr{B}}^{(1)}_{xy} + n_2\cdot \widetilde{\mathscr{B}}^{(1)}_{c^2} \\
& + n_3\cdot \widetilde{\beta}(A_{x+y})+ n_4\cdot Nw_2\,,
\end{aligned}
\eeq
with $n_1\in\{0,\dots,2N-1\}$, $n_2\in\{0,\dots,(2N,4)-1\}$, $n_{3,4}\in\{0,1\}$. Here $\widetilde{\mathscr{B}}^{(1)}_{xy}$, $\widetilde{\mathscr{B}}^{(1)}_{c^2}$, $\widetilde{\beta}(A_{x+y})$ and $Nw_2$ are generators of $\Z_{2N}$, $\Z_{(2N,4)}$ and two $\Z_2$ pieces, respectively (the representative cochains and the reason for the names of these generators are given in Appendix \ref{app:SFclasses}). For each symmetry fractionalization class, the $U$- and $\eta$-symbols can be obtained via Eqs. \eqref{eq:torsor}, \eqref{eq: microscopic U and eta} and \eqref{eq: U U(1) level 2N}.

Again, because symmetry fractionalization classes related by relabeling anyons are physically identical, different symmetry realizations on $U(1)_{2N}$ in this case are specified by $\{n_1, n_2, n_3, n_4\}$, where $\{n_1, n_2, n_3, n_4\}$ is identified with $\{[-n_1]_{2N}, [-n_2]_{(2N, 4)}, n_3, n_4\}$. Calculating the anomaly indicators for $U(1)_{2N}$ state with symmetry fractionalization class labeled by $\{n_1, n_2, n_3, n_4\}$, we get

\beq
\begin{aligned}
&\mathsf{I}_1 = (-1)^{n_2 n_4 }\,,\quad\mathsf{I}_2 = (-1)^{(n_1 + n_2) n_4 }\,,\\
&\mathsf{I}_3 = (-1)^{(n_2+n_3)n_4 }\,.\\
\end{aligned}
\eeq

Therefore, by matching these anomaly indicators with Table \ref{tab: anomaly indicators p4 x SO(3)}, we arrive at the classification in Table~\ref{tab:Anomaliesp4SO3_1}.
\begin{table}[!htbp]
\centering
\renewcommand{\arraystretch}{1.3}
\begin{tabular}{|c|c|}
\toprule[2pt]
\makecell[c]{Lattice homotopy\\ class} &  \makecell[c]{Symmetry fractionalization\\ class} \\ \hline
0 & \makecell[c]{$\{n_1, n_2, n_3, 0\}$,\\ or $\{n_1, n_2, 0, 1\}$ with $n_{1,2}$ even}\\ \hline
a & $\{n_1, n_2, 1, 1\}$ with $n_{1, 2}$ odd\\ \hline 
b & $\{n_1, n_2, 0, 1\}$ with $n_1$ odd, $n_2$ even\\ \hline 
c & $\{n_1, n_2, 1, 1\}$ with $n_{1,2}$ even\\ \hline 
a+b & $\{n_1, n_2, 1, 1\}$ with $n_1$ even, $n_2$ odd\\ \hline
a+c & $\{n_1, n_2, 0, 1\}$ with $n_{1,2}$ odd\\ \hline 
b+c & $\{n_1, n_2, 1, 1\}$ with $n_1$ odd, $n_2$ even\\ \hline 
a+b+c & $\{n_1, n_2, 0, 1\}$ with $n_1$ even, $n_2$ odd\\ \hline 
\bottomrule[2pt]	
\end{tabular}
\caption{Classification of symmetry-enriched $U(1)_{2N}$ topological quantum spin liquids in lattice systems with $p4\times SO(3)$ symmetry, if no symmetry permutes anyons.}
\label{tab:Anomaliesp4SO3_1}
\end{table}

\item Nontrivial $C_4$ action, trivial $T_{1,2}$ action. 

In this case, the possible symmetry fractionalization classes are given by 
\beq
H^2_{(2)}(p4\times SO(3), \Z_{2N}) = (\Z_2)^4\,.
\eeq
We can write these elements as 
\beq
\begin{aligned}
&w = n_1\cdot NB_{xy} + n_2\cdot NB_{c^2} \\
&+ n_3\cdot \widetilde{\beta}(A_{x+y})+ n_4\cdot Nw_2\,,
\end{aligned}
\eeq
with $n_{1,2,3,4}\in\{0,1\}$. Here $NB_{xy}$, $NB_{c^2}$, $\widetilde{\beta}(A_{x+y})$ and $Nw_2$ are generators of the four $\Z_2$ pieces respectively (the representative cochains and the reason for the names of these generators are given in Appendix \ref{app:SFclasses}). For each symmetry fractionalization class, the $U$- and $\eta$-symbols can be obtained via Eqs. \eqref{eq:torsor}, \eqref{eq: microscopic U and eta} and \eqref{eq: U U(1) level 2N}.

Calculating the anomaly indicators for $U(1)_{2N}$ with symmetry fractionalization class labeled by $(n_1, n_2, n_3, n_4)$ as before, we get
\beq
\begin{aligned}
&\mathsf{I}_1 = (-1)^{n_2 n_4 N}\,,\quad\mathsf{I}_2 = (-1)^{(n_1 + n_2) n_4 N}\,,\\
&\mathsf{I}_3 = (-1)^{(n_2 N + n_3)n_4 }\,.
\end{aligned}
\eeq

Therefore, by matching these anomaly indicators with Table \ref{tab: anomaly indicators p4 x SO(3)}, we arrive at the classification in Table~\ref{tab:Anomaliesp4SO3_2}.
\begin{table}[!htbp]
\centering
\renewcommand{\arraystretch}{1.3}
\begin{tabular}{|c|c|}
\toprule[2pt]
\makecell[c]{Lattice\\ homotopy \\ class} &  \makecell[c]{Symmetry fractionalization\\ class} \\ \hline
0 & \begin{tabular}{c}
$\{n_1, n_2, n_3, 0\}$ or $\{n_1, n_2, 0, 1\}$ for $N$ even, \\
$\{n_1, n_2, n_3, 0\}$ or $\{0,0,0,1\}$ for $N$ odd\end{tabular}\\ \hline
a & $\{1, 1, 1, 1\}$ for $N$ odd\\ \hline 
b & $\{1, 0, 0, 1\}$ for $N$ odd\\ \hline 
c & \begin{tabular}{c}
$\{n_1, n_2, 1, 1\}$ for $N$ even, \\
$\{0, 0, 1, 1\}$ for $N$ odd\end{tabular}\\ \hline 
a+b & $\{0, 1, 1, 1\}$ for $N$ odd\\ \hline
a+c & $\{1, 1, 0, 1\}$ for $N$ odd\\ \hline 
b+c & $\{1, 0, 1, 1\}$ for $N$ odd\\ \hline 
a+b+c & $\{0, 1, 0, 1\}$ for $N$ odd\\ \hline 
\bottomrule[2pt]	
\end{tabular}
\caption{Classification of symmetry-enriched $U(1)_{2N}$ topological quantum spin liquids in lattice systems with $p4\times SO(3)$ symmetry, if $C_4$ acts as charge conjugation.}
\label{tab:Anomaliesp4SO3_2}
\end{table}

\item Nontrivial $T_{1,2}$ action, trivial $C_4$ action. 

In this case, the possible symmetry fractionalization classes are given by 
\beq
H^2_{(3)}(p4\times SO(3), \Z_{2N}) = \Z_{(2N,4)}\oplus(\Z_2)^3\,.
\eeq
We can write these elements as 
\beq
\begin{aligned}
w &= n_1\cdot \widetilde{\mathscr{B}}^{(3)}_{xy} + n_2\cdot NB_{c^2} \\
&+ n_3\cdot NA_{x+y}^2+ n_4\cdot Nw_2\,,
\end{aligned}
\eeq
with $n_1\in\{0,\dots,(2N,4)-1\}$, $n_{2,3,4}\in\{0,1\}$. Here $ \widetilde{\mathscr{B}}^{(3)}_{xy}$, $NB_{c^2}$, $NA_{x+y}^2$ and $Nw_2$ are generators of the $\Z_{(2N,4)}$ and the three $\Z_2$ pieces, respectively (the representative cochains and the reason for the names of these generators are given in Appendix \ref{app:SFclasses}). For each symmetry fractionalization class, the $U$- and $\eta$-symbols can be obtained via Eqs. \eqref{eq:torsor}, \eqref{eq: microscopic U and eta} and \eqref{eq: U U(1) level 2N}.

Again, because symmetry fractionalization classes related by relabeling anyons are physically identical, different symmetry realizations on $U(1)_{2N}$ in this case are specified by  $\{n_1, n_2, n_3, n_4\}$, where $\{n_1, n_2, n_3, n_4\}$ is identified with $\{[-n_1]_{(2N, 4)}, n_2, n_3, n_4\}$. Calculating the anomaly indicators for $U(1)_{2N}$ with symmetry fractionalization class labeled by $(n_1, n_2, n_3, n_4)$ as before, we get
\beq
\begin{aligned}
&\mathsf{I}_1 = (-1)^{(n_1 + n_2 N) n_4}\,,\quad\mathsf{I}_2 = (-1)^{n_2 n_4 N}\,,\\
&\mathsf{I}_3 = (-1)^{(n_2 + n_3 + 1)n_4 N}\,.
\end{aligned}
\eeq

Therefore, by matching these anomaly indicators with Table \ref{tab: anomaly indicators p4 x SO(3)}, we arrive at the classification in Table~\ref{tab:Anomaliesp4SO3_3}.
\begin{table}[!htbp]
\centering
\renewcommand{\arraystretch}{1.3}
\begin{tabular}{|c|c|}
\toprule[2pt]
\makecell[c]{Lattice\\ homotopy\\ class} &  \makecell[c]{Symmetry fractionalization\\ class} \\ \hline
0 & \begin{tabular}{c}
$\{n_1, n_2, n_3, 0\}$, $\{0, n_2, n_3, 1\}$, \\
or $\{2, n_2, n_3, 1\}$ for $N$ even, \\
$\{n_1, n_2, n_3, 0\}$ or $\{0,0,1,1\}$ for $N$ odd\end{tabular}\\ \hline
a & \begin{tabular}{c}
$\{1, n_2, n_3, 1\}$  for $N$ even, \\
$\{1, 0, 1, 1\}$ for $N$ odd\end{tabular}\\ \hline 
b & $\{1, 1, 0, 1\}$ for $N$ odd\\ \hline 
c & $\{0, 0, 0, 1\}$ for $N$ odd\\ \hline 
a+b & $\{0, 1, 0, 1\}$ for $N$ odd\\ \hline
a+c & $\{1, 0, 0, 1\}$ for $N$ odd\\ \hline 
b+c & $\{1, 1, 1, 1\}$ for $N$ odd\\ \hline 
a+b+c & $\{0, 1, 1, 1\}$ for $N$ odd\\ \hline 
\bottomrule[2pt]	
\end{tabular}
\caption{Classification of symmetry-enriched $U(1)_{2N}$ topological quantum spin liquids in lattice systems with $p4\times SO(3)$ symmetry, if translations act as charge conjugation.}
\label{tab:Anomaliesp4SO3_3}
\end{table}

\item Nontrivial $T_{1,2}$ and $C_4$ action. 

In this case, the possible symmetry fractionalization classes are given by 
\beq
H^2_{(4)}(p4\times SO(3), \Z_{2N}) = \Z_{(2N,4)}\oplus(\Z_2)^3\,.
\eeq
We can label these elements as 
\beq
\begin{aligned}
w = &n_1\cdot  \widetilde{\mathscr{B}}^{(4)}_{xy} + n_2\cdot NB_{c^2} \\
& + n_3\cdot NA_{x+y}^2+ n_4\cdot Nw_2\,,
\end{aligned}
\eeq
with $n_1\in\{0,\dots,(2N,4)-1\}$, $n_{2,3,4}\in\{0,1\}$. Here $ \widetilde{\mathscr{B}}^{(4)}_{xy}$, $NB_{c^2}$, $\widetilde{A_{x+y}^2}$ and $Nw_2$ are generators of the $\Z_{(2N,4)}$ and the three $\Z_2$ pieces, respectively (the representative cochains and the reason for the names of these generators are given in Appendix \ref{app:SFclasses}). For each symmetry fractionalization class, the $U$- and $\eta$-symbols can be obtained via Eqs. \eqref{eq:torsor}, \eqref{eq: microscopic U and eta} and \eqref{eq: U U(1) level 2N}.

Again, because symmetry fractionalization classes related by relabeling anyons are physically identical, different symmetry realizations on $U(1)_{2N}$ in this case are specified by $\{n_1, n_2, n_3, n_4\}$, where $\{n_1, n_2, n_3, n_4\}$ is identified with $\{[-n_1]_{(2N, 4)}, n_2, n_3, n_4\}$. Calculating the anomaly indicators for $U(1)_{2N}$ with symmetry fractionalization class labeled by $(n_1, n_2, n_3, n_4)$ as before, we get
\beq
\begin{aligned}
&\mathsf{I}_1 = (-1)^{n_2 n_4 N}\,,\quad\mathsf{I}_2 = (-1)^{(n_1 + n_2 N) n_4}\,,\\
&\mathsf{I}_3 = (-1)^{(n_2  + n_3 + 1)n_4 N}\,.
\end{aligned}
\eeq

Therefore, by matching these anomaly indicators with Table \ref{tab: anomaly indicators p4 x SO(3)}, we arrive at the classification in Table~\ref{tab:Anomaliesp4SO3_4}.
\begin{table}[!htbp]
\centering
\renewcommand{\arraystretch}{1.3}
\begin{tabular}{|c|c|}
\toprule[2pt]
\makecell[c]{Lattice\\ homotopy\\ class} &  Symmetry fractionalization class \\ \hline
0 & \begin{tabular}{c}
$\{n_1, n_2, n_3, 0\}$, $\{0, n_2, n_3, 1\}$,\\
or $\{2, n_2, n_3, 1\}$ for $N$ even, \\
$\{n_1, n_2, n_3, 0\}$ or $\{0,0,1,1\}$ for $N$ odd\end{tabular}\\ \hline
a & $\{1, 1, 0, 1\}$ for $N$ odd\\ \hline 
b & \begin{tabular}{c}
$\{1, n_2, n_3, 1\}$ for $N$ even, \\
$\{1, 0, 1, 1\}$ for $N$ odd\end{tabular}\\ \hline 
c & $\{0, 0, 0, 1\}$ for $N$ odd\\ \hline 
a+b & $\{0, 1, 0, 1\}$ for $N$ odd\\ \hline
a+c & $\{1, 1, 1, 1\}$ for $N$ odd\\ \hline 
b+c & $\{1, 0, 0, 1\}$ for $N$ odd\\ \hline 
a+b+c & $\{0, 1, 1, 1\}$ for $N$ odd\\ \hline 
\bottomrule[2pt]	
\end{tabular}
\caption{Classification of symmetry-enriched $U(1)_{2N}$ topological quantum spin liquids in lattice systems with $p4\times SO(3)$ symmetry, if translations and $C_4$ both act as charge conjugation.}
\label{tab:Anomaliesp4SO3_4}
\end{table}
\end{enumerate}

Summarizing all cases, the total number of different $p4\times SO(3)$ symmetry-enriched $U(1)_{2N}$ topological quantum spin liquids is summarized in Table \ref{tab:RealizationNumber U(1) level 2N}. Note that this classification is complete for $N\leqslant 5$, but incomplete for $N\geqslant 6$, because we have assumed that the only way the symmetry can permute anyons is via charge conjugation, while for $N\geqslant 6$ a symmetry can in principle permute anyons in other manners.

\section{Ising$^{(\nu)}$ topological orders: Kitaev's non-Abelian chiral spin liquids} \label{sec: Ising}

Our next class of examples are non-Abelian chiral spin liquid states, which we dub the ``Ising$^{(\nu)}$  states", with $\nu$ an odd integer. Their topological properties are discussed in detail by Kitaev \cite{Kitaev2006} and will be reviewed below. The exactly solvable model in Ref. \cite{Kitaev2006} has triggered enormous interest in realizing the Ising$^{(1)}$ state in real materials \cite{Rau2015, Trebst2017, Winter2017}. We remark that usually these Kitaev quantum spin liquids are discussed in the context of spin-orbit coupled systems, but here we consider them in systems without spin-orbit coupling for simplicity. In particular, we will classify Ising$^{(\nu)}$ states in lattice systems with $p6\times SO(3)$ or $p4\times SO(3)$ symmetry.

The Ising$^{(\nu)}$ state has three anyons $\{I, \sigma, \psi\}$, where the trivial anyon here is denoted $I$, and the nontrivial fusion rules are given by
\begin{equation}
	\psi\times\psi=I, \quad \sigma\times\psi=\sigma, \quad \sigma\times\sigma=I+\psi.
\end{equation}
The nontrivial $F$-symbols are
\begin{equation}
	\begin{gathered}
	F^{\psi\sigma\psi}_\sigma=F^{\sigma\psi\sigma}_\psi=-1\\
	\left[F^{\sigma\sigma\sigma}_\sigma\right]_{ab}=
	\frac{\varkappa_{\sigma}}{\sqrt{2}} \left[
	\begin{matrix}
		1 & 1\\
		1 & -1
	\end{matrix}
\right]_{ab}.
	\end{gathered}
	\label{eqn:IsingF}
\end{equation}
Here, the column and row labels of the matrix take values $I$ and $\psi$ (in this order). All other $F$-symbols are 1 if it is compatible with the fusion rule and 0 if it is not. $\varkappa_{\sigma}=(-1)^{\frac{\nu^2-1}{8}}$ is the Frobenius-Schur indicator of $\sigma$.

The nontrivial $R$-symbols are
\begin{equation}
	\begin{gathered}
	R^{\psi\psi} = -1,\quad R^{\psi\sigma}_\sigma=R^{\sigma\psi}_\sigma=(-i)^{\nu}\\
	R^{\sigma\sigma}_{I}=\varkappa_{\sigma} e^{-i\frac{\pi}{8}\nu}, \quad R^{\sigma\sigma}_\psi=\varkappa_{\sigma} e^{i\frac{3\pi}{8}\nu}.
	\end{gathered}
\end{equation}
The topological spins are $\theta_\psi = -1$, $\theta_\sigma=e^{i\frac{\pi }{8}\nu}$, and the chiral central charge $c_{-} = \frac{\nu}{2}$.

The topological symmetry of Ising$^{(\nu)}$ is trivial and no symmetry of Ising$^{(\nu)}$ can permute anyons. The $U$-symbol and a set of $\eta$-symbol can all be chosen to be 1.

The symmetry fractionalization classes of $p6\times SO(3)$ are classified by 
\beq
H^2(p6\times SO(3), \Z_2)\cong (\Z_2)^3\,.
\eeq
We can label these elements as
\beq
w = n_1\cdot B_{xy} + n_2\cdot A_c^2 + n_3\cdot w_2\,,
\eeq
with $n_{1,2,3} \in \{0,1\}$. The symmetry fractionalization classes of $p4\times SO(3)$ are given by 
\beq
H^2(p4\times SO(3), \Z_2)\cong (\Z_2)^4\,.
\eeq
We can label these elements as
\beq
w = n_1\cdot B_{xy} + n_2\cdot B_{c^2} + n_3\cdot A_{x+y}^2 + n_4\cdot w_2\,,
\eeq
with $n_{1,2,3,4} \in \{0,1\}$. The representative cochains of these elements are presented in Appendix~\ref{app:SFclasses}. Physically, these generators can be viewed as detecting whether the non-Abelian anyon $\sigma$ carries a projective quantum number under these global symmetries. For each symmetry fractionalization class, the $U$- and $\eta$-symbols can be obtained via Eqs. \eqref{eq:torsor} and \eqref{eq: microscopic U and eta}.

The above discussion implies that, without considering anomaly matching, there are in total $2^3=8$ different $p6\times SO(3)$ symmetric Ising$^{(\nu)}$ states, and $2^4=16$ different $p4\times SO(3)$ symmetric Ising$^{(\nu)}$ states. Calculating the anomaly indicators for the Ising$^{(\nu)}$ state in a way similar to the the calculation for the $U(1)_{2N}$ state, we find that for any symmetry fractionalization class of either $p6\times SO(3)$ or $p4\times SO(3)$ symmetry, all anomaly indicators always evaluate to 1 and hence the anomaly is always absent.
Therefore, all $p6\times SO(3)$ or $p4\times SO(3)$ symmetry-enriched Ising$^{(\nu)}$ topological quantum spin liquids can emerge in lattice systems within lattice homotopy class 0 (including, for example, honeycomb lattice spin-1/2 system or spin-1 system on any lattice), but not other lattice homotopy class (including, for example, spin-1/2 system on triangular, kagome, square and checkerboard lattices). We notice that in most previous discussions of the Ising$^{(1)}$ state in spin-orbit coupled systems, the underlying lattice systems indeed have a trivial anomaly, since they can be obtained from the lattice homotopy class 0 here by breaking certain symmetries.

\section{$\Z_{N}$ topological orders: Generalized toric codes} \label{sec: Zn}

In this section, we consider the $\Z_{N}$ topological order, which is the $\mathbb{Z}_N$ generalization of the famous $\Z_2$ topological order~\cite{Kitaev2003,barkeshli2014, Kivelson1987, Rokhasar1988, Read1991, Wen1991, Senthil2000}. The case with $N=2$ has been studied extensively in many different types of lattice systems. However, as mentioned in the Introduction, when $N>2$ these states do not allow a description in terms of a simple parton mean field (instead, the partons have to be strongly interacting), and they are much less explored (see examples in Refs. \cite{Motrunich2002, Myers2017, Devakul2017, Dong2018, Kurecic2019, Giudice2022}). Our framework in Sec. \ref{sec: framework} allows us to classify a general $\Z_N$ topological quantum spin liquid enriched by a general symmetry. For concreteness, the symmetry we will consider below are one of these four: $p6m\times SO(3)\times\Z_2^T$, $p4m\times SO(3)\times\Z_2^T$, $p6m\times\Z_2^T$ and $p4m\times\Z_2^T$, where $p6m$ and $p4m$ are lattice symmetries, while $SO(3)$ and $\Z_2^T$ are on-site spin rotational symmetry and time reversal symmetry, respectively.

In the $\Z_N$ topological order, there are $N^2$ anyons in total, which can be labeled by two integers as $a=(a_e, a_m)$, with $a_e, a_m \in\{0,\dots, N-1\}$. Following the convention in the $\Z_2$ toric code, we will call the anyon labeled by $(1,0)$ as $e$, and the anyon labeled by $(0,1)$ as $m$. The fusion rules are element-wise addition modulo $N$, \ie
\begin{equation}
(a_{e},a_{m}) \times (b_e,b_{m}) = ([a_e + b_{e}]_N, [a_{m}+b_{m}]_N)\,,
\end{equation}
In a choice of gauge, the $F$-symbols of this topological order are all 1 and the $R$-symbols are given by
\begin{equation}
R^{ab}=e^{i \frac{2\pi}{N} a_{m} b_{e}}\,.
\end{equation}

The topological symmetry group is complicated to determine for general $N$ \cite{Delmastro2019, Geiko2022}. For $N=2$, the topological symmetry is $\Z_2\times \Z_2^T$, generated by the unitary electric-magnetic duality symmetry $S$ that exchanges $e$ and $m$, \ie $(a_e, a_m)\rightarrow (a_m, a_e)$,
and an anti-unitary symmetry $T$ which exchanges $e$ and $m$ the same way as $S$.
For this $\Z_2\times\Z_2^T$ symmetry, we can choose the $U$-symbols as 
\begin{equation}
U_{\bf g}(a,b;c)=
\begin{cases}
(-1)^{a_m b_e}, & \text{${\bf g}$ permutes anyons,}\\
1, & \text{otherwise,}
\end{cases}
\end{equation}
and a set of $\eta$-symbols as
\beq
\eta_a({\bf g}, {\bf h})=
\begin{cases}
(-1)^{a_e a_m}, & \text{${\bf g}$, ${\bf h}$ permute anyons,}\\
1, & \text{otherwise.}
\end{cases}
\eeq
with $(a_e, a_m), (b_e,b_m)$ the anyon labels of anyons $a, b$.

For $N\geqslant 3$, there is always a $\Z_4^T\rtimes \Z_2$ topological symmetry. The anti-unitary $\Z_4^T$ is generated by an action $T: (a_e, a_m)\rightarrow (a_m, [-a_e]_N)$, and the unitary $\Z_2$ is generated by an action $S: (a_e, a_m)\rightarrow (a_m, a_e)$. The two generators satisfy the relation
\beq
S^2 = \textbf{1}\,,\quad T^4 = \textbf{1}\,,\quad STS = T^{-1}\,.
\eeq
For this $\Z_4^T\rtimes\Z_2$ symmetry, writing a group element as ${\bf g}=T^{g_1}S^{g_2}$, with $g_1\in\{0, 1, 2, 3\}$ and $g_2\in\{0, 1\}$, we can choose the $U$-symbols as
\begin{equation}
U_{\bf g}(a,b;c)=
\begin{cases}
e^{\frac{2\pi i}{N}a_m b_e}, & \text{$g_1+g_2$ is odd,}\\
1, & \text{otherwise}
\end{cases},
\end{equation}
and a set of $\eta$-symbols as
\beq
\eta_a({\bf g}, {\bf h})=
\begin{cases}
e^{\frac{2\pi i}{N}a_e a_m}, & \text{$g_1+g_2$, $h_1+h_2$ are odd,}\\
1, & \text{otherwise}
\end{cases}
\eeq

For certain $N\geqslant 3$ there can be other topological symmetries, in addition to the above $\Z_4\rtimes\Z_2^T$ symmetry. For example, when $N=5$, the action $(a_e, a_m)\rightarrow ([3a_e]_5, [3a_m]_5)$ is an anti-unitary topological symmetry. For simplicity, below we will focus on the cases where $N=2, 3, 4$.

The analysis of the classification is similar to the previous cases. In the present case, we need to understand the anomaly indicators of the $p6m\times SO(3)\times\Z_2^T$, $p4m\times SO(3)\times\Z_2^T$, $p6m\times\Z_2^T$ and $p4m\times\Z_2^T$ symmetries. These anomaly indicators and their values for different lattice homotopy classes can be found in Appendix \ref{app: anomaly indicators}. Carrying out the procedure listed in Sec. \ref{sec: framework}, we can obtain the classification. In Tables \ref{tab:RealizationNumber}, we list the number of different symmetry-enriched $\Z_N$ topological quantum spin liquids in different lattice homotopy classes under these symmetries. The precise symmetry fractionalization classes in each case can be found in Appendix~\ref{app:SFclasses}. We also upload codes using which one can 1) see all symmetry fractionalization classes of the symmetry-enriched states within each lattice homotopy class, and 2) check which lattice homotopy class a given symmetry-enriched state belongs to \cite{code}. Below we comment on some of these results.

For the case with $N=2$ and the $p6m\times SO(3)\times\Z_2^T$ symmetry, the classification was carried out for spin-1/2 systems on a triangular, kagome and honeycomb lattice \cite{Qi2015b, Qi2016, Lu2011}, which belongs to the lattice homotopy class $a$, $c$ and 0, respectively. For lattice homotopy classes $a$ and $c$, our results agree with those in Refs. \cite{Qi2015b, Qi2016}. For the lattice homotopy class 0, using the parton-mean-field approach and assuming that one of $e$ and $m$ carries spin-1/2 under the $SO(3)$ symmetry, Ref. \cite{Lu2011} found 128 different states. We find 336 states in total, where 128 of them have one of $e$ and $m$ carrying half-odd-integer spin, and in the other 208 states both $e$ and $m$ carry integer spin, 9 of which also have symmetries permuting $e$ and $m$. For the case with $N=2$ and the $p4m\times SO(3)\times\Z_2^T$ symmetry, Ref. \cite{Lu2016} found 64 states on square lattice spin-1/2 system, which belongs to our lattice homotopy class $a$, agreeing with our results.

For the case with $N=2$ and the $p4m\times\Z_2^T$ symmetry, using the parton-mean-field approach, Ref. \cite{Yang2015} found 64 states on square lattice system with Kramers doublet spins, which can all be obtained from the $p4m\times SO(3)\times\Z_2^T$ symmetric $\Z_2$ topological quantum spin liquids by breaking the $SO(3)$ symmetry. Suppose in the $p4m\times SO(3)\times\Z_2^T$ symmetric version of these states, the anyon $e$ carries half-odd-integer spin under $SO(3)$, then projective quantum numbers of $m$ are fixed for all these 64 states \cite{Qi2015b}. In particular, $m$ experiences no nontrivial symmetry fractionalization pattern that simultaneously involves the time reversal and lattice symmetries. The absence of such symmetry fractionalization pattern still holds in the 64 ``within parton" $p4m\times\Z_2^T$ symmetric states obtained by breaking $SO(3)$. However, in addition to these 64 states, we have found $117-64=53$ other states, with their symmetry fractionalization classes presented in Appendix \ref{app: beyond parton} (anyons are not permuted by symmetries in all these 117 states). A common property of these 53 states is the presence of nontrivial symmetry fractionalization involving both the lattice symmetry and time reversal symmetry for the anyon $m$, \eg translation and time reversal may not commute for $m$. Furthermore, for all 117 states, the $C_2\equiv C_4^2$ symmetry fractionalizes on the $m$ anyon, \ie effectively $C_2^2=-1$ for $m$. Usually, the interpretation of this phenomenon is that there is a background $e$ anyon at each square lattice site (the $C_4$ center), and the mutual braiding statistics between $e$ and $m$ yields $C_2^2=-1$. However, for 16 of the 53 ``beyond-parton" states, $(T_1C_2)^2=(T_2C_2)^2=-1$ for $m$, which seems to suggest that there are also background $e$ anyons at the 2-fold rotation centers of $T_1C_2$ and $T_2C_2$, although microscopically there is no spin at those positions. So the analysis based on anomaly matching suggests that the simple picture where the fractionalization of rotational symmetries purely comes from background anyons is actually incomplete.

The above example shows that even for simple states like the $\Z_2$ topological order, the parton-mean-field approach may miss some of their symmetry enrichment patterns, and our framework in Sec. \ref{sec: framework} is more general. Note that here by ``parton mean field", we are referring to the ususal parton mean fields where the partons are non-interacting. If the partons are allowed to interact strongly, say, if they form nontrivial interacting symmetry-protected topological states under the projective symmetry group of the partons, symmetry-enriched states not captured by Ref. \cite{Yang2015} may arise, but it is technically complicated to study them. Also, by using parton constructions other than the one in Ref. \cite{Yang2015}, one may also obtain states beyond those in Ref. \cite{Yang2015}, but it is challenging to make this approach systematic.

We also notice that the number of $\Z_3$ topological quantum spin liquids is nonzero only in the lattice homotopy class 0. This phenomenon is actually true for general odd $N$. To see it, first notice all lattice homotopy classes except 0 have some mixed anomalies between the $SO(3)$ symmetry and the lattice symmetry \cite{Ye2021a}. In order to match this anomaly, it is impossible for both $e$ and $m$ to carry integer spin. Suppose that $e$ carries half-odd-integer spin, and consider threading an $SO(3)$ monopole through the system. The monopole will be viewed as a $\pi$ flux from the perspective of $e$. Then the local nature of the monopole implies that it must trap an anyon that has $\pi$ braiding statistics with $e$. For odd $N$, no such anyon exists, which leads to a contradiction. So $\Z_N$ topological quantum spin liquids with $N$ odd cannot possibly arise in lattice homotopy class other than 0. Note that the above argument does not rely on the time reversal symmetry, and it is valid no matter how the symmetries permute anyons.

For $\Z_N$ topological quantum spin liquids in systems belonging to a lattice homotopy class other than 0, which requires $N$ to be even, anyons $(1, 0)$ and $(0, N/2)$ cannot simutaneously carry half-odd-integer spin, otherwise there would be a mixed anomaly between the $SO(3)$ and time reversal symmetries \cite{Zou2017}.

\begin{widetext}

\begin{table}[!htbp]
\centering
\renewcommand{\arraystretch}{1.3}
\begin{tabular}{|c|c|c|c|c|c|c|c|}
\toprule[2pt]
Symmetry group & Lattice homotopy class &  $\Z_2$ & $\Z_3$ & $\Z_4$ & $\U_2\times \U_{-2}$ & $\U_4\times \U_{-4}$\\ \hline
\multirow{4}{*}{$p6m\times SO(3)\times\Z_2^T$} & 0 & 336 & 8 & 16453 & 32 & 144 \\ \cline{2-7}
&a & 8 & 0 & 70 & 0 & 0\\  \cline{2-7}
&c & 8 & 0 & 70 & 0 & 0\\  \cline{2-7}
&a+c & 4 & 0 & 82 & 0 & 0\\  \hline

\multirow{4}{*}{$p6m\times \Z_2^T$} & 0 & 208 & 8 & 4725 & 16 & 72 \\ \cline{2-7}
&a & 13 & 0 & 61 & 0 & 0\\  \cline{2-7}
&c & 13 & 0 & 61 & 0 & 0\\  \cline{2-7}
&a+c & 12 & 0 & 167 & 0 & 0\\  \hline

\multirow{8}{*}{$p4m\times SO(3)\times\Z_2^T$} & 0 & 3653 & 9 & 886740 & 128 & 1344 \\ \cline{2-7}
&a & 64 & 0 & 5008 & 0 & 0\\  \cline{2-7}
&b & 64 & 0& 5008 & 0 & 0\\  \cline{2-7}
&c & 64 & 0& 8872 & 0 & 0\\  \cline{2-7}
&a+b & 16 & 0& 636 & 0 & 0\\  \cline{2-7}
&a+c & 16 & 0& 656 & 0 & 0\\  \cline{2-7}
&b+c & 16 & 0& 656 & 0 & 0\\  \cline{2-7}
&a+b+c & 8 & 0& 318 & 0 & 0\\  \hline

\multirow{8}{*}{$p4m\times\Z_2^T$} & 0 & 2629 & 9 & 280852 & 64 & 672 \\ \cline{2-7}
&a & 117 & 0 & 3491 & 0 & 0\\  \cline{2-7}
&b & 117 & 0& 3491 & 0 & 0\\  \cline{2-7}
&c & 193 & 0& 12449 & 0 & 0\\  \cline{2-7}
&a+b & 33 & 0& 513 & 0 & 0\\  \cline{2-7}
&a+c & 34 & 0& 610 & 0 & 0\\  \cline{2-7}
&b+c & 34 & 0& 610 & 0 & 0\\  \cline{2-7}
&a+b+c & 21 & 0& 309 & 0 & 0\\  \cline{2-7}

\bottomrule[2pt]	
\end{tabular}
\caption{Number of various topological quantum spin liquids enriched by $p6m\times SO(3)\times\Z_2^T$, $p6m\times\Z_2^T$, $p4m\times SO(3)\times\Z_2^T$or $p4m\times\Z_2^T$ symmetry, where the third, fourth and fifth columns represent $\Z_2$, $\Z_3$ and $\Z_4$ topological orders, while the last two colomns represent the $U(1)_2\times U(1)_{-2}$ and $U(1)_4\times U(1)_{-4}$ topological orders, respectively. The details of the symmetry fractionalization classes of each state can be found in Appendix \ref{app:SFclasses}. For $\Z_2$ and $\Z_4$ topological orders, we also upload codes containing the symmetry fractionalization class for each state in each lattice homotopy class \cite{code}. For $\Z_3$, $U(1)_2\times U(1)_{-2}$, and $U(1)_4\times U(1)_{-4}$ topological orders, all symmetry-enriched states are anomaly-free.}
\label{tab:RealizationNumber}
\end{table}

\end{widetext}

\section{$\U_{2N}\times \U_{-2N}$ topological orders: Generalizations of the double-semion state} \label{sec: generalization of double semion}

In this section, we consider the $\U_{2N}\times \U_{-2N}$ topological order, which is the generalization of the double-semion state, \ie the case with $N=1$. Effectively, this state can be obtained by stacking a $U(1)_{2N}$ state, which is discussed in Sec. \ref{sec: U(1) level 2N}, on its time reversal partner, the $U(1)_{-2N}$ state. In addition, these states can also be constructed via the twisted quantum double models or the string-net models \cite{Kitaev1997, Levin2004}. We would like to classify the $\U_{2N}\times \U_{-2N}$ topological order enriched by one of these four symmetries: $p6m\times SO(3)\times\Z_2^T$, $p4m\times SO(3)\times\Z_2^T$, $p6m\times\Z_2^T$ and $p4m\times\Z_2^T$.

In a $U(1)_{2N}\times U(1)_{-2N}$ topological quantum spin liquid, there are $4N^2$ anyons in total, which can be labeled by two integers as $a=(a_s, a_{\bar{s}})$, with $a_s, a_{\bar{s}} \in\{0,\dots,2N-1\}$. Following the convention in the double-semion state, we will call the anyon labeled by $(1,0)$ as $s$, and the anyon labeled by $(0,1)$ as $\bar{s}$ (note in this convention $s$ and $\bar{s}$ are not anti-particles of each other). The fusion rules are element-wise addition modulo $2N$, \ie
\begin{equation}
(a_{s},a_{\bar{s}}) \times (b_s,b_{\bar{s}}) = ([a_s + b_{s}]_{2N}, [a_{\bar{s}}+b_{\bar{s}}]_{2N})\,,
\end{equation}
In a choice of gauge, the $F$-symbols of the theory are 
\begin{widetext}
\begin{equation}
    F^{abc}=\exp\left(i\frac{2\pi}{N}\left(a_s(b_s+c_s-[b_s+c_s]_{2N})-a_{\bar{s}}(b_{\bar{s}}+c_{\bar{s}}-[b_{\bar{s}}+c_{\bar{s}}]_{2N})\right)\right)\,
\end{equation}
\end{widetext}
and the $R$-symbols are 
\begin{equation}
R^{ab}=\exp\left(i \frac{2\pi}{N}(a_s b_s - a_{\bar{s}} b_{\bar{s}})\right)\,.
\end{equation}

The topological symmetry group is complicated to determine for general $N$, just like the $U(1)_{2N}$ state \cite{Delmastro2019, Geiko2022}. Here we list the topological symmetry group for $N=1,2$ here. For $N=1$, the topological symmetry is $\Z_2^T$, generated by $\tilde{S}$ exchanging $s$ and $\bar s$, \ie
\beq
(a_s, a_{\bar{s}})\rightarrow (a_{\bar{s}}, a_s)\,.
\eeq
We can choose the $U$-symbols and a set of $\eta$-symbols all equal to 1.

For $N=2$, the topological symmetry is $ \Z_4^T\rtimes\Z_2^T$, generated by an order 2 anti-unitary symmetry $\tilde{S}$ which exchanges $s$ and $\bar{s}$, $\tilde{S}\colon (a_s, a_{\bar{s}})\rightarrow (a_{\bar{s}}, a_s)$, and another order 4 anti-unitary symmetry $T$, which permutes anyons in the following way $T\colon (a_s, a_{\bar{s}})\rightarrow (a_{\bar{s}}, [-a_s]_{2N})$. The two generators satisfy the relation
\beq
\tilde{S}^2 = \textbf{1}\,,\quad T^4 = \textbf{1}\,,\quad \tilde{S}T\tilde{S} = T^{-1}\,.
\eeq
An element in $\Z_4^T\rtimes\Z_2^T$ can be written as $T^{g_1} \tilde{S}^{g_2}$, 
with $g_1\in\{0,\dots,3\}$ and $g_2\in\{0,1\}$. To define the $U$-symbols, first we define the following function
\beq
\tilde{U}(a_s, b_s) = 
\begin{cases}
(-1)^{a_s} & b_s\neq 0\\
1 & b_s=0
\end{cases}
\eeq
Given an element ${\bf g}\in \Z_4^T\rtimes \Z_2^T$, the $U$-symbols can be chosen such that 
\beq
U_{\bf g}(a,b;c) = 
\begin{cases}
1 & g_1=0 \\
\tilde{U}(a_{\bar{s}}, b_{\bar{s}}) & g_1 = 1 \\
\tilde{U}(a_s, b_s)\tilde{U}(a_{\bar{s}}, b_{\bar{s}}) & g_1 = 2 \\
\tilde{U}(a_s, b_s) & g_1 = 3 \\
\end{cases}
\eeq
And a set of $\eta$-symbols can be chosen to be all identity.

Carrying out the procedure in Sec. \ref{sec: framework} in a manner similar to the previous examples, we can obtain the classification of $U(1)_{2}\times U(1)_{-2}$ and $U(1)_4\times U(1)_{-4}$ topological quantum spin liquids enriched by $p6m\times SO(3)\times\Z_2^T$, $p4m\times SO(3)\times\Z_2^T$, $p6m\times\Z_2^T$ or $p4m\times\Z_2^T$ symmetry. The results are summarized in Table \ref{tab:RealizationNumber}.

We notice that in all symmetry groups considered here, $U(1)_{2}\times U(1)_{-2}$ and $U(1)_4\times U(1)_{-4}$ can only arise in the lattice homotopy class 0. Ref. \cite{Zaletel2014} presented a physical reason for this phenomenon. If we only consider the symmetry group $p6m\times SO(3)\times\Z_2^T$ and $p4m\times SO(3)\times\Z_2^T$, the following simpler argument can explain it. To be concrete, suppose the symmetry group is $p6m\times SO(3)\times\Z_2^T$, and a similar argument can be made if the symmetry group is $p4m\times SO(3)\times\Z_2^T$. Now suppose breaking the symmetry to $p6\times SO(3)$. Then the system can be viewed as a $p6\times SO(3)$ symmetric $U(1)_{2N}$ state on top of a $p6\times SO(3)$ symmetric $U(1)_{-2N}$ state, and these two states must have the opposite anomalies under the $p6\times SO(3)$ symmetry, otherwise they cannot be connected by time reversal to form the original $p6m\times SO(3)\times\Z_2^T$ symmetric state. Namely, after breaking $p6m\times SO(3)\times\Z_2^T$ to $p6\times SO(3)$, there is no remaining anomaly and the state is in lattice homotopy class 0. Now we ask which lattice homotopy class with a $p6m\times SO(3)\times\Z_2^T$ symmetry becomes the lattice homotopy class 0 with a $p6\times SO(3)$ symmetry after this symmetry breaking. From the representative configurations of all lattice homotopy classes in Sec. \ref{sec: anomaly of lattice}, clearly only the lattice homotopy class 0 does.

\section{Discussion} \label{sec: discussion}

In this paper, we have presented a general framework in Sec. \ref{sec: framework} to classify symmetry-enriched topological quantum spin liquids in two spatial dimensions. This framework applies to all topological quantum spin liquids, which may be Abelian or non-Abelian, and chiral or non-chiral. The symmetry we consider may include both lattice symmetry and internal symmetry, may contain anti-unitary symmetry, and may permute anyons. We then apply this framework to various examples in Secs. \ref{sec: U(1) level 2N}, \ref{sec: Ising}, \ref{sec: Zn} and \ref{sec: generalization of double semion}. As argued in the Introduction, our framework combines the advantages of the previous approaches in the literature, while avoiding their disadvantages. Indeed, we are able to identify symmetry-enriched topological quantum spin liquids that are not easily captured by the usual parton-mean-field approach (see examples in Sec. \ref{sec: Zn}), and we can systematically distinguish different lattice systems with the same symmetry group using their quantum anomalies.

We finish this paper by discussing some open questions.

\begin{itemize}
    
    \item In this paper, we characterize a topological quantum spin liquid with a lattice symmetry by one with an internal symmetry via the crystalline equivalence principle in Sec. \ref{subsec: crystalline equivalence principle}. However, it is more ideal to have a theory that directly describes topological quantum spin liquids with lattice symmetries. 
    
    Such a theory should be able to tell how an arbitrary symmetry acts on a state obtained by creating some anyons from the ground state and putting them at arbitrary positions. The symmetry action should be some analog of Eq. \eqref{eq: symmetry localization}, but it is subtle to understand what constraints the analogs of $U_{\bf g}(a, b; c)$ and $\eta_a({\bf g}, {\bf h})$ should satisfy. So far this question has been answered if the lattice symmetry only contains translation symmetry \cite{Cheng2015}, but for cases with point group symmetries it is answered in a very specific case, where the lattice symmetry is reflection, and the state only contains two anyons that are 1) anti-particles of each other, 2) transformed into each other under the reflection symmetry, and 3) located at two reflection related positions \cite{walker2019}. It is useful to have a complete theory that can answer this question in full generality. Such a theory is also helpful for the purpose of identifying observable signatures of different symmetry-enriched topological quantum spin liquids.

    \item Strictly speaking, our classification is a classification of different patterns of how symmetries permute anyons and the symmetry fractionalization patterns. In principle, one should further consider how the classification is modified upon stacking an invertible state on the topological quantum spin liquid with the same symmetry. This question is subtle because some nontrivial invertible states can be trivialized in the presence of a long-range entangled state \cite{barkeshli2014, Wang2015c, Zou2017}. We leave this problem for future study.

    \item In this paper, we focus on how symmetry permutes anyons and the symmetry fractionalization classes, which can be viewed as the bulk properties of different symmetry-enriched topological quantum spin liquids. It is also interesting to explore their boundary properties in the future. In particular, sometimes the symmetry enrichment pattern may enforce the boundary of the topological quantum spin liquid to be gapless, even if it is non-chiral \cite{Wang2017a, Wen2022}. Similarly, it is intriguing to study the properties of defects in different symmetry-enriched topological quantum spin liquids, and examine their potential to perform quantum computation \cite{Barkeshli2012, barkeshli2014}.

    \item It is useful to find numerical algorithms to identify the symmetry enrichement pattern of a topological quantum spin liquid that emerges in a lattice model that is not fine-tuned, and find experimental methods to detect the symmetry enrichment pattern in experiments. Some previous proposals for various specific cases include Refs.~\cite{Essin2014, Qi2015a, Zaletel2015, Sun2018, Lu2022}, but it is useful to find algorithms and methods applicable to the general setting.

    \item After classifying different symmetry-enriched topological quantum spin liquids and finding methods to detect them numerically and experimentally, it is important to construct explicit models that realize these topological orders.  For many topological orders enriched by internal symmetries, Refs. \cite{Heinrich2016, Cheng2017, Bulmash2020, Wang2021} construct their exactly solvable models with explicit Hamiltonians and ground-state wavefunctions. Moreover, there are many proposals for realizing symmetry-protected and symmetry-enriched topological states with lattice symmetries in the literature, including Refs. \cite{Song2017, Huang2017, Song2018a, Song2018, Zhang2022,Zhang2020, Qi2017, Else2019}. We anticipate that we can combine the above constructions to obtain exactly solvable models with concrete Hamiltonians that realize the symmetry-enriched topological quantum spin liquids discussed in this paper.
    
    It will be also interesting to find quantum materials and develop quantum simulators to realize these different phases, and explore interesting continuous quantum phase transitions out of them, which are beyond the conventional Landau-Ginzburg-Wilson-Fisher paradigm.

    \item In this paper, our focus is topological quantum spin liquids in two spatial dimensions. It is interesting to generalize our work to other systems, such as fermionic systems, systems in higher dimensions, systems with spin-orbit coupling, gapless systems and fractonic systems. 
    
    In particular, many interesting experimental systems feature spin-orbit couplings, and the general framework in the present paper can be straightforwardly extended to such systems. Because systems with spin-orbit coupling can often be obtained by breaking some symmetries in systems without spin-orbit coupling, there can be two main differences between these two types of systems. First, compared to systems without spin-orbit coupling, the distinction between some quantum phases may disappear in systems with spin-orbit coupling, since the latter has a smaller symmetry compared to the former and the relevant distinction may only be well defined in the presence of a larger symmetry. Second, there can be quantum phases that can be realized in systems with spin-orbit coupling, but not in systems without spin-orbit coupling, \ie they are incompatible with a larger symmetry. These are both intriguing phenomena that deserve future investigation.
    
    Also, there are many experimental candidates of $(3+1)$-dimensional symmetry-enriched gapless $U(1)$ quantum spin liquids in pyrochlores \cite{Gingras2013}, and their classification has been discussed within the framework of projective symmetry groups \cite{Chern2021, Liu2021, Desrochers2021, Chern2022, Desrochers2022, Chauhan2023}. As discussed in the present paper, the classification based on projective symmetry groups may be incomplete. Using more general approaches, $U(1)$ quantum spin liquids with only internal symmetries have been classified \cite{Wang2015c, Zou2017, Ning2019, Hsin2019}, and some examples of their lattice symmetry enriched versions have been constructed \cite{Zou2017a}. However, a systematic classification of $(3+1)$-dimensional $U(1)$ quantum spin liquids enriched by both lattice and internal symmetries is lacking, and it is interesting to apply the idea in the present paper to those settings in the future.
    
\end{itemize}

{\it Note: } Codes for checking anomaly matching and details of realizations for $\Z_2$ and $\Z_4$ topological order available at \url{https://github.com/Weicheng-Ye/Classification-of-QSL.git}.

\begin{acknowledgements}

We thank Maissam Barkeshli, Dominic Else, Meng Guo, Yin-Chen He, T. Senthil, Chong Wang, Fa Wang and Qing-Rui Wang for helpful discussion. WY acknowledges the hospitality of Yu-An Chen and Qing-Rui Wang during his stay at Peking University and Yau Mathematical Sciences Center, where a significant part of the draft was written. Research at Perimeter Institute is supported in part by the Government of Canada through the Department of Innovation, Science and Industry Canada and by the Province of Ontario through the Ministry of Colleges and Universities.

\end{acknowledgements}

\onecolumngrid

\begin{appendices}
\section{Translation between the characterization of reflection symmetry and time-reversal symmetry} \label{app: reflection and time reversal} 

It is a common folklore that ``reflection symmetry $=$ time-reversal symmetry $\times$ charge-conjugation''. However, one precise formulation of the statement is based on the CPT theorem, which is formulated in relativistic quantum field theory and requires Lorentz symmetry as premise \cite{weinberg2005quantum}. In the context of topological order, even though Lorentz symmetry is not explicitly present, it is also widely believed that the statement also holds true. However, the precise correspondence between reflection symmetry and time-reversal symmetry, especially the matching between the  data $\{\rho_{\bf g}; U_{\bf g}(a, b; c), \eta_a({\bf g}, {\bf h})\}$ for these two symmetries, is little known in the literature. We summarize this correspondence in this appendix.

For this purpose, we need more formal treatments of the topological order in terms of a unitary modular tensor category (UMTC), which go beyond what is reviewed in Sec. \ref{sec: characterization}, and we refer the interested readers to Sec. II of Ref. \cite{Ye2022} for the basics. Following the convention of Refs.~\cite{barkeshli2014,walker2019,WalkerLecture}, we model the time-reversal symmetry action on the UMTC as a $\mathbb{C}$-anti-linear functor and the (unitary) reflection symmetry action as an \textit{anti-monoidal} functor (also see Ref.~\cite{Kong2022} for slightly different treatments). Therefore, mathematically speaking, in this appendix we establish a precise correspondence between the data of a $\mathbb{C}$-anti-linear functor and the data of an {anti-monoidal} functor. We believe that such correspondence will imply the correspondence of the data $\{\rho_{\bf g}; U_{\rm g}(a, b; c), \eta_a({\bf g}, {\bf h})\}$ for these two symmetries on the explicit wavefunctions of the topological order, and we defer it to future study. 

Throughout the appendix, we assume that the reflection symmetry is unitary. We can also consider anti-unitary reflection symmetry, which in the crystalline equivalence principle should correspond to a unitary symmetry which does not reflect spacetime. Following the treatment in this appendix, we can similarly establish a correspondence between a $\mathbb{C}$-linear functor for the unitary symmetry and a $\mathbb{C}-$anti-linear anti-monoidal functor for the anti-unitary reflection symmetry. The details can be worked out by following closely the treatment in this appendix and we omit them.  

Recall that anyon lines may be ``bent" using the $A$ and $B$ symbols, given diagrammatically by
\begin{equation}
 \raisebox{-0.5\height}{
\begin{pspicture}[shift=-0.65](-0.1,-0.2)(1.5,1.2)
  \small
  \psset{linewidth=0.9pt,linecolor=black,arrowscale=1.5,arrowinset=0.15}
  \psline{->}(0,0)(0,0.55)
  \psline(0,0)(0,0.85)
  \psarc{->}(0.3,0.85){0.3}{25}{115}
  \psarc{-}(0.3,0.85){0.3}{25}{180}
  \psline{->}(1,0)(1,0.45)
  \psline(1,0)(1,0.55)
  \psline(1,0.55) (0.55,1)
  \psline{->}(1,0.55)(0.6,0.95)
  \psline(1,0.55) (1.45,1)	
  \psline{->}(1,0.55)(1.4,0.95)
  \rput[bl]{0}(0.7,0){$c$}
  \rput[br]{0}(1.7,0.8){$b$}
  \rput[bl]{0}(0.4,0.7){$a$}
  \rput[bl]{0}(0.15,0.1){$\bar{a}$}
 \scriptsize
  \rput[bl]{0}(1.15,0.35){$\mu$}
  \end{pspicture}
  }
 = \sum_{\nu} \left[A^{ab}_c\right]_{\mu \nu}
   \raisebox{-0.5\height}{
\begin{pspicture}(-0.1,-0.2)(1.5,-1.2)
  \small
  \psset{linewidth=0.9pt,linecolor=black,arrowscale=1.5,arrowinset=0.15}
  \psline{-<}(0.7,0)(0.7,-0.35)
  \psline(0.7,0)(0.7,-0.55)
  \psline(0.7,-0.55) (0.25,-1)
  \psline{-<}(0.7,-0.55)(0.35,-0.9)
  \psline(0.7,-0.55) (1.15,-1)	
  \psline{-<}(0.7,-0.55)(1.05,-0.9)
  \rput[tl]{0}(0.4,0){$b$}
  \rput[br]{0}(1.4,-0.95){$c$}
  \rput[bl]{0}(0,-0.95){$\bar{a}$}
 \scriptsize
  \rput[bl]{0}(0.85,-0.5){$\nu$}
  \end{pspicture}
  },\\
  \label{eqn:Abend}
\end{equation}

\begin{equation}
 \raisebox{-0.5\height}{
\begin{pspicture}[shift=-0.65](-0.1,-0.2)(2.15,1.2)
  \small
  \psset{linewidth=0.9pt,linecolor=black,arrowscale=1.5,arrowinset=0.15}
  \psline{->}(2,0)(2,0.55)
  \psline(2,0)(2,0.85)
  \psarc{<-}(1.7,0.85){0.3}{65}{155}
  \psarc{-}(1.7,0.85){0.3}{0}{155}
  \psline{->}(1,0)(1,0.45)
  \psline(1,0)(1,0.55)
  \psline(1,0.55) (0.55,1)
  \psline{->}(1,0.55)(0.6,0.95)
  \psline(1,0.55) (1.45,1)	
  \psline{->}(1,0.55)(1.4,0.95)
  \rput[bl]{0}(0.7,0){$c$}
  \rput[br]{0}(1.55,0.65){$b$}
  \rput[bl]{0}(0.4,0.7){$a$}
  \rput[br]{0}(1.8,0.1){$\bar{b}$}
 \scriptsize
  \rput[bl]{0}(1.15,0.35){$\mu$}
  \end{pspicture}
  }
  = \sum_{\nu} \left[B^{ab}_c\right]_{\mu \nu}
   \raisebox{-0.5\height}{
\begin{pspicture}(-0.1,-0.2)(1.5,-1.2)
  \small
  \psset{linewidth=0.9pt,linecolor=black,arrowscale=1.5,arrowinset=0.15}
  \psline{-<}(0.7,0)(0.7,-0.35)
  \psline(0.7,0)(0.7,-0.55)
  \psline(0.7,-0.55) (0.25,-1)
  \psline{-<}(0.7,-0.55)(0.35,-0.9)
  \psline(0.7,-0.55) (1.15,-1)	
  \psline{-<}(0.7,-0.55)(1.05,-0.9)
  \rput[tl]{0}(0.4,0){$a$}
  \rput[br]{0}(1.4,-0.95){$\bar{b}$}
  \rput[bl]{0}(0,-0.95){$c$}
 \scriptsize
  \rput[bl]{0}(0.85,-0.5){$\nu$}
  \end{pspicture}
  }.
\label{eqn:Bbend}
\end{equation}
They can be expressed in terms of $F$-symbols by
\begin{align}
\left[A_c^{ab}\right]_{\mu \nu} &= \sqrt{\frac{d_a d_b}{d_c}}\varkappa_a^* \left[F^{\bar{a}ab}_b\right]^{\ast}_{1,(c,\mu, \nu)}\\
\left[B^{ab}_c\right]_{\mu \nu} &= \sqrt{\frac{d_a d_b}{d_c}}\left[F^{ab\bar{b}}_a\right]_{(c,\mu, \nu),1}
\end{align}
where the phase $\varkappa_a$ is the Frobenius-Schur indicator
\begin{equation}
\varkappa_a = d_a F^{a\bar{a}a}_{a11} .
\end{equation}

Let us start with a time-reversal symmetry $\bf\mc{T}$ and construct a unitary reflection symmetry $\bf \mc{R}$ from $\bf\mc{T}$ as follows. The action of $\bf\mc{R}$ on anyon $a$ is
\beq
{\bf\mc{R}}:\quad a\rightarrow \,^{\bf\mc{T}}\overline{a},
\eeq
and the action of $\bf\mc{R}$ on the topological state $|a,b;c\rangle_\mu$ is
\beq \label{eq: anti monoidal}
F_{\bf \mc{R}}|a,b;c\rangle_\mu \equiv \sum_{\nu} U_{\bf \mc{R}}\left(^{\bf\mc{R}}b, ^{\bf\mc{R}}a; ^{\bf\mc{R}}c\right)_{\mu\nu} |^{\bf\mc{R}}b, ^{\bf\mc{R}}a; ^{\bf\mc{R}}c\rangle_\nu,
\eeq
where $U_{\bf \mc{R}}\left(a, b; c\right)$ is an $N^c_{ab}\times N^c_{ab}$ matrix that is defined in terms of matrix multiplication as follows
\beq\label{eq:U_symbol_map}
U_{\bf \mc{R}}\left(a, b; c\right) \equiv U_{\bf \mc{T}}\left(\overline{b}, \overline{a}; \overline{c}\right)^*\left(B_{\overline{c}}^{\overline{b},\overline{a}}\right)^*\left(A_{\overline{b}}^{\overline{c},a}\right)\left(B_{a}^{c,\overline{b}}\right)^*,
\eeq
and we have suppressed indices when $N^c_{ab}>1$. Note that the positions of $^{\bf\mc{R}}a$ and $^{\bf\mc{R}}b$ are flipped compared to $a$ and $b$ on the two sides of Eq. \eqref{eq: anti monoidal}. Therefore, we call $F_{\bf \mc{R}}$ an \emph{anti-monoidal} functor instead of a monoidal functor. The extra factors take account of the ``flipping'' of anyons after charge-conjugation.

Now we explicitly check that various consistency conditions for anti-monoidal functors can indeed be satisfied. To preserve the structure of braiding and fusion, under the action of ${\bf \mc{R}}$, the $F$ and $R$ symbols should transform according to:
\begin{align}
F_{\bf \mc{R}}[ F^{abc}_{def}] &= U_{\bf \mc{R}}(\,^{\bf \mc{R}}b, \,^{\bf \mc{R}}a; \,^{\bf \mc{R}}e) U_{\bf \mc{R}}(\,^{\bf \mc{R}}c, \,^{\bf \mc{R}}e; \,^{\bf \mc{R}}d) \left(F^{\,^{\bf \mc{R}}c \,^{\bf \mc{R}}b \,^{\bf \mc{R}}a }_{\,^{\bf \mc{R}}d } \right)^{-1}_{\,^{\bf \mc{R}}e \,^{\bf \mc{R}}f}
U^{-1}_{\bf \mc{R}}(\,^{\bf \mc{R}}c, \,^{\bf \mc{R}}b; \,^{\bf \mc{R}}f) U^{-1}_{\bf \mc{R}}(\,^{\bf \mc{R}}f, \,^{\bf \mc{R}}a; \,^{\bf \mc{R}}d) = F^{abc}_{def}
\nonumber \\
F_{\bf \mc{R}} [R^{ab}_c] &= U_{\bf \mc{R}}(\,^{\bf \mc{R}}a, \,^{\bf \mc{R}}b; \,^{\bf \mc{R}}c)  \left(R^{\,^{\bf \mc{R}}a \,^{\bf \mc{R}}b}_{\,^{\bf \mc{R}}c}\right)^{-1} U_{\bf \mc{R}}(\,^{\bf \mc{R}}b, \,^{\bf \mc{R}}a; \,^{\bf \mc{R}}c)^{-1} = R^{ab}_c
\label{eqn:UFURConsistencyforR}
\end{align}
This is indeed satisfied if we let $U_{\bf \mc{R}}$ be the expression in Eq.~\eqref{eq:U_symbol_map}, which can be proven by a straightforward diagrammatic manipulation.

Under a vertex basis gauge transformation, $\Gamma^{ab}_c \colon V^{ab}_c \rightarrow V^{ab}_c$, according to the right hand side of Eq.~\eqref{eq:U_symbol_map}, $[U_{\bf \mc{R}}(a, b; c )]_{\mu\nu} $ transforms in the following way
\begin{equation}\label{eq:vertex_trans_R}
\tilde{U}_{{\bf \mc{R}}}(a,b,c)_{\mu \nu} = \sum_{\mu', \nu'} \left[\Gamma^{^{\bf \overline{\mc{R}}}{b} ^{\bf \overline{\mc{R}}}{a}}_{^{\bf\overline{\mc{R}}}{c}}\right]_{\mu, \mu'} U_{{\bf g}}(a,b,c)_{\mu' \nu'}\left[(\Gamma^{ab}_c)^{-1}\right]_{\nu'\nu},
\end{equation}
with the shorthand $\overline{{\bf \mc{R}}}={\bf \mc{R}}^{-1}$. This is indeed what we expect from an anti-monoidal functor. Here we use the gauge-fixing condition $\Gamma^{a1}_a = \Gamma^{1a}_a = \Gamma^{\bar{a}a}_1 = 1$ regarding the vertex basis gauge transformation. Under a symmetry action gauge transformation,  $[U_{\bf \mc{R}}(a, b; c )]_{\mu\nu} $ transforms in the following way
\begin{equation}\label{eq:sa_trans_R}
\tilde{U}_{{\bf \mc{R}}}(a,b,c)_{\mu \nu} = \frac{\gamma_{\bf \mc{T}}(\bar{a})^*\gamma_{\bf \mc{T}}(\bar{b})^*}{\gamma_{\bf \mc{T}}(\bar{c})^*}U_{{\bf \mc{R}}}(a,b,c)_{\mu \nu}\,,
\end{equation}
which is also what we expect.

Finally, we should write down how the $\eta$-symbols match. This can be done by considering the consistency equation between $U$-symbols and $\eta$-symbols. For example, suppose $\bf g$ is some unitary symmetry that does not reverse orientation, we have

\beq
\kappa_{ {\bf \mc{R}}, {\bf g}}(a, b; c) \equiv U_{\bf \mc{R}}(a,b;c)^{-1} U_{\bf g}( \,^{\overline{\bf \mc{R}}}b, \,^{\overline{\bf \mc{R}}}a; \,^{\overline{\bf \mc{R}}}c  )^{-1} U_{\bf \mc{R}g}(a,b;c) = \kappa_{\bf \mc{T}, \bf g}\left(\overline{b}, \overline{a};\overline{c}\right)^* = \frac{\eta_{\overline{a}}\left(\bf \mc{T}, \bf g\right)^*\eta_{\overline{b}}\left(\bf \mc{T}, \bf g\right)^*}{\eta_{\overline{c}}\left(\bf \mc{T}, \bf g\right)^*}
\eeq
Hence, the correspondence between $\eta$-symbols should be given by the following equation,
\beq
\eta_a({\bf \mc{R}}, {\bf g}) = \eta_{\bar{a}}({\bf \mc{T}}, {\bf g})^*\,.
\eeq
Following similar derivation, we have
\begin{align}
\eta_a({\bf g}, {\bf \mc{R}}) &= \frac{\eta_{\bar{a}}({\bf g}, {\bf \mc{T}})^*}{U_{\bf g}(\bar{a}, a;1)}\,,\\
\eta_a({\bf \mc{R}_1}, {\bf \mc{R}_2}) &= \eta_{a}({\bf \mc{T}_1}, {\bf \mc{T}_2})U_{\bf \mc{T}_1}(a, \bar{a}; 1)\,.
\end{align}
It is straightforward to check that the desired consistency conditions for the $\eta$-symbols of reflection symmetries are also satisfied.

\section{Wallpaper group symmetries: group structure and $\Z_2$ cohomology}\label{app:cohomology}

In this appendix, for the readers' convenience, we collect necessary information of the wallpaper group symmetries appearing in this paper, $p6$, $p6m$, $p4$ and $p4m$, together with its $\Z_2$ cohomology. This will be important in the identification of symmetry fractionalization classes and the calculation of anomaly matching. A complete list of the $\Z_2$ cohomology for all the 17 wallpaper group symmetries is collected in Ref.~\cite{Ye2021a}.

The $\Z_2$ cohomology of wallpaper group symmetries is presented in terms of its $\Z_2$ cohomology ring. The product in the $\Z_2$ cohomology ring is understood as the cup product. Namely, given $\omega\in H^k(G, \Z_2)$ and $\eta\in H^n(G, \Z_2)$, we can define $\omega \cup \eta\in H^{k+n}(G, \Z_2)$, which is abbreviated to  $\omega \eta$ in this paper, such that
\beq
(\omega \cup \eta)(g_1,\dots, g_{k+n}) = \omega(g_1, \dots,g_k) \cdot \eta(g_{k+1}, \dots, g_l)\,.
\eeq
Here $\cdot$ is simply the multiplication in $\Z_2$. By identifying a set of generators $A_\bullet, B_\bullet$, etc, we can identify all elements in the $\Z_2$ cohomology with the help of addition and cup product.

We define a set of functions that take integers as their arguments:
\beq
\begin{aligned}
&P(x)=
\left\{
\begin{array}{lr}
1, & x{\rm\ is\ odd}\\
0, & x{\rm\ is\ even}
\end{array}
\right.,
\  
P_c(x)=1-P(x),\\
& [x]_a=\{y=x\ ({\rm mod\ }a)|0\leqslant y<a\},
\   
P_{ab}(x)=
\left\{
\begin{array}{lr}
1, & x=b\ ({\rm mod\ }a)\\
0, & {\rm otherwise}
\end{array}
\right.
\end{aligned}
\eeq

When writing down the cohomology corresponding to the LSM constraints, we also need the cohomology of $SO(3)$ and $ SO(3)\times \Z_2^T$. We use $w_2\in H^2(SO(3), \Z_2)$ to denote the second Stiefel-Whitney class of $SO(3)$, and $t\in H^1(\Z_2, \Z_2)$ to denote the generator for the $\Z_2$ cohomology of the time-reversal $\Z_2^T$ symmetry.

\subsection{$p6$}

This group is generated by $T_1$, $T_2$ and $C_6$, two translations with translation vectors that have the same length and an angle of $2\pi/3$, and a 6-fold rotational symmetry, such that 
\beq
C_6^6=1, ~~~~~~C_6T_1C_6^{-1}=T_1T_2, ~~~~~~C_6T_2C_6^{-1}=T_1^{-1}, ~~~~~~T_1 T_2 = T_2 T_1.
\eeq
An arbitrary element in $p6$ can be written as $g=T_1^x T_2^y C_6^c$, with $x, y\in\Z$ and $c\in\{0, 1, 2, 3, 4, 5\}$. 

The $\Z_2$ cohomology ring of $p6$ is 
\beq
\Z_2[A_c; B_{xy}]/(B_{xy}^2= B_{xy}A_c^2)
\eeq
Here $H^1(p6, \Z_2) = \Z_2$, with generator $\xi_1=A_c$, which have a representative cochain,
\beq
\xi_1(g) = c.
\eeq
$H^2(p6, \Z_2) = \Z_2^2$, with generators $\lambda_1 = B_{xy}, \quad \lambda_2 =  A_{c}^2$, and we can choose the representative cochains to be
\begin{align}\label{eq:cochain_p6_Bxy lattice}
&B_{xy}(g_1, g_2)=    P_{60}(c_1)y_1x_2+P_{61}(c_1)\left(\frac{x_2(x_2-1)}{2}+y_1x_2-y_2(x_2+y_1)\right)\nonumber\\
&\qquad\qquad\ 
+P_{62}(c_1)\left(\frac{y_2(y_2+1)}{2}-x_2-y_2(x_2+y_1)\right)+P_{63}(c_1)(-x_2+y_2-y_1x_2)\nonumber\\
&\qquad\qquad\ 
+P_{64}(c_1)\left(\frac{x_2(x_2-1)}{2}+y_2-y_1x_2-y_2(x_2-y_1)\right)+P_{65}(c_1)\left(\frac{y_2(y_2+1)}{2}-y_2(x_2-y_1)\right)\\
\label{eq:cochain_p6_Ac2 lattice}
&A_c^2(g_1, g_2)=c_1 c_2
\end{align}

According to Ref. \cite{Ye2021a}, the anomalies of $p6\times SO(3)$ symmetric lattice systems in lattice homotopy classes $a$, $c$ and $a+c$ can be respectively written as
\begin{align}
&\exp\left(\pi i (B_{xy}+A_c^2) w_2\right)\,,\\
&\exp\left(\pi i B_{xy} w_2 \right)\,,\\
&\exp\left(\pi i A_c^2 w_2\right)\,.
\end{align}

\subsection{$p6m$}

This group is generated by $T_1$, $T_2$, $C_6$ and $M$, where the first three generators have the same properties as those in $p6$, and the last one is a mirror symmetry whose mirror axis passes through the $C_6$ center and bisects $T_1$ and $T_2$, such that 
\beq
\begin{aligned}
&M^2=1, ~~~~~~ MC_6M = C_6^{-1}, ~~~~~~ MT_1M=T_2,  ~~~~~~MT_2M=T_1,\\
&C_6^6=1, ~~~~~~ C_6T_1C_6^{-1}=T_1T_2,  ~~~~~~C_6T_2C_6^{-1}=T_1^{-1},  ~~~~~~T_1 T_2 = T_2 T_1.
\end{aligned}
\eeq
An arbitrary element in $p6m$ can be written as $g=T_1^x T_2^y C_6^c M^m$, with $x, y\in\Z$, $c\in\{0, 1, 2, 3, 4, 5\}$ and $m\in\{0, 1\}$. 

The $\Z_2$ cohomology ring of $p6m$ is 
\beq
\Z_2[A_c, A_m; B_{xy}]/\left(B_{xy}^2=B_{xy}(A_c^2+A_c A_m) \right)
\eeq
Here $H^1(p6m, \Z_2) = \Z_2^2$, with generators $\xi_1=A_c,\quad\xi_2=A_m$, which have representative cochains,
\beq
\xi_1(g) = c,\quad\quad\xi_2(g)=m.
\eeq
$H^2(p6m, \Z_2) = \Z_2^4$, with generators $\lambda_1 = B_{xy}, \quad \lambda_2 =  A_{c}^2,\quad \lambda_3 = A_c A_m, \quad \lambda_4 = A_m^2$, and we can choose the representative cochains to be
\begin{align}\label{cochain_p6m_Bxy_Z2}
    &B_{xy}(g_1, g_2) = P_{60}(c_1)\left[P_c(m_1)y_1x_2+m_1y_2(x_2+y_1)\right]\nonumber\\
    &\qquad\qquad\ 
    +P_{61}(c_1)\left[P_c(m_1)\left(\frac{x_2(x_2-1)}{2}+y_1x_2-y_2(x_2+y_1)\right)+m_1\left(\frac{y_2(y_2-1)}{2}+y_1(-x_2+y_2)\right)\right]\nonumber\\
    &\qquad\qquad\ 
    +P_{62}(c_1)\left[P_c(m_1)\left(\frac{y_2(y_2+1)}{2}-x_2-y_2(x_2+y_1)\right)+m_1\left(\frac{x_2(x_2+1)}{2}-y_2-y_1x_2\right)\right]\nonumber\\
    &\qquad\qquad\ 
    +P_{63}(c_1)\left[P_c(m_1)(-x_2+y_2-y_1x_2)+m_1(x_2-y_2+y_2(x_2-y_1)\right]\nonumber\\
    &\qquad\qquad\ 
    +P_{64}(c_1)\left[P_c(m_1)\left(\frac{x_2(x_2-1)}{2}+y_2-y_1x_2-y_2(x_2-y_1)\right)+m_1\left(\frac{y_2(y_2-1)}{2}+x_2+y_1(x_2-y_2)\right)\right]\nonumber\\
    &\qquad\qquad\ 
    +P_{65}(c_1)\left[P_c(m_1)\left(\frac{y_2(y_2+1)}{2}-y_2(x_2-y_1)\right)+m_1\left(\frac{x_2(x_2+1)}{2}+y_1x_2\right)\right]\\
    \label{cochain_p6m_Bc2_Z2}
    &A_c^2(g_1, g_2)=c_1c_2,\quad\quad A_cA_m(g_1, g_2) = m_1 c_2,\quad\quad A_m^2(g_1, g_2) = m_1 m_2
\end{align}

According to Ref. \cite{Ye2021a}, the anomalies of $p6m\times SO(3)\times\Z_2^T$ symmetric lattice systems in lattice homotopy classes $a$, $c$ and $a+c$ can be respectively written as
\begin{align}
&\exp\left(\pi i (B_{xy}+A_c^2 + A_cA_m) (w_2 + t^2)\right)\,,\\
&\exp\left(\pi i B_{xy} (w_2 + t^2) \right)\,,\\
&\exp\left(\pi i (A_c^2 + A_cA_m) (w_2 + t^2)\right)\,.
\end{align}

\subsection{$p4$}

This group is generated by $T_1$, $T_2$ and $C_4$, two translations with perpendicular translation vectors that have equal length, and a 4-fold rotational symmetry, such that 
\beq
C_4^4=1,~~~~~~ C_4T_1C_4^{-1}=T_2,~~~~~~C_4T_2C_4^{-1}=T_1^{-1}, ~~~~~~T_1 T_2 = T_2 T_1.
\eeq
An arbitrary element in $p4$ can be written as $g=T_1^x T_2^y C_4^c$, with $x, y\in\Z$ and $c\in\{0, 1, 2, 3\}$.

The $\Z_2$ cohomology ring of $p4$ is 
\beq
\begin{aligned}
\mathbb{Z}_2[A_{c}, A_{x+y}; B_{c^2}, B_{xy}]/\big(&A_{c}^2=0,~ A_{x+y}A_{c} = 0, \\
&B_{xy}A_{x+y} = B_{xy}A_{c}, ~B_{c^2} A_{x+y} = B_{xy}A_{x+y} + A_{x+y}^3,\\
&B_{xy}^2 = B_{xy}B_{c^2}\big)
\end{aligned}
\eeq
Here $H^1(p4, \Z_2) = \Z_2^2$, with generators $\xi_1=A_{x+y},\quad\xi_2=A_c$, which have representative cochains,
\beq
\xi_1(g) = x+y,\quad\quad\xi_2(g)=c.
\eeq
$H^2(p4, \Z_2) = \Z_2^3$, with generators $\lambda_1 = B_{xy}, \quad \lambda_2 = B_{c^2}, \quad \lambda_3 = A_{x+y}^2$, and we can choose the representative cochains to be 
\begin{align}
&B_{xy}(g_1, g_2) = P_c(c_1)y_1 x_2 + P(c_1) y_2(y_1 + x_2) \\
&B_{c^2}(g_1, g_2) = \frac{[c_1]_4+[c_2]_4-[c_1+c_2]_4}{4}\\
&A_{x+y}^2(g_1, g_2) = (x_1 + y_1)(x_2 + y_2)
\end{align}

According to Ref. \cite{Ye2021a}, the anomalies of $p4\times SO(3)$ symmetric lattice systems in lattice homotopy classes $a$, $b$, $c$, $a+b$, $a+c$, $b+c$ and $a+b+c$ can be respectively written as
\begin{align}
&\exp\left(\pi i (B_{xy}+B_{c^2}+A_{x+y}^2) w_2\right)\,,\\
&\exp\left(\pi i B_{xy} w_2 \right)\,,\\
&\exp\left(\pi i A_{x+y}^2 w_2\right)\,,\\
&\exp\left(\pi i (B_{c^2}+A_{x+y}^2) w_2\right)\,,\\
&\exp\left(\pi i (B_{xy}+B_{c^2}) w_2\right)\,,\\
&\exp\left(\pi i (B_{xy}+A_{x+y}^2) w_2\right)\,,\\
&\exp\left(\pi i B_{c^2} w_2\right)\,.
\end{align}

\subsection{$p4m$}

This group is generated by $T_1$, $T_2$, $C_4$ and $M$, where the first three generators have the same properties as those in $p4$, and the last generator $M$ is a mirror symmetry that flips the translation vector of $T_1$, such that 
\beq
\begin{aligned}
&M^2=1, ~~~~~~ MC_4 M =C_4^{-1},~~~~~~  MT_1M=T_1^{-1}, ~~~~~~ MT_2M=T_2,\\
&C_4^4=1,~~~~~~  C_4T_1C_4^{-1}=T_2, ~~~~~~ C_4T_2C_4^{-1}=T_1^{-1},~~~~~~  T_1 T_2 = T_2 T_1.
\end{aligned}
\eeq
An arbitrary element in $p4m$ can be written as $g=T_1^xT_2^yC_4^cM^m$, with $x, y\in\Z$, $c\in\{0, 1, 2, 3\}$ and $m\in\{0, 1\}$. 

The $\Z_2$ cohomology ring of $p4m$ is 
\beq
\begin{aligned}
\mathbb{Z}_2[A_{c}, A_m, A_{x+y}; B_{c^2}, B_{xy}]/\big(&A_{c}(A_c+A_m) = 0,~ A_{x+y}A_{c} = 0, \\
&B_{xy}A_{x+y} = B_{xy}(A_{c} + A_m), ~B_{c^2} A_{x+y} = B_{xy}A_{x+y} + A_{x+y}^3 + A_{x+y}^2 A_m, \\
&B_{xy}^2 = B_{xy}B_{c^2}\big)
\end{aligned}
\eeq
Here $H^1(p4m, \Z_2) = \Z_2^3$, with generators $\xi_1=A_{x+y},\quad\xi_2=A_c,\quad\xi_3=A_m$, which have representative cochains,
\beq
\xi_1(g) = x+y,\quad\quad\xi_2(g)=c,\quad\quad\xi_3(g)=m.
\eeq
$H^2(p4m, \Z_2) = \Z_2^6$, with generators $\lambda_1 = B_{xy}, \quad \lambda_2 = B_{c^2}, \quad \lambda_3 = A_{x+y}^2,\quad\lambda_4 = A_{x+y}A_m, \quad \lambda_5 = A_c^2, \quad \lambda_6 = A_m^2$, and we can choose the representative cochains to be
\begin{align}
&B_{xy}(g_1, g_2) = P_c(c_1)y_1 x_2 + P(c_1) y_2(y_1 + x_2) \\
&B_{c^2}(g_1, g_2) = \frac{[c_1]_4+(-1)^{m_1}[c_2]_4-[c_1+(-1)^{m_1}c_2]_4}{4}\\
&A_{x+y}^2(g_1, g_2) = (x_1 + y_1)(x_2 + y_2),\quad A_{x+y}A_m(g_1, g_2) = m_1(x_2+y_2), \\
&A_c^2(g_1, g_2) = c_1 c_2,\quad A_m^2(g_1, g_2) = m_1 m_2
\end{align}

According to Ref. \cite{Ye2021a}, the anomalies of $p4m\times SO(3)\times\Z_2^T$ symmetric lattice systems in lattice homotopy classes $a$, $b$, $c$, $a+b$, $a+c$, $b+c$ and $a+b+c$ can be respectively written as
\begin{align}
&\exp\left(\pi i (B_{xy}+ B_{c^2} + A_{x+y}(A_{x+y}+A_m)) (w_2 + t^2)\right)\,,\\
&\exp\left(\pi i B_{xy} (w_2 + t^2) \right)\,,\\
&\exp\left(\pi i A_{x+y}(A_{x+y} + A_m) (w_2 + t^2) \right)\,,\\
&\exp\left(\pi i (B_{c^2} + A_{x+y}(A_{x+y}+A_m)) (w_2 + t^2)\right)\,,\\
&\exp\left(\pi i (B_{xy}+ B_{c^2}) (w_2 + t^2)\right)\,,\\
&\exp\left(\pi i (B_{xy} + A_{x+y}(A_{x+y}+A_m)) (w_2 + t^2)\right)\,,\\
&\exp\left(\pi i B_{c^2} (w_2 + t^2)\right)\,.
\end{align}

\section{Details of realizations: Anyon permutation patterns and symmetry fractionalization classes}\label{app:SFclasses}

In this appendix, for the topological orders appearing in this paper, we give the full details of all possible symmetry fractionalization classes given different anyon permutation patterns, including the explicit representative cochain for each generator of the symmetry fractionalization classes. For $\Z_2$ and $\Z_4$ topological orders, we also upload codes using which one can 1) see all symmetry fractionalization classes of the symmetry-enriched states within each lattice homotopy class, and 2) check which lattice homotopy class a given symmetry-enriched state belongs to \cite{code}. As for $\Z_3$, $U(1)_{2}\times U(1)_{-2}$ and $U(1)_{4}\times U(1)_{-4}$ topological orders, all symmetry enrichment patterns lead to anomaly-free states.

As reviewed in Sec. \ref{subsec: global symmetry}, given a topological order and how the symmetry $G$ permutes the anyons of this topological order, all possible symmetry fractionalization classes form a torsor over $H^2(G, \mc{A})$, where $\mc{A}$ is the group formed by abelian anyons in this topological order (to simplify the notation, in this appendix we will not write down the subscript $\rho$ of $H^2_\rho(G, \mc{A})$). Given a reference set of $\eta$-symbols for $G$, which can be chosen to come from the pullback of the $\eta$-symbols of the topological symmetry using Eq. \eqref{eq: microscopic U and eta}, we can identify all other symmetry fractionalization classes from Eq.~\eqref{eq:torsor}. Hence, all we need to do is to identify all elements in $H^2(G, \mc{A})$. More preciesely, we need to write down the representative cochains of all generators of $H^2(G, \mc{A})$. It turns out that elements in $H^2(G, \mc{A})$ can usually be determined by its relation to $H^2(G, \Z_2)$ and $H^2(G, \Z)$. The $\Z_2$ cohomology of the wallpaper group symmetries involved in this paper has been collected in Appendix~\ref{app:cohomology}, and the $\Z_2$ cohomology of all wallpaper group symmetries are worked out in Ref.~\cite{Ye2021a}. Also recall that we use $t$ to denote the generator of $H^1(\Z_2^T, \Z_2)$, as in Appendix \ref{app:cohomology}.

Now we explain some technical tricks to identify the elements in $H^2(G, \mc{A})$. Let us specialize to the case where $\mc{A}=\Z_{N}$ (with a potential nontrivial $G$-action on $\mc{A}$). Consider the projection map $p$ from $\Z$ to $\Z_{N}$,
\beq 
p\colon \Z\rightarrow \Z_{N}\,,
\eeq
 which induces the map between $\Z$ cohomology and $\Z_{N}$ cohomology 
\beq \label{eq: map from Z to Z2N}
p_*\colon H^k(G, \Z)\rightarrow H^k(G, \Z_{N})\,.
\eeq
Given an element $[\omega]\in H^k(G, \Z)$ with some representative cochain $\omega$, the representative cochain of $[p_*(\omega)]$ is identically $\omega$ with the outcome understood as an element in $\Z_{N}$ instead of $\Z$. We will use $\widetilde{\omega}$ to label the obtained element in $H^k(G, \Z_{N})$.  

To identify an element in $H^k(G, \Z)$, usually it is helpful to consider the Bochstein homomorphism \cite{brown2012cohomology} associated to the short exact sequence $1\rightarrow\Z\rightarrow\Z\rightarrow\Z_2\rightarrow 1$,
\beq
\beta \colon H^{k-1}(G, \Z_2)\rightarrow H^k(G, \Z)\,,
\eeq
In particular, for $k=2$, consider an element $[x]\in H^1(G, \Z_2)$ with representative cochain $x$, the representative cochain of $[\beta(x)]$ is given by 
\beq\label{eq: Bochstein}
\beta(x)({\bf g}, {\bf h}) = \frac{x({\bf g}) + (-1)^{q({\bf g})}x({\bf h}) - x({\bf gh})}{2}\,,
\eeq
where we demand that $x({\bf g})$ takes values only in $\{0, 1\}$, and $q({\bf g})$ denotes whether the $\bf g$ action on $\Z$ is trivial ($q({\bf g})=0$) or nontrivial ($q({\bf g})=1$).

If $N=2N'$ is even, we can also consider the map $i$ from $\Z_2$ to $\Z_{2N'}$ defined by multiplication by $N'$, \ie
\beq
i\colon \Z_2\rightarrow \Z_{2N'}\,.
\eeq
It induces the map from $\Z_2$ cohomology to $\Z_{2N'}$ cohomology 
\beq \label{eq: map from Z2 to Z2N}
i_*\colon H^k(G, \Z_2)\rightarrow H^k(G, \Z_{2N'})\,.
\eeq
Utilizing the map $i_*$, we can use elements in $H^k(G, \Z_2)$ to identify elements in $H^k(G, \Z_{2N'})$. In particular, given an element $[\omega]\in H^k(G, \Z_2)$ with some representative cochain $\omega$, the representative cochain of $[i_*(\omega)]$ is simply $N'\omega$. For clarity purposes, later we will omit the bracket and use $N'\omega$ to label the obtained element in $H^k(G, \Z_{2N'})$. 

The symmetry group $G$ we consider usually takes the form $G_1\times G_2$. In this situation, we can specify an element in the cohomology of $G$ by specifying an element in the cohomology of $G_1$ or $G_2$. Namely, we can consider the projection,
\beq
f\colon G\rightarrow G_1\,.
\eeq
It induces the map from the cohomology of $G_1$ to the cohomology of $G$
\beq
f^*\colon H^k(G_1, \mc{A})\rightarrow H^k(G, \mc{A})\,.
\eeq
Hence, given an element $[\omega]\in H^k(G_1, \mc{A})$ with some representative cochain $\omega$, we can use it to specify $[f^*\omega]$. Writing an element in $g\in G$ as $g_1 g_2$ with $g_1\in G_1$ and $g_2\in G_2$, the representative cochain of $[f^*\omega]$ can be identified as 
\beq
f^*\omega(g, h, \dots) = \omega(g_1, h_1, \dots)\,.
\eeq

In this appendix, to simplify the notation, we will not explicitly write down the $f^*$ symbol in the front and use the cohomology and cochain of a subgroup to implicitly refer to the cohomology and cochain of the total group. For example, we may specify an element $[\omega_1]\in H^k(G_1, \mc{A})$ and an element $[\omega_2]\in H^k(G_2, \mc{A})$, then an element in $H^k(G, \mc{A})$ written as $\omega_1+\omega_2$ really means $[f_1^*\omega_1]+[f_2^*\omega_2]$, where $f_{1, 2}: G\rightarrow G_{1, 2}$ is the projection from $G$ to $G_{1, 2}$.

It turns out that the above is enough to determine almost all symmetry fractionalization classes of our interest.

In this paper, for chiral topological orders, the symmetry groups we consider are $p6\times SO(3)$ and $p4\times SO(3)$. For non-chiral topological orders, we explicitly discuss the symmetry fractionalization classes for $p6m\times SO(3)\times \Z_2^T$ and $p4m\times SO(3)\times \Z_2^T$ in this appendix, and we can simply ignore all terms involving $w_2$ to get the corresponding symmetry fractionalization classes for $p6m \times \Z_2^T$ and $p4m \times \Z_2^T$, where $w_2$ will be defined later and detects whether certain anyon carries an half-odd-integer spin under $SO(3)$.

\subsection{$\U_{2N}$}

In this case, we have $\mc{A} = \Z_{2N}$. The topological symmetry of $\U_{2N}$ is complicated for general $N$. For $N=1$, there is no nontrivial topological symmetry. For $N\geqslant 2$, there is always a $\Z_2$ subgroup of topological symmetry generated by the charge conjugation symmetry $C$ such that anyon $(a) \rightarrow ([-a]_{2N})$ under $C$. For this topological symmetry, we can take
\begin{align}
U_{C}((a),(b);([a+b]_{2N})) = 
\begin{cases}
(-1)^a & b>0\\
1 & b=0
\end{cases}
\end{align}
and a set of $\eta$-symbols equal to 1. When $2\leqslant N \leqslant 5$, this is the whole topological symmetry group. In the latter discussion we consider general $N\geqslant 2$, but limit to the cases where $G$ can only act as charge conjugation. For $N=1$, we just need to ignore the cases where $G$ can permute anyons nontrivially.

We start with explaining the symmetry fractionalization classes of the warmup example in Sec.~\ref{subsec: Z2}, $\Z_2\times SO(3)$. Denote the generator of the $\Z_2$ group as $C_2$. Depending on whether $C_2$ permutes anyons or not, we have two possibilities.

\begin{enumerate}
    \item Trivial $C_2$ action.

    The possible symmetry fractionalization classes are given by 
    \beq
    H^2(\Z_2\times SO(3), \Z_{2N}) = \Z_2\oplus\Z_2\,.
    \eeq

    We denote the generator of the first $\Z_2$ piece by $\widetilde{\beta}(x)$, where $x\in H^1(\Z_2, \Z_2)$ is the nontrivial generator, and the tilde  is because it comes from the image of $p_*\colon H^2(\Z_2\times SO(3), \Z)\rightarrow H^2(\Z_2\times SO(3), \Z_{2N})$, with trivial $\Z_2\times SO(3)$ action on $\Z$. We can explicitly write down the representative cochain of $\widetilde{\beta}(x)$ according to Eq.~\eqref{eq: Bochstein},
    \beq
    \widetilde{\beta}(x)(C_2^i, C_2^j) = \frac{i + j - [i+j]_2}{2} = ij\,,
    \eeq
    where $i,j\in \{0, 1\}$.

    We denote the generator of the second $\Z_2$ piece by $Nw_2$, where $w_2\in H^2(SO(3), \Z_2)$ is the second Stiefel-Whitney class and the $N$ in the front is because it comes from the image of $i_*\colon H^2(\Z_2\times SO(3), \Z_2)\rightarrow H^2(\Z_2\times SO(3), \Z_{2N})$. The explicit representative cochain of $w_2$ when restricted to the subgroup generated by the $\pi$-rotations about two orthogonal axes is
    \beq\label{eq:rep_cochain_w2 app}
    w_2(U_\pi^{i_1} {U'}_\pi^{i_2}, U_\pi^{j_1} {U'}_\pi^{j_2}) = N(i_1j_1 + i_2j_2 + i_1j_2) \mod 2\,.
    \eeq
    
    \item   Nontrivial $C_2$ action.
    
    The possible symmetry fractionalization classes are given by 
    \beq
    H^2(\Z_2\times SO(3), \Z_{2N}) = \Z_2\oplus\Z_2\,.
    \eeq

    We denote the generator of the two $\Z_2$ pieces by $Nx^2$ and $Nw_2$, because both come from the image of $i_*\colon H^2(\Z_2\times SO(3), \Z_2)\rightarrow H^2(\Z_2\times SO(3), \Z_{2N})$. In particular, the representative cochain of $x^2\in H^2(\Z_2\times SO(3), \Z_2)$ is still
    \beq\label{eq:cochain_x2}
    x^2(C_2^i, C_2^j) = \frac{i + j - [i+j]_2}{2} = ij\,,
    \eeq
    and the representative cochain of $Nx^2\in H^2(\Z_2\times SO(3), \Z_{2N})$ is simply multiplication of Eq.~\eqref{eq:cochain_x2} by $N$.

\end{enumerate}

For $p6\times SO(3)$, there are two possible anyon permutation patterns, determined by whether $C_6$ permutes anyons or not. The classification of symmetry fractionalization classes and the generators for the two possibilities are listed in Table~\ref{tab:SF U(1)2N}. The generators with tilde come from the image of $p_*\colon H^2(p6\times SO(3), \Z)\rightarrow H^2(p6\times SO(3), \Z_{2N})$, with different actions of $p6\times SO(3)$ on $\Z$. Now we present the information about $H^2(p6, \Z)$ and the generators for completeness.

\begin{enumerate}
\item Trivial action on $\Z$.
\beq
H^2(p6, \Z) = \Z\oplus\Z_6\,.
\eeq
We denote the generator of the $\Z$ piece and the $\Z_6$ piece by $\mathscr{B}^{(1)}_{xy}$ and $\mathscr{B}^{(1)}_{c^2}$, respectively, which have representative cochains,
\begin{align}\label{eq:cochain_p6_Bxy}
&\mathscr{B}^{(1)}_{xy}(g_1, g_2)=    P_{60}(c_1)y_1x_2+P_{61}(c_1)\left(\frac{x_2(x_2-1)}{2}+y_1x_2-y_2(x_2+y_1)\right)\nonumber\\
&\qquad\qquad\ 
+P_{62}(c_1)\left(\frac{y_2(y_2+1)}{2}-x_2-y_2(x_2+y_1)\right)+P_{63}(c_1)(-x_2+y_2-y_1x_2)\nonumber\\
&\qquad\qquad\ 
+P_{64}(c_1)\left(\frac{x_2(x_2-1)}{2}+y_2-y_1x_2-y_2(x_2-y_1)\right)+P_{65}(c_1)\left(\frac{y_2(y_2+1)}{2}-y_2(x_2-y_1)\right)\\
\label{eq:cochain_p6_Ac2}
&\mathscr{B}^{(1)}_{c^2}(g_1, g_2)=\frac{[c_1]_6+[c_2]_6-[c_1+c_2]_6}{6}
\end{align}

The representative cochain of $\mathscr{B}^{(1)}_{xy}$ has identically the same expression as Eq.~\eqref{eq:cochain_p6_Bxy lattice}. Note that if we think of the expression as a representative cochain of $\Z_2$ cohomology, it does not matter whether we have $+$ sign or $-$ sign in front of an integer, because we only care about the mod 2 value of the expression. However, we carefully choose the sign in Eq.~\eqref{eq:cochain_p6_Bxy lattice} or Eq.~\eqref{eq:cochain_p6_Bxy} such that the expression is a $\Z$ cochain as well. Hence, we immediately see that the $\Z_2$ reduction of $\mathscr{B}^{(1)}_{xy}$ is $B_{xy}$ in Eq. \eqref{eq:cochain_p6_Bxy lattice}. Likewise, the $\Z_2$ reduction of $\mathscr{B}^{(1)}_{c^2}$ is $A_c^2$ in Eq. \eqref{eq:cochain_p6_Ac2 lattice}.

\item $C_6$ acts nontrivially on $\Z$.
\beq
H^2(p6, \Z) = 0\,.
\eeq
\end{enumerate}

\begin{table}[!htbp]
\centering
\renewcommand{\arraystretch}{1.3}
\begin{tabular}{|c|c|c|c|c|}
\toprule[2pt]
Symmetry Group & Action & $H^2(G, \mc{A})$ & Realizations & Generators \\ \hline
\multirow{2}{*}{$p6\times SO(3)$} & Trivial & $\Z_{2N}\oplus\Z_{(2N, 6)}\oplus(\Z_2)$ & $4(N(N,3)+1)$ & $\widetilde{\mathscr{B}}^{(1)}_{xy}$, $\widetilde{\mathscr{B}}^{(1)}_{c^2}$, $Nw_2$ \\ \cline{2-5}  
     & $C_6\colon (a)\rightarrow([-a]_{2N})$ & $(\Z_2)^3$ & 8 & $NB_{xy}$, $NA_{c}^2$, $Nw_2$ \\ \hline
\multirow{4}{*}{$p4\times SO(3)$} & Trivial & $\Z_{2N}\oplus\Z_{(2N, 4)}\oplus(\Z_2)^2$ & $8(N(N,2)+1)$ & $\widetilde{\mathscr{B}}^{(1)}_{xy}$, $\widetilde{\mathscr{B}}^{(1)}_{c^2}$, $\widetilde{\beta}(A_{x+y})$, $Nw_2$ \\ \cline{2-5}  
     & $C_4\colon (a)\rightarrow([-a]_{2N})$ & $(\Z_2)^4$ & 16 & $NB_{xy}$, $NB_{c^2}$, $\widetilde{\beta}(A_{x+y})$, $Nw_2$ \\ \cline{2-5}
     & $T_{1,2}\colon (a)\rightarrow([-a]_{2N})$ & $\Z_{(2N,4)}\oplus (\Z_2)^3$ & $8((N,2)+1)$ & $\widetilde{\mathscr{B}}^{(3)}_{xy}$, $NB_{c^2}$, $NA_{x+y}^2$, $Nw_2$ \\ \cline{2-5}
     & $T_{1,2}, C_4\colon (a)\rightarrow([-a]_{2N})$ & $\Z_{(2N,4)}\oplus (\Z_2)^3$ &  $8((N,2)+1)$ & $\widetilde{\mathscr{B}}^{(4)}_{xy}$, $NB_{c^2}$, $N A_{x+y}^2$, $Nw_2$ \\ \cline{2-5}
\bottomrule[2pt]	
\end{tabular}
\caption{All possible symmetry fractionalization classes of $G=p6\times SO(3)$ and $G=p4\times SO(3)$ for $\U_{2N}$, given all possible anyon permutation patterns. All generators with a tilde come from $H^2(G, \Z)$ via Eq. \eqref{eq: map from Z to Z2N}, and all generators with $N$ in the front come from $H^2(G, \Z_2)$ via Eq. \eqref{eq: map from Z2 to Z2N}. When counting the number of realizations in each case, overcounts due to the equivalence from relabeling anyons have been taken care of. To simplify the notation, in this table sometimes a single symbol can have different meanings. For example, $\widetilde{\mathscr{B}}^{(1)}_{xy}$ for $p6\times SO(3)$ is different from $\widetilde{\mathscr{B}}^{(1)}_{xy}$ for $p4\times SO(3)$, and their precise meanings and expressions can be found in the discussion regarding the $p6\times SO(3)$ and $p4\times SO(3)$ symmetries in this appendix.}
\label{tab:SF U(1)2N}
\end{table}

For $p4\times SO(3)$, there are four possible anyon permutation patterns, determined by whether $C_4$ and $T_{1,2}$ permute anyons or not. The classification of symmetry fractionalization classes and the generators are listed in Table~\ref{tab:SF U(1)2N}. Here we present the information about $H^2(p4, \Z)$ and the generators.

\begin{enumerate}
\item Trivial action on $\Z$.
\beq
H^2(p4, \Z) = \Z\oplus\Z_4\oplus\Z_2\,.
\eeq
We denote the generators of the $\Z$, $\Z_4$ and $\Z_2$ piece by $\mathscr{B}^{(1)}_{xy}$, $\mathscr{B}^{(1)}_{c^2}$ and $\beta^{(1)}(A_{x+y})$, respectively, which have representative cochains,

\begin{align}\label{eq:cochain_p4_Bxy_1}
&\mathscr{B}^{(1)}_{xy}(g_1, g_2)=   P_{40}(c_1) y_1 x_2 -
  P_{41}(c_1)y_2(x_2 + y_1) -
  P_{42}(c_1)y_1 x_2 + 
  P_{43}(c_1) y_2(y_1-x_2)\\
\label{eq:cochain_p4_Ac2_1}
&\mathscr{B}^{(1)}_{c^2}(g_1, g_2)=\frac{[c_1]_4+[c_2]_4-[c_1+c_2]_4}{4}\\
\label{eq:cochain_p4_Ax+y2_1}
&\beta^{(1)}(A_{x+y})(g_1, g_2)=\frac{[x_1+y_1]_2+[x_2+y_2]_2-[x_1+x_2 + y_1 + y_2]_2}{2}
\end{align}

\item $C_4$ acts nontrivially on $\Z$.
\beq
H^2(p6, \Z) = \Z_2\,.
\eeq
We denote the generator by $\beta^{(2)}(A_{x+y})$, which has a representative cochain,
\beq\label{eq:cochain_p6_Ax+y2}
\beta^{(2)}(A_{x+y})(g_1, g_2)=\frac{[x_1+y_1]_2+(-1)^{c_1}[x_2+y_2]_2-[x_1+x_2 + y_1 + y_2]_2}{2}
\eeq

\item $T_{1,2}$ act nontrivially on $\Z$.
\beq
H^2(p6, \Z) = \Z_4\,.
\eeq
We denote the generator by $\mathscr{B}^{(3)}_{xy}$, which has a representative cochain,
\beq\label{eq:cochain_p4_Bxy_3}
\begin{aligned}
&\mathscr{B}^{(3)}_{xy}(g_1, g_2) = (-1)^{\tilde{x}_1}(P_c(c_1) P(\tilde{y}_1) P(\tilde{x}_2) +  P(c_1) P(\tilde{y}_1 + \tilde{x}_2) P(\tilde{y}_2) ) \\
 &\quad\quad\quad\quad\text{with }\tilde{x} = x + P_{41}(c) + P_{42}(c) ,\quad\quad \tilde{y} = y + P_{42}(c) + P_{43}(c) \\
\end{aligned}
\eeq
Note that the $\Z_2$ reduction of $\mathscr{B}^{(3)}_{xy}$ is actually $B_{xy}+B_{c^2}+A_{x+y}^2$.

\item Both $T_{1,2}$ and $C_4$ act nontrivially on $\Z$.
\beq
H^2(p6, \Z) = \Z_4\,.
\eeq
We denote the generator by $\mathscr{B}^{(4)}_{xy}$, which has a representative cochain,
\beq\label{eq:cochain_p4_Bxy_4}
\mathscr{B}^{(4)}_{xy}(g_1, g_2) = (-1)^{x_1}(P_c(c_1) P(y_1) P(x_2) +  P(c_1) P(y_1 + x_2) P(y_2) )
\eeq

\end{enumerate}

\subsection{Ising$^{(\nu)}$}

In this case, we have $\mc{A}=\Z_2$, with no nontrivial topological symmetry. The classification of symmetry fractionalization classes and the generators are listed in Table~\ref{tab:SF Ising}.

\begin{table}[!htbp]
\centering
\renewcommand{\arraystretch}{1.3}
\begin{tabular}{|c|c|c|c|c|}
\toprule[2pt]
Symmetry Group & Action & $H^2(G, \mc{A})$ & Realizations & Generators \\ \hline
$p6\times SO(3)$ & Trivial & $(\Z_2)^3$ & 8 & $B_{xy}$, $A_c^2$, $w_2$ \\ \hline
$p4\times SO(3)$ & Trivial & $(\Z_2)^4$ & 16 & $B_{xy}$, $B_{c^2}$, $A_{x+y}^2$, $w_2$ \\ \hline
\bottomrule[2pt]	
\end{tabular}
\caption{All possible symmetry fractionalization classes of $p6\times SO(3)$ and $p4\times SO(3)$ for Ising$^{(\nu)}$, where $\nu$ is an odd integer.}
\label{tab:SF Ising}
\end{table}

\subsection{$\Z_2$ topological order}

In this case, we have $\mc{A}=(\Z_2)^2$. The topological symmetry is  $\Z_2\times \Z_2^T$. The unitary generator of the topological symmetry is the unitary electric-magnetic duality symmetry $S$ that exchanges $e$ and $m$, \ie
\beq
S\colon(a_e, a_m)\rightarrow (a_m, a_e)\,,
\eeq
while the anti-unitary generator is simply the anti-unitary electric-magnetic duality symmetry that permutes anyons in the same way as $S$.
We can choose the $U$-symbol such that 
\begin{equation}
U_{\bf g}(a,b;c)=
\begin{cases}
(-1)^{a_m b_e} & \text{${\bf g}$ permutes anyons,}\\
1 & \text{otherwise.}
\end{cases}
\end{equation}
And a set of $\eta$-symbols can be chosen such that
\beq
\eta_a({\bf g}_1, {\bf g}_2)=
\begin{cases}
(-1)^{a_e a_m} & \text{${\bf g}_1$, ${\bf g}_2$ permute anyons,}\\
1 & \text{otherwise.}
\end{cases}
\eeq

\begin{table}
\centering
\resizebox{\columnwidth}{!}{%
\begin{tabular}{|c|c|c|c|c |}
\toprule[2pt]
Symmetry Group & Action & $H^2(G, \mc{A})$ & Realizations & Generators \\ \hline
\multirow{8}{*}{$p6m\times SO(3)\times\Z_2^T$} & Trivial & $(\Z_2)^{16}$ & 32896 &
\begin{tabular}{c}  
$(B_{xy}, 0)$, $(A_c^2, 0)$, $(A_cA_m, 0)$, $(A_m^2, 0)$,\\ 
$(A_ct, 0)$, $(A_mt, 0)$, $(t^2, 0)$, $(w_2, 0)$, \\ 
$(0, B_{xy})$, $(0, A_c^2)$, $(0, A_cA_m)$, $(0, A_m^2)$, \\
$(0, A_ct)$, $(0, A_mt)$, $(0, t^2)$,$(0, w_2)$
\end{tabular} \\ \cline{2-5}  

& $\mc{T}\colon (a_e, a_m)\rightarrow(a_m, a_e)$ 
& \multirow{4}{*}{$(\Z_2)^5$} &
\multirow{4}{*}{32} 
& \multirow{4}{*}{\begin{tabular}{c}
$(B_{xy}, B_{xy})$, $(A_{c}^2, A_c^2)$, \\
$(A_cA_m, A_cA_m)$, $(A_m^2, A_m^2)$, $(w_2, w_2)$ \end{tabular}} 
\\  \cline{2-2} 
& $M, \mc{T}\colon (a_e, a_m)\rightarrow(a_m, a_e)$ &  &  \\ \cline{2-2} 
& $C_6, \mc{T}\colon (a_e, a_m)\rightarrow(a_m, a_e)$ &  & \\ \cline{2-2} 
& $C_6, M, \mc{T}\colon (a_e, a_m)\rightarrow(a_m, a_e)$ &  & \\ \cline{2-5}
& $M\colon (a_e, a_m)\rightarrow(a_m, a_e)$ & $(\Z_2)^5$ & 32 &  $(B_{xy}, B_{xy})$, $(A_{c}^2, A_c^2)$, $(A_ct, A_ct)$, $(t^2, t^2)$, $(w_2, w_2)$ \\ \cline{2-5}
& $C_6\colon (a_e, a_m)\rightarrow(a_m, a_e)$ & \multirow{2}{*}{$(\Z_2)^5$} & \multirow{2}{*}{32} & \multirow{2}{*}{$(B_{xy}, B_{xy})$, $(A_{m}^2, A_m^2)$, $(A_mt, A_mt)$, $(t^2, t^2)$, $(w_2, w_2)$}\\\cline{2-2}
& $C_6, M\colon (a_e, a_m)\rightarrow(a_m, a_e)$ &  & \\ \hline

\multirow{24}{*}{$p4m\times SO(3)\times\Z_2^T$} & Trivial & $(\Z_2)^{22}$ & 2098176 & 
\begin{tabular}{c}  
$(B_{xy}, 0)$, $(B_{c^2}, 0)$, $(A_{x+y}^2, 0)$, \\
$(A_{x+y}A_m, 0)$, $(A_c^2, 0)$, $(A_m^2, 0)$,\\ 
$(A_{x+y}t,0)$, $(A_ct, 0)$, $(A_mt, 0)$, $(t^2, 0)$, $(w_2, 0)$, \\ 
$(0, B_{xy})$, $(0, B_{c^2})$, $(0, A_{x+y}^2)$, \\
$(0, A_{x+y}A_m)$, $(0, A_c^2)$, $(0, A_m^2)$,\\ 
$(0, A_{x+y}t)$, $(0, A_ct)$, $(0, A_mt)$, $(0, t^2)$, $(0, w_2)$\\ 
\end{tabular} \\ \cline{2-5}  
& $\mc{T}\colon (a_e, a_m)\rightarrow(a_m, a_e)$ & \multirow{8}{*}{$(\Z_2)^7$} & \multirow{8}{*}{128}  & \multirow{8}{*}{\begin{tabular}{c}$(B_{xy}, B_{xy})$, $(B_{c^2}, B_{c^2})$, $(A_{x+y}^2, A_{x+y}^2)$,\\
$(A_{x+y}A_m, A_{x+y}A_m)$, $(A_c^2, A_c^2)$, $(A_m^2, A_m^2)$, $(w_2, w_2)$ \end{tabular}} \\  \cline{2-2} 
& $M, \mc{T}\colon (a_e, a_m)\rightarrow(a_m, a_e)$ &  &  \\ \cline{2-2} 
& $C_4, \mc{T}\colon (a_e, a_m)\rightarrow(a_m, a_e)$ &  & \\ \cline{2-2} 
& $C_4, M, \mc{T}\colon (a_e, a_m)\rightarrow(a_m, a_e)$ &  & \\ \cline{2-2}
& $T_{1,2}, \mc{T}\colon (a_e, a_m)\rightarrow(a_m, a_e)$ &  & \\ \cline{2-2} 
& $T_{1,2}, M, \mc{T}\colon (a_e, a_m)\rightarrow(a_m, a_e)$ &  & \\ \cline{2-2}
& $T_{1,2}, C_4, \mc{T}\colon (a_e, a_m)\rightarrow(a_m, a_e)$ &  & \\ \cline{2-2} 
& $T_{1,2}, C_4, M, \mc{T}\colon (a_e, a_m)\rightarrow(a_m, a_e)$ &  & \\ \cline{2-5}
& $M \colon (a_e, a_m)\rightarrow(a_m, a_e)$ 
& $(\Z_2)^7$ & 128 &
\begin{tabular}{c}$(B_{xy}, B_{xy})$, $(B_{c^2}, B_{c^2})$, $(A_{x+y}^2, A_{x+y}^2)$,\\
$(A_{x+y}t, A_{x+y}t)$, $(A_ct, A_ct)$, $(t^2, t^2)$, $(w_2, w_2)$ \end{tabular} 
\\  \cline{2-5} 
& $C_4 \colon (a_e, a_m)\rightarrow(a_m, a_e)$ 
& $(\Z_2)^{14}$ & 8448 &
\begin{tabular}{c}
$(B_{xy}, B_{xy})$, $(B_{c^2}, B_{c^2})$, $(t^2, t^2)$, $(w_2, w_2)$ \\
$\widetilde{(\mathtt{A}_x\mathtt{A}_m, \mathtt{A}_y\mathtt{A}_m))}$, $\widetilde{(\mathtt{A}_c\mathtt{A}_m, (\mathtt{A}_c+\mathtt{A}_m)\mathtt{A}_m)}$,\\
$\widetilde{(\mathtt{A}_x \mathtt{A}_c, \mathtt{A}_y(\mathtt{A}_c+\mathtt{A}_m))}$, $\widetilde{(\mathtt{A}_xt, \mathtt{A}_yt)}$, 
$\widetilde{(\mathtt{A}_ct, (\mathtt{A}_c+\mathtt{A}_m)t)}$,\\ $\widetilde{(\mathtt{A}_y\mathtt{A}_m, \mathtt{A}_x\mathtt{A}_m)}$, $\widetilde{((\mathtt{A}_c+\mathtt{A}_m)\mathtt{A}_m, \mathtt{A}_c\mathtt{A}_m)}$, \\
$\widetilde{(\mathtt{A}_y(\mathtt{A}_c+\mathtt{A}_m) , \mathtt{A}_x\mathtt{A}_c)}$, 
$\widetilde{(\mathtt{A}_yt, \mathtt{A}_xt)}$, 
$\widetilde{((\mathtt{A}_c+\mathtt{A}_m)t, \mathtt{A}_ct)}$\end{tabular}\\  \cline{2-5} 
& $C_4, M \colon (a_e, a_m)\rightarrow(a_m, a_e)$ 
& $(\Z_2)^{10}$ & 640 &
\begin{tabular}{c}$(B_{xy}, B_{xy})$, $(B_{c^2}, B_{c^2})$, $(A_{x+y}^2, A_{x+y}^2)$, \\
$(A_{x+y}t, A_{x+y}t)$, $(t^2, t^2)$, $(w_2, w_2)$, \\
$\widetilde{(\mathtt{A}_c^2, \mathtt{A}_c^2+\mathtt{A}_m^2)}$, 
$\widetilde{(\mathtt{A}_ct, (\mathtt{A}_c+\mathtt{A}_m)t)}$, \\
$\widetilde{(\mathtt{A}_c^2 + \mathtt{A}_m^2, \mathtt{A}_c^2)}$, $\widetilde{((\mathtt{A}_c+\mathtt{A}_m)t, \mathtt{A}_ct)}$\\
\end{tabular}  \\  \cline{2-5} 
& $T_{1,2} \colon (a_e, a_m)\rightarrow(a_m, a_e)$ 
& $(\Z_2)^{11}$ & 1152 & 
\begin{tabular}{c}
$(B_{c^2}, B_{c^2})$, $(A_m^2, A_m^2)$, $(A_mt, A_mt)$, $(t^2, t^2)$, $(w_2, w_2)$, \\
$\widetilde{(\mathtt{B}_{xy}, \mathtt{B}_{xy} + \mathtt{B}_{c^2} + \mathtt{A}_{x+y}( \mathtt{A}_{x+y} + \mathtt{A}_m))}$, \\
$\widetilde{(\mathtt{A}_{x+y}^2, \mathtt{A}_{x+y}^2+\mathtt{A}_c^2+\mathtt{A}_m^2)}$,$\widetilde{(\mathtt{A}_{x+y}t, (\mathtt{A}_{x+y}+\mathtt{A}_c+\mathtt{A}_m)t)}$,  \\
$\widetilde{(\mathtt{B}_{xy} + \mathtt{B}_{c^2} + \mathtt{A}_{x+y}( \mathtt{A}_{x+y} + \mathtt{A}_m), \mathtt{B}_{xy})}$, \\
 $\widetilde{(\mathtt{A}_{x+y}^2+\mathtt{A}_c^2 + \mathtt{A}_m^2, \mathtt{A}_{x+y}^2)}$, $\widetilde{((\mathtt{A}_{x+y}+\mathtt{A}_c+\mathtt{A}_m)t, \mathtt{A}_{x+y}t)}$
\end{tabular} 
\\  \cline{2-5} 
& $T_{1,2}, M \colon (a_e, a_m)\rightarrow(a_m, a_e)$ 
& $(\Z_2)^7$ & 128 &
\begin{tabular}{c}$(B_{xy}, B_{xy})$, $(B_{c^2}, B_{c^2})$, $(A_m^2, A_m^2)$\\
$(A_ct, A_ct)$, $(A_mt, A_mt)$, $(t^2, t^2)$, $(w_2, w_2)$ \end{tabular} 
\\  \cline{2-5} 
& $T_{1,2}, C_4 \colon (a_e, a_m)\rightarrow(a_m, a_e)$ 
& $(\Z_2)^7$ & 128 & 
\begin{tabular}{c}$(B_{xy}, B_{xy})$, $(B_{c^2}, B_{c^2})$, $(A_m^2, A_m^2)$\\
$(A_ct, A_ct)$, $(A_mt, A_mt)$, $(t^2, t^2)$, $(w_2, w_2)$ \end{tabular}  \\  \cline{2-5} 
& $T_{1,2}, C_4, M \colon (a_e, a_m)\rightarrow(a_m, a_e)$ 
& $(\Z_2)^{11}$ & 1152 &
\begin{tabular}{c}
$(B_{c^2}, B_{c^2})$, $(A_m^2, A_m^2)$, $(A_mt, A_mt)$, $(t^2, t^2)$, $(w_2, w_2)$, \\
$\widetilde{(\mathtt{B}_{xy}, \mathtt{B}_{xy} + \mathtt{B}_{c^2} + \mathtt{A}_{x+y}( \mathtt{A}_{x+y} + \mathtt{A}_m))}$, \\
$\widetilde{(\mathtt{A}_{x+y}^2, \mathtt{A}_{x+y}^2+\mathtt{A}_c^2+\mathtt{A}_m^2)}$,$\widetilde{(\mathtt{A}_{x+y}t, (\mathtt{A}_{x+y}+\mathtt{A}_c+\mathtt{A}_m)t)}$,  \\
$\widetilde{(\mathtt{B}_{xy} + \mathtt{B}_{c^2} + \mathtt{A}_{x+y}( \mathtt{A}_{x+y} + \mathtt{A}_m), \mathtt{B}_{xy})}$, \\
 $\widetilde{(\mathtt{A}_{x+y}^2+\mathtt{A}_c^2 + \mathtt{A}_m^2, \mathtt{A}_{x+y}^2)}$, $\widetilde{((\mathtt{A}_{x+y}+\mathtt{A}_c+\mathtt{A}_m)t, \mathtt{A}_{x+y}t)}$
\end{tabular}  \\  \hline

\bottomrule[2pt]	
\end{tabular}
}
\caption{All possible symmetry fractionalization classes of $p6m\times SO(3)\times\Z_2^T$ and $p4m\times SO(3)\times\Z_2^T$ for the $\Z_2$ topological order. When counting the number of realizations in each case, overcounts due to the equivalence from relabeling anyons have been taken care of. Removing all generators involving $w_2$, we obtain the symmetry fractionalization classes of $p6m\times\Z_2^T$ and $p4m\times\Z_2^T$ for the $\Z_2$ topological order. In the codes \cite{code}, the anyon permutation patterns are indexed according to the order of this table. For example, for symmetry group $p6m\times SO(3)\times\Z_2^T$, anyon permutation pattern 2 represents $\mc{T}: (a_e, a_m)\rightarrow (a_m, a_e)$, and anyon permutation 
 pattern 4 represents $C_6, \mc{T}: (a_e, a_m)\rightarrow (a_m , a_e)$.}
\label{tab:SF Z2 TO}
\end{table}

The classification of symmetry fractionalization classes and the generators are listed in Table~\ref{tab:SF Ising}. In the following, we explicitly comment on the cases involving symmetries that permute $e$ and $m$.

Given a group $G$ with some element that permutes $e$ and $m$, we have the following short exact sequence,
\beq
1\rightarrow \tilde G \rightarrow G \rightarrow \Z_2 \rightarrow 1\,,
\eeq
where $\tilde G$ is the subgroup of $G$ that does not permute anyons. From the Serre spectral sequence \cite{brown2012cohomology}, we immediately see that
\beq
H^k(G, \Z_2\oplus\Z_2) \cong H^k(\tilde G, \Z_2)\,.
\eeq
Specifically, given an element $[\tilde \omega]\in H^2(\tilde G, \Z_2)$ with a representative cochain $\omega$, we can write down the representative cochain $\omega \in H^2(\tilde G, \Z_2\oplus \Z_2)$ as follows. First, choose an element not in $\tilde G$ such that $x^2 = {\bf 1}$. Then every $g\in G$ can be decomposed as $g=\tilde{g}x^i,i\in\{0,1\}$, where $i=0$ if $g$ is an element in $G$ and $i=1$ otherwise, and $\tilde g$ is an element in $\tilde G$. Then we can write down the represetative cochain $w(g_1, g_2)$ of $G$ from the representative cochain $\tilde w$ of $\tilde G$, \ie
\beq\label{eq:G tildeG coho map}
\omega(g_1, g_2) = \left(\tilde{w}(\tilde{g}_1, x^{i_1}\tilde{g}_2x_1^{i_1}), \tilde{w}(x\tilde{g}_1x, xx^{i_1}\tilde{g}_2xx^{i_1})\right)\,.
\eeq
We can think of the second term as the representative cochain obtained from the conjugation action of $x$ on $\tilde w$. It is straightforward to check that $w$ satisfies the cocycle equation and thus is the desired representative cochain. Therefore, for each case where some symmetry actions permute anyons, to identify the symmetry fractionalization classes, we need to identify $\tilde G$ that does not permute anyons. By simply calculating the cohomology of $\tilde G$ we can identify all the possible symmetry fractionalization classes. 

Still, usually there can be some simplification, because sometimes we can identify $\omega$ and write down its representative cochain directly in terms of the $\Z_2$ cohomology of $G$. When this happens, to keep notations consistent, we will still label $\omega$ using the $\Z_2$ cohomology of $G$. When we have to use the cohomology of $\tilde G$, we will use $\mathtt{A}$ or $\mathtt{B}$ to emphasize that it refers to an element in the cohomology of the subgroup $\tilde G$.

When the time-reversal symmetry $\mc{T}$ permutes anyons, no matter how other symmetries act on anyons, $\tilde G$ will be isomorphic to $p6m\times SO(3)$ or $p4m\times SO(3)$.{\footnote{To see it, suppose any generator of $p6m$ or $p4m$ permutes anyons, we can combine this generator with $\mc{T}$ to form a generator of $\tilde G$. It is easy to check that the resulting group is indeed isomorphic to $p6m\times SO(3)$ or $p4m\times SO(3)$.}} The symmetry fractionalization classes can be identified accordingly.

When the time-reversal symmetry does not permute anyons, $\tilde G$ will be the product of a subgroup $\tilde{G}_s$ of $p6m$ or $p4m$ and $SO(3)\times\Z_2^T$. For the cases involving $p6m$, we have the following three possibilities:
\begin{enumerate}
    \item $M\colon (a_e, a_m)\rightarrow(a_m, a_e)$. Then $\tilde{G}_s = p6$, generated by $T_1$, $T_2$ and $C_6$.
    \item  $C_6\colon (a_e, a_m)\rightarrow(a_m, a_e)$. Then $\tilde{G}_s = p31m$, generated by $T_1$, $T_2$, $C_6^2$ and $M$.
    \item $C_6, M\colon (a_e, a_m)\rightarrow(a_m, a_e)$. Then $\tilde{G}_s = p3m1$, generated by $T_1$, $T_2$, $C_6^2$ and $C_6^3M$.
\end{enumerate} 
In these cases, 
it turns out that we can write down the representative cochains directly in terms of the representative cochains of the $\Z_2$ cohomology of $p6m$, and this is what we present in Table~\ref{tab:SF Z2 TO}.

For the cases involving $p4m$, we have the following seven possibilities. In these cases, the explicit representative cochains may not come from the representative cochains of the $\Z_2$ cohomology of $p4m$, and we really need the expressions of the representative cochains for each $\tilde{G}_s$. 
The $\Z_2$ cohomology of these $\tilde{G}_s$ and the representative cochains of their generators can all be found in Appendix E of Ref. \cite{Ye2021a}, and we follow the notation  there, except we change \eg $A_x$ to $\mathtt{A}_x$ and use a wilde tilde to emphasize that we are refering to the cochain and cohomology of a subgroup.
\begin{enumerate}
    \item $M\colon (a_e, a_m)\rightarrow(a_m, a_e)$. Then $\tilde{G}_s = p4$, generated by $T_1$, $T_2$ and $C_4$.
    \item  $C_4\colon (a_e, a_m)\rightarrow(a_m, a_e)$. Then $\tilde{G}_s = pmm$, generated by $T_1$, $T_2$, $C_4^2$ and $M$. Now we have 10 elements in the cohomology that are ``asymmetric'', and we can write down the representative cochains of them with the help of the representative cochains of $pmm$ and Eq.~\eqref{eq:G tildeG coho map}.
    \item $C_4, M\colon (a_e, a_m)\rightarrow(a_m, a_e)$. Then $\tilde{G}_s = cmm$, generated by $T_1$, $T_2$, $C_4^2$ and $C_4^3M$. Now we have four elements in the cohomology that are asymmetric. 
    \item $T_{1,2}\colon (a_e, a_m)\rightarrow(a_m, a_e)$. Then $\tilde{G}_s = p4m$, generated by $T_1T_2$, $T_1^{-1}T_2$, $C_4$ and $C_4M$. Now we have six elements in the cohomology that are asymmetric. 
    \item  $T_{1,2}, M\colon (a_e, a_m)\rightarrow(a_m, a_e)$. Then $\tilde{G}_s = p4g$, generated by $T_1T_2$, $T_1^{-1}T_2$, $C_4$ and $T_2M$. 
    \item $T_{1,2}, C_4\colon (a_e, a_m)\rightarrow(a_m, a_e)$. Then $\tilde{G}_s = p4g$, generated by $T_1T_2$, $T_1^{-1}T_2$, $T_1C_4$ and $T_1T_2M$. 
    \item $T_{1,2}, C_4, M\colon (a_e, a_m)\rightarrow(a_m, a_e)$. Then $\tilde{G}_s = p4m$, generated by $T_1T_2$, $T_1^{-1}T_2$, $T_1C_4$ and $T_1T_2C_4M$. Now we have six elements in the cohomology that are asymmetric. 
\end{enumerate}

\subsection{$\Z_N$ topological order ($N\geqslant 3$)}\label{subapp: Z4 TO}

In this case, we have $\mc{A} = (\Z_N)^2$. The topological symmetry of $\Z_N$ is complicated for general $N$. For $N\geqslant 3$, there is always a subgroup $ \Z_4^T\rtimes \Z_2$. The unitary $\Z_2$ is generated by the electric-magnetic duality symmetry $S$ that exchanges $e$ and $m$, \ie
\beq
S\colon(a_e, a_m)\rightarrow (a_m, a_e)\,,
\eeq
and the anti-unitary generator $T$ of $\Z_4^T$ permutes anyons in the following way
\beq
T\colon (a_e, a_m)\rightarrow (a_m, [-a_e]_N)\,.
\eeq
The two generators satisfy the relation
\beq
S^2 = \textbf{1}\,,\quad T^4 = \textbf{1}\,,\quad STS = T^{-1}\,.
\eeq
An element in ${\bf g}\in \Z_4^T\rtimes \Z_2$ can be labeled by $(g_1, g_2)$ with $g_1\in\{0,\dots,3\},g_2\in\{0,1\}$, which corresponds to the element $T^{g_1} S^{g_2}$. Given such element ${\bf g}$, the $U$-symbols can be chosen such that 
\beq
U_{\bf g}(a,b;c) = 
\begin{cases}
e^{i\frac{2\pi}{N}a_m b_e} & g_1+g_2 \equiv 1 \mod 2\\
1 & g_1+g_2 \equiv 0 \mod 2\\
\end{cases}
\eeq
A specific choice of $\eta$-symbols can be chosen such that 
\beq
\eta_a({\bf g}, {\bf h}) = 
\begin{cases}
e^{i\frac{2\pi}{N}a_m b_e} & g_1+g_2\equiv h_1+h_2 \equiv 1 \mod 2\\
1 & \text{otherwise}\\
\end{cases}
\eeq
For $N=3,4$, this is the full topological symmetry group.

To determine the anyon permutation patterns of $p6m\times SO(3)\times\Z_2^T$ and $p4m\times SO(3)\times\Z_2^T$, we just need to specify how the generators of the symmetry groups permute anyons. Because $\mc{T}^2=\textbf{1}$, $\mc{T}$ can act on anyons in two ways: either $\mc{T}\colon (a_e, a_m)\rightarrow ([-a_e]_N, a_m)$, or $\mc{T}\colon (a_e, a_m)\rightarrow (a_e, [-a_m]_N)$. Because these two cases are related by relabling anyons using the electic-magnetic duality $S$, we can specialize to the cases $\mc{T}\colon (a_e, a_m)\rightarrow (a_e, [-a_m]_N)$, and we only need to consider how $p6m$ or $p4m$ permutes anyons.
For $p6m\times SO(3)\times\Z_2^T$ there are four possible anyon permutation patterns, while for $p4m\times SO(3)\times\Z_2^T$, there are eight possible anyon permutation patterns. The corresponding classification of symmetry fractionalization classes and the generators for $N=3,4$ are listed in Tables~\ref{tab:SF Z3 TO} and \ref{tab:SF Z4 TO}, respectively. Specifically, since the symmetries cannot permute $e$ and $m$, $H^2(G, \Z_{N}\times\Z_{N})$ simply becomes the direct sum of two $H^2(G, \Z_{N})$ pieces, with the actions on two $\Z_{N}$ pieces corresponding to symmetry actions on $e$ or $m$, respectively. As discussed at the beginning of the appendix,  $H^2(G, \Z_{N})$ can all be obtained from the $\Z$ cohomology or $\Z_2$ cohomology of the symmetry groups. In particular, to obtain the full data of symmetry fractionalization classes, we need the cohomology and representative cochains of $H^2(p6m, \Z)$ and $H^2(p4m, \Z)$ with all possible actions on $\Z$, which we present here for completeness.

\begin{table}[!htbp]
\centering
\renewcommand{\arraystretch}{1.3}
\begin{tabular}{|c|c|c|c|c|}
\toprule[2pt]
Symmetry Group & Action & $H^2(G, \mc{A})$ & Realizations & Generators \\ \hline
\multirow{6}{*}{$p6m\times SO(3)\times\Z_2^T$} & 
 $M\colon (a_e, a_m)\rightarrow(a_e, [-a_m]_3)$
& $\Z_1$ & 1 & $(0,0)$\\ \cline{2-5}  
 &  $M\colon (a_e, a_m)\rightarrow([-a_e]_3, a_m)$
& $(\Z_3)^{2}$ & 5 & 
$(\widetilde{\mathscr{B}}^{(2)}_{xy}, 0)$, $(\widetilde{\mathscr{B}}^{(2)}_{c^2}, 0)$
\\ \cline{2-5}  
  & \begin{tabular}{c}
 $C_6\colon (a_e, a_m)\rightarrow([-a_e]_N, [-a_m]_3)$  \\
 $M\colon (a_e, a_m)\rightarrow(a_e, [-a_m]_3)$
\end{tabular} & $\Z_1$ &1 &  $(0,0)$\\ \cline{2-5}  
 & \begin{tabular}{c}
 $C_6\colon (a_e, a_m)\rightarrow([-a_e]_3, [-a_m]_3)$  \\
 $M\colon (a_e, a_m)\rightarrow([-a_e]_3, a_m)$
\end{tabular} & $\Z_1$ & 1 &  $(0, 0)$\\ \hline

\multirow{14}{*}{$p4m\times SO(3)\times\Z_2^T$} &
 $M\colon (a_e, a_m)\rightarrow(a_e, [-a_m]_3)$
& $\Z_1$ & 1 &$(0,0)$\\ \cline{2-5}  
 &
 $M\colon (a_e, a_m)\rightarrow([-a_e]_3, a_m)$
 & $\Z_3$ &2 &  
 $(\widetilde{\mathscr{B}}^{(2)}_{xy}, 0)$\\ \cline{2-5}  
  & \begin{tabular}{c}
 $C_4\colon (a_e, a_m)\rightarrow([-a_e]_3, [-a_m]_3)$  \\
 $M\colon (a_e, a_m)\rightarrow(a_e, [-a_m]_3)$
\end{tabular} & $\Z_1$ & 1 & $(0,0)$\\ \cline{2-5}  
 & \begin{tabular}{c}
 $C_4\colon (a_e, a_m)\rightarrow([-a_e]_3, [-a_m]_3)$  \\
 $M\colon (a_e, a_m)\rightarrow([-a_e]_3, a_m)$
\end{tabular} & $\Z_1$ & 1 &  $(0, 0)$\\ \cline{2-5}
& \begin{tabular}{c}
$T_{1,2}\colon(a_e, a_m)\rightarrow([-a_e]_3, [-a_m]_3)$ \\
 $M\colon (a_e, a_m)\rightarrow(a_e, [-a_m]_3)$
\end{tabular} & $\Z_1$ &1 & $(0,0)$\\ \cline{2-5}  
 & \begin{tabular}{c}
$T_{1,2}\colon(a_e, a_m)\rightarrow([-a_e]_3, [-a_m]_3)$ \\
 $M\colon (a_e, a_m)\rightarrow([-a_e]_3, a_m)$
\end{tabular} & $\Z_1$ & 1 & $(0, 0)$\\ \cline{2-5}  
  & \begin{tabular}{c}
 $T_{1,2}, C_4\colon (a_e, a_m)\rightarrow([-a_e]_3, [-a_m]_3)$  \\
 $M\colon (a_e, a_m)\rightarrow(a_e, [-a_m]_3)$
\end{tabular} & $\Z_1$ & 1 & $(0,0)$\\ \cline{2-5}  
 & \begin{tabular}{c}
$T_{1,2}, C_4\colon (a_e, a_m)\rightarrow([-a_e]_3, [-a_m]_3)$  \\
 $M\colon (a_e, a_m)\rightarrow([-a_e]_3, a_m)$
\end{tabular} & $\Z_1$ & 1 &  $(0, 0)$\\ \hline
\bottomrule[2pt]	
\end{tabular}
\caption{All possible symmetry fractionalization classes of $p6m\times SO(3)\times\Z_2^T$ and $p4m\times SO(3)\times\Z_2^T$ for $\Z_3$ topological order, when the action of $\mc{T}$ is specified by $\mc{T}\colon (a_e, a_m)\rightarrow (a_e, [-a_m]_3)$. The cases where the action of $\mc{T}$ is $\mc{T}\colon (a_e, a_m)\rightarrow ([-a_e]_3, a_m)$ can easily obtained by duality. When counting the number of realizations in each case, overcounts due to the equivalence from relabeling anyons have been taken care of.}
\label{tab:SF Z3 TO}
\end{table}

\begin{table}[!htbp]
\centering
\renewcommand{\arraystretch}{1.3}
\resizebox{\columnwidth}{!}{%
\begin{tabular}{|c|c|c|c|c|}
\toprule[2pt]
Symmetry Group & Action & $H^2(G, \mc{A})$ & Realizations & Generators \\ \hline
\multirow{12}{*}{$p6m\times SO(3)\times\Z_2^T$} &  $M\colon (a_e, a_m)\rightarrow(a_e, [-a_m]_4)$ & $(\Z_2)^{16}$ & $2^{16}$ & \begin{tabular}{c}  
$(2B_{xy}, 0)$, $(\widetilde{\beta}(A_c), 0)$, $(2A_cA_m, 0)$, $(\widetilde{\beta}(A_m), 0)$,\\ 
$(2A_ct, 0)$, $(2A_mt, 0)$, $(\widetilde{\beta}(t), 0)$, $(2w_2, 0)$,\\
$(0, 2B_{xy})$, $(0, 2A_c^2)$, $(0, 2A_cA_m)$, $(0, 2A_m^2)$,\\ 
$(0, \widetilde{\beta}(A_c))$, $(0, \widetilde{\beta}(A_m))$, $(0, 2t^2)$, $(0, 2w_2)$\\
\end{tabular}\\ \cline{2-5}  
& $M\colon (a_e, a_m)\rightarrow([-a_e]_4, a_m)$
& $\Z_4\oplus(\Z_2)^{15}$ & $3\times 2^{15}$ &  
\begin{tabular}{c}  
$(\widetilde{\mathscr{B}}^{(2)}_{xy}, 0)$, $(\widetilde{\mathscr{B}}^{(2)}_{c^2}, 0)$, $(2A_cA_m, 0)$, $(2A_m^2, 0)$,\\ 
$(2A_ct, 0)$, $(2A_mt, 0)$, $(\widetilde{\beta}(t), 0)$, $(2w_2, 0)$,\\
$(0, 2B_{xy})$, $(0, 2A_c^2)$, $(0, 2A_cA_m)$, $(0, 2A_m^2)$,\\ 
$(0, \widetilde{\beta}(A_c))$, $(0, \widetilde{\beta}(A_m))$, $(0, 2t^2)$, $(0, 2w_2)$\\
\end{tabular}\\ \cline{2-5}  
  & \begin{tabular}{c}
 $C_6\colon (a_e, a_m)\rightarrow([-a_e]_4, [-a_m]_4)$  \\
 $M\colon (a_e, a_m)\rightarrow(a_e, [-a_m]_4)$
\end{tabular} & $(\Z_2)^{16}$ &  $2^{16}$ &  \begin{tabular}{c}  
$(2B_{xy}, 0)$, $(2A_{c}^2, 0)$, $(2A_cA_m, 0)$, $(\widetilde{\beta}(A_m), 0)$,\\ 
$(2A_ct, 0)$, $(2A_mt, 0)$, $(\widetilde{\beta}(t), 0)$, $(2w_2, 0)$,\\
$(0, 2B_{xy})$, $(0, 2A_c^2)$, $(0, 2A_cA_m)$, $(0, 2A_m^2)$,\\ 
$(0, \widetilde{\beta}(A_c))$, $(0, \widetilde{\beta}(A_m))$, $(0, 2t^2)$, $(0, 2w_2)$\\
\end{tabular}\\ \cline{2-5}  
 & \begin{tabular}{c}
 $C_6\colon (a_e, a_m)\rightarrow([-a_e]_4, [-a_m]_4)$  \\
 $M\colon (a_e, a_m)\rightarrow([-a_e]_4, a_m)$
\end{tabular} & $(\Z_2)^{16}$ & $2^{16}$ &   \begin{tabular}{c}  
$(2B_{xy}, 0)$, $(2A_{c}^2, 0)$, $(\widetilde{\beta}(A_m), 0)$, $(2A_m^2, 0)$,\\ 
$(2A_ct, 0)$, $(2A_mt, 0)$, $(\widetilde{\beta}(t), 0)$, $(2w_2, 0)$,\\
$(0, 2B_{xy})$, $(0, 2A_c^2)$, $(0, 2A_cA_m)$, $(0, 2A_m^2)$,\\ 
$(0, \widetilde{\beta}(A_c))$, $(0, \widetilde{\beta}(A_m))$, $(0, 2t^2)$, $(0, 2w_2)$\\
\end{tabular}\\ \hline

\multirow{20}{*}{$p4m\times SO(3)\times\Z_2^T$} & 
$M\colon (a_e, a_m)\rightarrow(a_e, [-a_m]_4)$
& $(\Z_2)^{22}$ & $2^{22}$ & 
\begin{tabular}{c}  
$(2B_{xy},0)$, $(2B_{c^2},0)$, $(\widetilde{\beta}(A_{x+y}),0)$, $(2A_{x+y}A_m,0)$, $(\widetilde{\beta}(A_c),0)$, $(\widetilde{\beta}(A_m), 0)$, \\
$(2A_{x+y}t, 0)$, $(2A_ct, 0)$, $(2A_mt, 0)$, $(\widetilde{\beta}(t), 0)$, $(2w_2, 0)$,\\
$(0, 2B_{xy})$, $(0, 2B_{c^2})$, $(0, 2A_{x+y}^2)$, $(0, 2A_{x+y}A_m)$, $(0, 2A_c^2)$, $(0, 2A_m^2)$, \\
$(0, \widetilde{\beta}(A_{x+y}))$, $(0, \widetilde{\beta}(A_c))$, $(0, \widetilde{\beta}(A_m))$, $(0, 2t^2)$, $(0, 2w_2)$\\
\end{tabular}
\\ \cline{2-5}  
 &  $M\colon (a_e, a_m)\rightarrow([-a_e]_4, a_m)$
& $(\Z_4)^2\oplus(\Z_2)^{20}$ &   $10\times 2^{20}$ & 
\begin{tabular}{c}  
$(\widetilde{\mathscr{B}}^{(2)}_{xy}, 0)$, $(\widetilde{\mathscr{B}}^{(2)}_{c^2}, 0)$, $(\widetilde{\beta}(A_{x+y}), 0)$, $(2A_{x+y}A_m, 0)$, $(2A_c^2, 0)$, $(2A_m^2, 0)$, \\ 
$(2A_{x+y}t, 0)$, $(2A_ct, 0)$, $(2A_mt, 0)$, $(\widetilde{\beta}(t), 0)$, $(2w_2, 0)$,\\
$(0, 2B_{xy})$, $(0, 2B_{c^2})$, $(0, 2A_{x+y}^2)$, $(0, 2A_{x+y}A_m)$, $(0, 2A_c^2)$, $(0, 2A_m^2)$, \\
$(0, \widetilde{\beta}(A_{x+y}))$, $(0, \widetilde{\beta}(A_c))$, $(0, \widetilde{\beta}(A_m))$, $(0, 2t^2)$, $(0, 2w_2)$\\
\end{tabular}
\\ \cline{2-5}  
  & \begin{tabular}{c}
 $C_4\colon (a_e, a_m)\rightarrow([-a_e]_4, [-a_m]_4)$  \\
 $M\colon (a_e, a_m)\rightarrow(a_e, [-a_m]_4)$
\end{tabular} 
& $(\Z_2)^{22}$ &  $2^{22}$ & 
\begin{tabular}{c}  
$(2B_{xy}, 0)$, $(2B_{c^2}, 0)$, $(\widetilde{\beta}(A_{x+y}), 0)$, $(2A_{x+y}A_m, 0)$, $(\widetilde{\beta}(A_m), 0)$, $(2A_m^2, 0)$, \\
$(2A_{x+y}t, 0)$, $(2A_ct, 0)$, $(2A_mt, 0)$, $(\widetilde{\beta}(t), 0)$, $(2w_2, 0)$,\\
$(0, 2B_{xy})$, $(0, 2B_{c^2})$, $(0, 2A_{x+y}^2)$, $(0, 2A_{x+y}A_m)$, $(0, 2A_c^2)$, $(0, 2A_m^2)$, \\
$(0, \widetilde{\beta}(A_{x+y}))$, $(0, \widetilde{\beta}(A_c))$, $(0, \widetilde{\beta}(A_m))$, $(0, 2t^2)$, $(0, 2w_2)$\\
\end{tabular}
\\ \cline{2-5}  
 & \begin{tabular}{c}
 $C_4\colon (a_e, a_m)\rightarrow([-a_e]_4, [-a_m]_4)$  \\
 $M\colon (a_e, a_m)\rightarrow([-a_e]_4, a_m)$
\end{tabular} & $(\Z_2)^{22}$ & $2^{22}$ &   \begin{tabular}{c}  
$(2B_{xy}, 0)$, $(2B_{c^2}, 0)$, $(\widetilde{\beta}(A_{x+y}), 0)$, $(2A_{x+y}A_m, 0)$, $(\widetilde{\beta}(A_m), 0)$, $(2A_m^2, 0)$, \\ 
$(2A_{x+y}t, 0)$, $(2A_ct, 0)$, $(2A_mt, 0)$, $(\widetilde{\beta}(t), 0)$, $(2w_2, 0)$,\\
$(0, 2B_{xy})$, $(0, 2B_{c^2})$, $(0, 2A_{x+y}^2)$, $(0, 2A_{x+y}A_m)$, $(0, 2A_c^2)$, $(0, 2A_m^2)$, \\
$(0, \widetilde{\beta}(A_{x+y}))$, $(0, \widetilde{\beta}(A_c))$, $(0, \widetilde{\beta}(A_m))$, $(0, 2t^2)$, $(0, 2w_2)$\\
\end{tabular}\\ \cline{2-5}
& \begin{tabular}{c}
$T_{1,2}\colon(a_e, a_m)\rightarrow([-a_e]_4, [-a_m]_4)$ \\
 $M\colon (a_e, a_m)\rightarrow(a_e, [-a_m]_4)$
\end{tabular} 
& $(\Z_2)^{22}$ &  $2^{22}$ & 
\begin{tabular}{c}  
$(2B_{xy}, 0)$, $(2B_{c^2}, 0)$, $(2A_{x+y}^2, 0)$, $(2A_{x+y}A_m, 0)$, $(\widetilde{\beta}(A_c), 0)$, $(\widetilde{\beta}(A_m), 0)$, \\
$(2A_{x+y}t, 0)$, $(2A_ct, 0)$, $(2A_mt, 0)$, $(\widetilde{\beta}(t), 0)$, $(2w_2, 0)$,\\
$(0, 2B_{xy})$, $(0, 2B_{c^2})$, $(0, 2A_{x+y}^2)$, $(0, 2A_{x+y}A_m)$, $(0, 2A_c^2)$, $(0, 2A_m^2)$, \\
$(0, \widetilde{\beta}(A_{x+y}))$, $(0, \widetilde{\beta}(A_c))$, $(0, \widetilde{\beta}(A_m))$, $(0, 2t^2)$, $(0, 2w_2)$\\
\end{tabular}
\\ \cline{2-5}  
 & \begin{tabular}{c}
$T_{1,2}\colon(a_e, a_m)\rightarrow([-a_e]_4, [-a_m]_4)$ \\
 $M\colon (a_e, a_m)\rightarrow([-a_e]_4, a_m)$
\end{tabular} & $\Z_4\oplus(\Z_2)^{21}$ & $3\times 2^{21}$ &   \begin{tabular}{c}  
$(\widetilde{\mathscr{B}}^{(6)}_{xy}, 0)$, $(2B_{c^2}, 0)$, $(2A_{x+y}^2, 0)$, $(\widetilde{\beta}(A_m), 0)$, $(2A_c^2,0)$, $(2A_m^2,0)$, \\
$(2A_{x+y}t, 0)$, $(2A_ct, 0)$, $(2A_mt, 0)$, $(\widetilde{\beta}(t), 0)$, $(2w_2, 0)$,\\
$(0, 2B_{xy})$, $(0, 2B_{c^2})$, $(0, 2A_{x+y}^2)$, $(0, 2A_{x+y}A_m)$, $(0, 2A_c^2)$, $(0, 2A_m^2)$, \\
$(0, \widetilde{\beta}(A_{x+y}))$, $(0, \widetilde{\beta}(A_c))$, $(0, \widetilde{\beta}(A_m))$, $(0, 2t^2)$, $(0, 2w_2)$\\
\end{tabular}
\\ \cline{2-5}  
  & \begin{tabular}{c}
 $T_{1,2}, C_4\colon (a_e, a_m)\rightarrow([-a_e]_4, [-a_m]_4)$  \\
 $M\colon (a_e, a_m)\rightarrow(a_e, [-a_m]_4)$
\end{tabular} 
& $\Z_4\oplus(\Z_2)^{21}$ & $3\times 2^{21}$ &   
\begin{tabular}{c}  
$(\widetilde{\mathscr{B}}^{(7)}_{xy}, 0)$, $(2B_{c^2}, 0)$, $(2A_{x+y}^2, 0)$, $(2A_{x+y}A_m, 0)$, $(2A_c^2,0)$, $(\widetilde{\beta}(A_m), 0)$, \\
$(2A_{x+y}t, 0)$, $(2A_ct, 0)$, $(2A_mt, 0)$, $(\widetilde{\beta}(t), 0)$, $(2w_2, 0)$,\\
$(0, 2B_{xy})$, $(0, 2B_{c^2})$, $(0, 2A_{x+y}^2)$, $(0, 2A_{x+y}A_m)$, $(0, 2A_c^2)$, $(0, 2A_m^2)$, \\
$(0, \widetilde{\beta}(A_{x+y}))$, $(0, \widetilde{\beta}(A_c))$, $(0, \widetilde{\beta}(A_m))$, $(0, 2t^2)$, $(0, 2w_2)$\\
\end{tabular}
\\ \cline{2-5}  
 & \begin{tabular}{c}
$T_{1,2}, C_4\colon (a_e, a_m)\rightarrow([-a_e]_4, [-a_m]_4)$  \\
 $M\colon (a_e, a_m)\rightarrow([-a_e]_4, a_m)$
\end{tabular} 
& $(\Z_2)^{22}$ & $2^{22}$ &   
\begin{tabular}{c}  
$(2B_{xy}, 0)$, $(2B_{c^2}, 0)$, $(2A_{x+y}^2, 0)$, $(\widetilde{\beta}(A_m), 0)$, $(\widetilde{\beta}(A_c), 0)$, $(2A_m^2, 0)$,\\
$(2A_{x+y}t, 0)$, $(2A_ct, 0)$, $(2A_mt, 0)$, $(\widetilde{\beta}(t), 0)$, $(2w_2, 0)$,\\
$(0, 2B_{xy})$, $(0, 2B_{c^2})$, $(0, 2A_{x+y}^2)$, $(0, 2A_{x+y}A_m)$, $(0, 2A_c^2)$, $(0, 2A_m^2)$, \\
$(0, \widetilde{\beta}(A_{x+y}))$, $(0, \widetilde{\beta}(A_c))$, $(0, \widetilde{\beta}(A_m))$, $(0, 2t^2)$, $(0, 2w_2)$\\
\end{tabular}
\\ \hline
\bottomrule[2pt]	
\end{tabular}
}
\caption{All possible symmetry fractionalization classes of $p6m\times SO(3)\times\Z_2^T$ and $p4m\times SO(3)\times\Z_2^T$ for $\Z_4$ topological order, when the action of $\mc{T}$ is specified by $\mc{T}\colon (a_e, a_m)\rightarrow (a_e, [-a_m]_4)$. The cases where the action of $\mc{T}$ is $\mc{T}\colon (a_e, a_m)\rightarrow ([-a_e]_4, a_m)$ is related to these cases by relabeling, hence are not distinct physical realizations. When counting the number of realizations in each case, overcounts due to the equivalence from relabeling anyons have been taken care of. Removing all generators involving $w_2$, we obtain the symmetry fractionalization classes of $p6m\times\Z_2^T$ and $p4m\times\Z_2^T$ for the $\Z_4$ topological order. In the codes \cite{code}, the anyon permutation patterns are indexed according to the order of this table. For example, for symmetry group $p6m\times SO(3)\times\Z_2^T$, anyon permutation pattern 1 represents $M: (a_e, a_m)\rightarrow (a_e, [-a_m]_4)$, and anyon permutation 
 pattern 2 represents $M: (a_e, a_m)\rightarrow ([-a_e]_4 , a_m)$.}
\label{tab:SF Z4 TO}
\end{table}

For $p6m$, we have
\begin{enumerate}
\item Trivial action on $\Z$.
\beq
H^2(p6m, \Z) = (\Z_2)^2\,.
\eeq
We denote the generators of the two $\Z_2$ pieces by $\beta^{(1)}(A_c)$ and $\beta^{(1)}(A_m)$, respectively, which have representative cochains,

\begin{align}\label{eq:cochain_p6m_Ac2}
&\beta^{(1)}(A_c)(g_1, g_2)=\frac{[c_1]_2+[c_2]_2-[c_1+c_2]_2}{2}\\
\label{eq:cochain_p6m_Am2_1}
&\beta^{(1)}(A_m)(g_1, g_2)=\frac{[m_1]_2+[m_2]_2-[m_1 + m_2]_2}{2}
\end{align}

\item $M$ acts nontrivially on $\Z$.
\beq
H^2(p6m, \Z) = \Z\oplus \Z_6\,.
\eeq
We denote the generators of the $\Z$ piece and the $\Z_6$ piece by $\mathscr{B}^{(2)}_{xy}$ and $\mathscr{B}^{(2)}_{c^2}$, respectively, which have representative cochains,
\begin{align}\label{cochain_p6m_Bxy_2}
    &\mathscr{B}^{(2)}_{xy}(g_1, g_2) = P_{60}(c_1)\left[P_c(m_1)y_1x_2+m_1y_2(x_2+y_1)\right]\nonumber\\
    &\qquad\qquad\ 
    +P_{61}(c_1)\left[P_c(m_1)\left(\frac{x_2(x_2-1)}{2}+y_1x_2-y_2(x_2+y_1)\right)+m_1\left(\frac{y_2(y_2-1)}{2}+y_1(-x_2+y_2)\right)\right]\nonumber\\
    &\qquad\qquad\ 
    +P_{62}(c_1)\left[P_c(m_1)\left(\frac{y_2(y_2+1)}{2}-x_2-y_2(x_2+y_1)\right)+m_1\left(\frac{x_2(x_2+1)}{2}-y_2-y_1x_2\right)\right]\nonumber\\
    &\qquad\qquad\ 
    +P_{63}(c_1)\left[P_c(m_1)(-x_2+y_2-y_1x_2)+m_1(x_2-y_2+y_2(x_2-y_1)\right]\nonumber\\
    &\qquad\qquad\ 
    +P_{64}(c_1)\left[P_c(m_1)\left(\frac{x_2(x_2-1)}{2}+y_2-y_1x_2-y_2(x_2-y_1)\right)+m_1\left(\frac{y_2(y_2-1)}{2}+x_2+y_1(x_2-y_2)\right)\right]\nonumber\\
    &\qquad\qquad\ 
    +P_{65}(c_1)\left[P_c(m_1)\left(\frac{y_2(y_2+1)}{2}-y_2(x_2-y_1)\right)+m_1\left(\frac{x_2(x_2+1)}{2}+y_1x_2\right)\right]\\
    \label{cochain_p6m_Bc2_2}
    &\mathscr{B}^{(2)}_{c^2}(g_1, g_2)=\frac{[c_1]_6+(-1)^{m_1}[c_2]_6-[c_1+(-1)^{m_1}c_2]_6}{6}
\end{align}
Note that the $\Z_2$ reduction of $\mathscr{B}^{(2)}_{c^2}$ is actually $A_c^2 + A_cA_m$.

\item $C_6$ acts nontrivially on $\Z$.
\beq
H^2(p6m, \Z) = \Z_2\,.
\eeq
We denote the generator by $\beta^{(3)}(A_m)$, which has a representative cochain,
\beq\label{eq:cochain_p6_Am2_3}
\beta^{(3)}(A_m)(g_1, g_2)=\frac{[m_1]_2+(-1)^{c_1}[m_2]_2-[m_1+m_2]_2}{2}
\eeq

\item Both $C_6$ and $M$ act nontrivially on $\Z$.
\beq
H^2(p6m, \Z) = \Z_2\,.
\eeq
We denote the generator by $\beta^{(4)}(A_m)$, which has a representative cochain,
\beq\label{eq:cochain_p6_Am2_4}
\beta^{(4)}(A_m)(g_1, g_2)=\frac{[m_1]_2+(-1)^{c_1+m_1}[m_2]_2-[m_1+m_2]_2}{2}
\eeq

\end{enumerate}

For $p4m$, we have
\begin{enumerate}
\item Trivial action on $\Z$.
\beq
H^2(p4m, \Z) = (\Z_2)^3\,.
\eeq
We denote the generators of the three $\Z_2$ pieces by $\beta^{(1)}(A_{x+y})$, $\beta^{(1)}(A_{c})$ and $\beta^{(1)}(A_{m})$, respectively, which have representative cochains,

\begin{align}
\label{eq:cochain_p4m_Ax+y2_1}
&\beta^{(1)}(A_{x+y})(g_1, g_2)=\frac{[x_1+y_1]_2+[x_2+y_2]_2-[x_1+x_2 + y_1 + y_2]_2}{2}\\
\label{eq:cochain_p4m_Ac2_1}
&\beta^{(1)}(A_{c})(g_1, g_2)=\frac{[c_1]_2+[c_2]_2-[c_1+(-1)^{m_1}c_2]_2}{2}\\
\label{eq:cochain_p4m_Am2_1}
&\beta^{(1)}(A_{m})(g_1, g_2)=\frac{[m_1]_2+[m_2]_2-[m_1+m_2]_2}{2}
\end{align}

\item $M$ acts nontrivially on $\Z$.
\beq
H^2(p4m, \Z) = \Z\oplus\Z_4\oplus\Z_2\,.
\eeq
We denote the generators of the $\Z$, $\Z_4$ and $\Z_2$ piece by $\mathscr{B}^{(2)}_{xy}$, $\mathscr{B}^{(2)}_{c^2}$ and $\beta^{(2)}(A_{x+y})$, respectively, which have representative cochains,
\begin{align}\label{eq:cochain_p4m_Bxy_2}
&\mathscr{B}^{(2)}_{xy}(g_1, g_2)=   P_{40}(c_1) (-1)^{m_1}y_1 x_2   -  P_{41}(c_1)y_2(y_1 + (-1)^{m_1}x_2) \nonumber \\
&\quad\quad\quad\quad\quad \quad - P_{42}(c_1)(-1)^{m_1}y_1 x_2 +   P_{43}(c_1) y_2(y_1-(-1)^{m_1}x_2)\\
\label{eq:cochain_p4m_Bc2_2}
&\mathscr{B}^{(2)}_{c^2}(g_1, g_2)=\frac{[c_1]_4+(-1)^{m_1}[c_2]_4-[c_1+(-1)^{m_1}c_2]_4}{4}\\
\label{eq:cochain_p4m_Ax+y2_2}
&\beta^{(2)}(A_{x+y})(g_1, g_2)=\frac{[x_1+y_1]_2+(-1)^{m_1}[x_2+y_2]_2-[x_1+x_2 + y_1 + y_2]_2}{2}
\end{align}

\item $C_4$ acts nontrivially on $\Z$.
\beq
H^2(p4m, \Z) = (\Z_2)^2\,.
\eeq
We denote the generators of the two $\Z_2$ pieces by $\beta^{(3)}(A_{x+y})$ and $\beta^{(3)}(A_{m})$, respectively, which have representative cochains,
\begin{align}
\label{eq:cochain_p4m_Ax+y2_3}
&\beta^{(3)}(A_{x+y})(g_1, g_2)=\frac{[x_1+y_1]_2+(-1)^{c_1}[x_2+y_2]_2-[x_1+x_2 + y_1 + y_2]_2}{2}\\
\label{eq:cochain_p4m_Am2_3}
&\beta^{(3)}(A_{m})(g_1, g_2)=\frac{[m_1]_2+(-1)^{c_1}[m_2]_2-[m_1+m_2]_2}{2}
\end{align}

\item Both $C_4$ and $M$ act nontrivially on $\Z$.
\beq
H^2(p4m, \Z) = (\Z_2)^2\,.
\eeq
We denote the generators of the two $\Z_2$ pieces by $\beta^{(4)}(A_{x+y})$ and $\beta^{(4)}(A_{m})$, respectively, which have representative cochains,
\begin{align}
\label{eq:cochain_p4m_Ax+y2_4}
&\beta^{(4)}(A_{x+y})(g_1, g_2)=\frac{[x_1+y_1]_2+(-1)^{c_1+m_1}[x_2+y_2]_2-[x_1+x_2 + y_1 + y_2]_2}{2}\\
\label{eq:cochain_p4m_Am2_4}
&\beta^{(4)}(A_{m})(g_1, g_2)=\frac{[m_1]_2+(-1)^{c_1+m_1}[m_2]_2-[m_1+m_2]_2}{2}
\end{align}

\item $T_{1,2}$ acts nontrivially on $\Z$.
\beq
H^2(p4m, \Z) = (\Z_2)^2\,.
\eeq
We denote the generator of the two $\Z_2$ pieces by $\beta^{(5)}(A_c)$ and $\beta^{(5)}(A_{m})$, respectively, which have representative cochains,

\begin{align}
\label{eq:cochain_p4m_Ac2_5}
&\beta^{(5)}(A_{c})(g_1, g_2)=\frac{[c_1]_2+(-1)^{x_1+y_1}[c_2]_2-[c_1+(-1)^{m_1}c_2]_2}{2}\\
\label{eq:cochain_p4m_Am2_5}
&\beta^{(5)}(A_{m})(g_1, g_2)=\frac{[m_1]_2+(-1)^{x_1+y_1}[m_2]_2-[m_1+m_2]_2}{2}
\end{align}

\item Both $T_{1,2}$ and $M$ act nontrivially on $\Z$.
\beq
H^2(p4m, \Z) = \Z_4\oplus\Z_2\,.
\eeq
We denote the generators of the $\Z_4$ and $\Z_2$ piece by $\mathscr{B}^{(6)}_{xy}$ and $\beta^{(6)}(A_{c})$, respectively, which have representative cochains,
\begin{align}
\label{eq:cochain_p4m_Bxy_6}
&\mathscr{B}^{(6)}_{xy}(g_1, g_2) = (-1)^{\tilde{x}_1}(P_c(c_1) P(\tilde{y}_1) P(\tilde{x}_2) +  P(c_1) P(\tilde{y}_1 + \tilde{x}_2) P(\tilde{y}_2) ) \\
 &\text{with }\tilde{x} = x + P_{41}(c) + P_c(m)P_{42}(c) + P(m)P_{40}(c),\quad\quad \tilde{y} = y + P_{42}(c) + P_c(m)P_{43}(c) + P(m)P_{41}(c)\nonumber\\
\label{eq:cochain_p4m_Am2_6}
&\beta^{(6)}(A_{m})(g_1, g_2)=\frac{[m_1]_2+(-1)^{x_1+y_1+m_1}[m_2]_2-[m_1+m_2]_2}{2}
\end{align}
Note that the $\Z_2$ reduction of $\mathscr{B}^{(7)}_{xy}$ is actually $B_{xy} + B_{c^2} + A_{x+y}(A_{x+y}+A_m)$.

\item Both $T_{1,2}$ and $C_4$ act nontrivially on $\Z$.
\beq
H^2(p4m, \Z) = \Z_4\oplus\Z_2\,.
\eeq
We denote the generators of the $\Z_4$ and $\Z_2$ pieces by $\beta^{(7)}(A_{x+y})$ and $\beta^{(7)}(A_{m})$, respectively, which have representative cochains,
\begin{align}
\label{eq:cochain_p4m_Bxy_7}
&\mathscr{B}^{(7)}_{xy}(g_1, g_2)=(-1)^{x_1}(P_c(c_1) P(y_1) P(x_2) +  P(c_1) P(y_1+x_2) P(y_2) )\\
\label{eq:cochain_p4m_Am2_7}
&\beta^{(7)}(A_{m})(g_1, g_2)=\frac{[m_1]_2+(-1)^{x_1+y_1+c_1}[m_2]_2-[m_1+m_2]_2}{2}
\end{align}

\item All of $T_{1,2}$, $C_4$ and $M$ act nontrivially on $\Z$.
\beq
H^2(p4m, \Z) = (\Z_2)^2\,.
\eeq
We denote the generators of the two $\Z_2$ pieces by $\beta^{(8)}(A_c)$ and $\beta^{(8)}(A_{m})$, respectively, which have representative cochains,
\begin{align}
\label{eq:cochain_p4m_Ac2_8}
&\beta^{(8)}(A_c)(g_1, g_2)=\frac{[c_1]_2+(-1)^{x_1+y_1+c_1+m_1}[c_2]_2-[c_1 + (-1)^{m_1}c_2]_2}{2}\\
\label{eq:cochain_p4m_Am2_8}
&\beta^{(8)}(A_{m})(g_1, g_2)=\frac{[m_1]_2+(-1)^{x_1+y_1+c_1+m_1}[m_2]_2-[m_1+m_2]_2}{2}
\end{align}

\end{enumerate}

\subsection{$\U_2\times \U_{-2}$ (Double Semion)}

In this case, we have $\mc{A}=(\Z_2)^2$. The topological symmetry group is $\Z_2^T$, generated by $\tilde{S}$ exchanging $s$ and $\bar s$, \ie
\beq
\tilde S\colon (a_s, a_{\bar{s}})\rightarrow (a_{\bar{s}}, a_s)\,.
\eeq
We can choose the $U$-symbols and a set of $\eta$-symbols all equal to 1. Therefore, the anyon permutation patterns are completely fixed. The classification of symmetry fractionalization patterns and the generators are listed in Table~\ref{tab:SF Double Semion}. It turns out that all symmetry fractionalization classes lead to anomaly-free states.

\begin{table}[!htbp]
\centering
\renewcommand{\arraystretch}{1.3}
\resizebox{\columnwidth}{!}{%
\begin{tabular}{|c|c|c|c|c|}
\toprule[2pt]
Symmetry Group & Action & $H^2(G, \mc{A})$ & Realizations & Generators \\ \hline
$p6m\times SO(3)\times\Z_2^T$ & $M, \mc{T}\colon (a_s, a_{\bar{s}})\rightarrow (a_{\bar{s}}, a_s)$ & $(\Z_2)^5$ & 32 & \begin{tabular}{c}$(B_{xy}, B_{xy})$, $(A_{c}^2, A_c^2)$, \\
$(A_cA_m, A_cA_m)$, $(A_m^2, A_m^2)$, $(w_2, w_2)$ \end{tabular} \\ \hline
$p4m\times SO(3)\times\Z_2^T$ & $M, \mc{T}\colon (a_s, a_{\bar{s}})\rightarrow (a_{\bar{s}}, a_s)$ & $(\Z_2)^7$ & 128 & \begin{tabular}{c}$(B_{xy}, B_{xy})$, $(B_{c^2}, B_{c^2})$, $(A_{x+y}^2, A_{x+y}^2)$, \\
$(A_{x+y}A_m, A_{x+y}A_m)$, $(A_c^2, A_c^2)$, $(A_m^2, A_m^2)$, $(w_2, w_2)$ \end{tabular}\\ \hline
\bottomrule[2pt]	
\end{tabular}
}
\caption{All possible symmetry fractionalization classes of $p6m\times SO(3)\times \Z_2^T$ and $p4m\times SO(3)\times \Z_2^T$ for $\U_2\times \U_{-2}$ (double semion theory).}
\label{tab:SF Double Semion}
\end{table}

\subsection{$\U_4\times \U_{-4}$}

\begin{table}[!htbp]
\centering
\renewcommand{\arraystretch}{1.3}
\resizebox{\columnwidth}{!}{%
\begin{tabular}{|c|c|c|c|c|}
\toprule[2pt]
Symmetry Group & Action & $H^2(G, \mc{A})$ & Realizations & Generators \\ \hline
\multirow{8}{*}{$p6m\times SO(3)\times\Z_2^T$} 
&  $M\colon (a_s, a_{\bar{s}})\rightarrow(a_{\bar{s}}, a_s)$
& $(\Z_2)^5$ & 32 & 
\begin{tabular}{c}  
$(2B_{xy}, 2B_{xy})$, $(\widetilde{\beta}(A_c), \widetilde{\beta}(A_c))$, \\
$(2A_cA_m, 2A_cA_m)$, $(\widetilde{\beta}(A_m), \widetilde{\beta}(A_m))$, $(2w_2, 2w_2)$
\end{tabular}
\\ \cline{2-5}
 &  $M\colon (a_s, a_{\bar{s}})\rightarrow([-a_{\bar{s}}]_4, [-a_s]_4)$
& $\Z_4\oplus(\Z_2)^4$ & 48 &
\begin{tabular}{c}  
$(\widetilde{\mathscr{B}}^{(2)}_{xy}, \widetilde{\mathscr{B}}^{(2)}_{xy})$, $(\widetilde{\mathscr{B}}^{(2)}_{c^2}, \widetilde{\mathscr{B}}^{(2)}_{c^2})$, \\
$(2A_cA_m, 2A_cA_m)$, $(2A_m^2, 2A_m^2)$, $(2w_2, 2w_2)$
\end{tabular}
\\ \cline{2-5} 
  & \begin{tabular}{c}
 $C_6\colon (a_s, a_{\bar{s}})\rightarrow([-a_s]_4, [-a_{\bar{s}}]_4)$  \\
 $M\colon (a_s, a_{\bar{s}})\rightarrow(a_{\bar{s}}, a_s)$
\end{tabular} 
& $(\Z_2)^5$ & 32 &
 \begin{tabular}{c}  
$(2B_{xy}, 2B_{xy})$, $(2A_{c}^2, 2A_c^2)$, \\
$(2A_cA_m, 2A_cA_m)$, $(\widetilde{\beta}(A_m), \widetilde{\beta}(A_m))$, $(2w_2, 2w_2)$
\end{tabular} 
\\ \cline{2-5}  
 & \begin{tabular}{c}
 $C_6\colon (a_s, a_{\bar{s}})\rightarrow([-a_s]_4, [-a_{\bar{s}}]_4)$  \\
 $M\colon (a_s, a_{\bar{s}})\rightarrow([-a_{\bar{s}}]_4, [-a_s]_4)$
\end{tabular} 
& $(\Z_2)^5$ & 32 &
 \begin{tabular}{c}  
$(2B_{xy}, 2B_{xy})$, $(2A_{c}^2, 2A_c^2)$, \\
$(\widetilde{\beta}(A_m), \widetilde{\beta}(A_m))$, $(2A_{m}^2, 2A_m^2)$, $(2w_2, 2w_2)$
\end{tabular} 
\\ \hline

\multirow{20}{*}{$p4m\times SO(3)\times\Z_2^T$} & 
 $M\colon (a_s, a_{\bar{s}})\rightarrow(a_{\bar{s}}, a_s)$
& $(\Z_2)^7$ & 128 &
 \begin{tabular}{c}  
$(2B_{xy}, 2B_{xy})$, $(2B_{c^2}, 2B_{c^2})$, \\
$(\widetilde{\beta}(A_{x+y}), \widetilde{\beta}(A_{x+y}))$, $(2A_{x+y}A_m, 2A_{x+y}A_m)$, \\
$(\widetilde{\beta}(A_c), \widetilde{\beta}(A_c))$, $(\widetilde{\beta}(A_m), \widetilde{\beta}(A_m))$, $(2w_2, 2w_2)$
\end{tabular} 
\\ \cline{2-5}  
 & $M\colon (a_s, a_{\bar{s}})\rightarrow([-a_{\bar{s}}]_4, [-a_s]_4)$
& $(\Z_4)^2\oplus(\Z_2)^5$ & 320 &
 \begin{tabular}{c}  
$(\widetilde{\mathscr{B}}^{(2)}_{xy}, \widetilde{\mathscr{B}}^{(2)}_{xy})$, $(\widetilde{\mathscr{B}}^{(2)}_{c^2}, \widetilde{\mathscr{B}}^{(2)}_{c^2})$, \\
$(\widetilde{\beta}(A_{x+y}), \widetilde{\beta}(A_{x+y}))$, $(2A_{x+y}A_m, 2A_{x+y}A_m)$, \\
$(2A_c^2, 2A_c^2)$, $(2A_m^2, 2A_m^2)$, $(2w_2, 2w_2)$
\end{tabular} 
\\ \cline{2-5}  
  & \begin{tabular}{c}
 $C_4\colon (a_s, a_{\bar{s}})\rightarrow([-a_s]_4, [-a_{\bar{s}}]_4)$  \\
 $M\colon (a_s, a_{\bar{s}})\rightarrow(a_{\bar{s}}, a_s)$
\end{tabular} 
& $(\Z_2)^7$ & 128 &
 \begin{tabular}{c}  
$(2B_{xy}, 2B_{xy})$, $(2B_{c^2}, 2B_{c^2})$, \\
$(\widetilde{\beta}(A_{x+y}), \widetilde{\beta}(A_{x+y}))$, $(2A_{x+y}A_m, 2A_{x+y}A_m)$, \\
$(\widetilde{\beta}(A_m), \widetilde{\beta}(A_m))$, $(2A_m^2, 2A_m^2)$, $(2w_2, 2w_2)$
\end{tabular} 
\\ \cline{2-5}  
 & \begin{tabular}{c}
 $C_4\colon (a_s, a_{\bar{s}})\rightarrow([-a_s]_4, [-a_{\bar{s}}]_4)$  \\
 $M\colon (a_s, a_{\bar{s}})\rightarrow([-a_{\bar{s}}]_4, [-a_s]_4)$
\end{tabular} 
& $(\Z_2)^7$ & 128 &
 \begin{tabular}{c}  
$(2B_{xy}, 2B_{xy})$, $(2B_{c^2}, 2B_{c^2})$, \\
$(\widetilde{\beta}(A_{x+y}), \widetilde{\beta}(A_{x+y}))$, $(2A_{x+y}A_m, 2A_{x+y}A_m)$, \\
$(\widetilde{\beta}(A_m), \widetilde{\beta}(A_m))$, $(2A_m^2, 2A_m^2)$, $(2w_2, 2w_2)$
\end{tabular} 
\\ \cline{2-5}
& \begin{tabular}{c}
$T_{1,2}\colon(a_s, a_{\bar{s}})\rightarrow([-a_s]_4, [-a_{\bar{s}}]_4)$ \\ $M\colon (a_s, a_{\bar{s}})\rightarrow(a_{\bar{s}}, a_s)$
\end{tabular} 
& $(\Z_2)^7$ & 128 &
 \begin{tabular}{c}  
$(2B_{xy}, 2B_{xy})$, $(2B_{c^2}, 2B_{c^2})$, \\
$(2A_{x+y}^2, 2A_{x+y}^2)$, $(2A_{x+y}A_m, 2A_{x+y}A_m)$, \\
$(\widetilde{\beta}(A_c), \widetilde{\beta}(A_c))$, $(\widetilde{\beta}(A_m), \widetilde{\beta}(A_m))$, $(2w_2, 2w_2)$
\end{tabular} 
\\ \cline{2-5}  
 & \begin{tabular}{c}
$T_{1,2}\colon(a_s, a_{\bar{s}})\rightarrow([-a_s]_4, [-a_{\bar{s}}]_4)$ \\ $M\colon (a_s, a_{\bar{s}})\rightarrow([-a_{\bar{s}}]_4, [-a_s]_4)$
\end{tabular} 
& $\Z_4\oplus(\Z_2)^6$ & 192 &
 \begin{tabular}{c}  
$(\widetilde{\mathscr{B}}^{(6)}_{xy}, \widetilde{\mathscr{B}}^{(6)}_{xy})$, $(2B_{c^2}, 2B_{c^2})$, \\
$(2A_{x+y}^2, 2A_{x+y}^2)$, $(\widetilde{\beta}(A_m), \widetilde{\beta}(A_m))$, \\
$(2A_c^2, 2A_c^2)$, $(2A_m^2, 2A_m^2)$, $(2w_2, 2w_2)$
\end{tabular} 
\\ \cline{2-5}  
  & \begin{tabular}{c}
 $T_{1,2}, C_4\colon (a_s, a_{\bar{s}})\rightarrow([-a_s]_4, [-a_{\bar{s}}]_4)$  \\
 $M\colon (a_s, a_{\bar{s}})\rightarrow(a_{\bar{s}}, a_s)$
\end{tabular}
& $\Z_4\oplus(\Z_2)^6$ & 192 &
 \begin{tabular}{c}  
$(\widetilde{\mathscr{B}}^{(7)}_{xy}, \widetilde{\mathscr{B}}^{(7)}_{xy})$, $(2B_{c^2}, 2B_{c^2})$, \\
$(2A_{x+y}^2, 2A_{x+y}^2)$, $(2A_{x+y}A_m, 2A_{x+y}A_m)$, \\
$(2A_c^2, 2A_c^2)$, $(\widetilde{\beta}(A_m), \widetilde{\beta}(A_m))$, $(2w_2, 2w_2)$
\end{tabular} 
\\ \cline{2-5}  
 & \begin{tabular}{c}
$T_{1,2}, C_4\colon (a_s, a_{\bar{s}})\rightarrow([-a_s]_4, [-a_{\bar{s}}]_4)$  \\
 $M\colon (a_s, a_{\bar{s}})\rightarrow([-a_{\bar{s}}]_4, [-a_s]_4)$
\end{tabular} 
& $(\Z_2)^7$ & 128 &
 \begin{tabular}{c}  
$(2B_{xy}, 2B_{xy})$, $(2B_{c^2}, 2B_{c^2})$, \\
$(2A_{x+y}^2, 2A_{x+y}^2)$, $(\widetilde{\beta}(A_m), \widetilde{\beta}(A_m))$, \\
$(\widetilde{\beta}(A_c), \widetilde{\beta}(A_c))$, $(2A_m^2, 2A_m^2)$, $(2w_2, 2w_2)$
\end{tabular} 
\\ \hline
\bottomrule[2pt]	
\end{tabular}
}
\caption{All possible symmetry fractionalization classes of $p6m\times SO(3)\times\Z_2^T$ and $p4m\times SO(3)\times\Z_2^T$ for $\U_4\times \U_{-4}$, when the action of $\mc{T}$ is specified by $\mc{T}\colon (a_s, a_{\bar{s}})\rightarrow (a_{\bar{s}}, a_s)$. The cases where the action of $\mc{T}$ is $\mc{T}\colon (a_s, a_{\bar{s}})\rightarrow ([-a_{\bar{s}}]_4, [-a_s]_4)$ is related to these cases by relabeling, hence are not distinct physical realizations. When counting the number of realizations in each case, overcounts due to the equivalence from relabeling anyons have been taken care of.}
\label{tab:U(1) 4 U(1) -4}
\end{table}

In this case, we have $\mc{A}=(\Z_{4})^2$. For $N=2$, the topological symmetry is $ \Z_4^T\rtimes\Z_2^T$, generated by an order 2 anti-unitary symmetry $\tilde{S}$ which exchanges $s$ and $\bar{s}$, \ie
\beq
\tilde{S}\colon (a_s, a_{\bar{s}})\rightarrow (a_{\bar{s}}, a_s)\,,
\eeq
and another order 4 anti-unitary symmetry $T$, which permutes anyons in the following way
\beq
T\colon (a_s, a_{\bar{s}})\rightarrow (a_{\bar{s}}, [-a_s]_{2N})\,.
\eeq
The two generators satisfy the relation
\beq
\tilde{S}^2 = \textbf{1}\,,\quad T^4 = \textbf{1}\,,\quad \tilde{S}T\tilde{S} = T^{-1}\,.
\eeq
An element in $\Z_4^T\rtimes \Z_2^T$ can be written as $T^{g_1} \tilde{S}^{g_2}$, with $g_1\in\{0,\dots,3\}$ and $g_2\in\{0,1\}$. To define the $U$-symbols, first we define the following function
\beq
\tilde{U}(a_s, b_s) = 
\begin{cases}
(-1)^{a_s} & b_s\neq 0\\
1 & b_s=0
\end{cases}
\eeq
Given an element ${\bf g}\in \Z_4^T\rtimes \Z_2^T$, the $U$-symbols can be chosen such that 
\beq
U_{\bf g}(a,b;c) = 
\begin{cases}
1 & g_1=0 \\
\tilde{U}(a_{\bar{s}}, b_{\bar{s}}) & g_1 = 1 \\
\tilde{U}(a_s, b_s)\tilde{U}(a_{\bar{s}}, b_{\bar{s}}) & g_1 = 2 \\
\tilde{U}(a_s, b_s) & g_1 = 3 \\
\end{cases}
\eeq
And a set of $\eta$-symbols can be chosen to be all identity.

Because $\mc{T}^2=\textbf{1}$, $\mc{T}$ can act on anyons in two ways: either $\mc{T}\colon (a_s, a_{\bar{s}})\rightarrow (a_{\bar{s}}, a_s)$, or $\mc{T}\colon (a_s, a_{\bar{s}})\rightarrow ([-a_{\bar{s}}]_4, [-a_s]_4)$. Because these two cases are related by relabling anyons using $T\tilde{S}$, we can specialize to the cases $\mc{T}\colon (a_s, a_{\bar{s}})\rightarrow (a_{\bar{s}}, a_s)$, and we only need to consider how $p6m$ or $p4m$ permutes anyons. For the $p6m\times SO(3)\times\Z_2^T$ case, there are four possible anyon permutation patterns, while for the $p4m\times SO(3)\times\Z_2^T$ case, there are eight possible anyon permutation patterns. 

The corresponding classification of symmetry fractionalization classes and the generators are listed in Table~\ref{tab:U(1) 4 U(1) -4}. Specifically, since $\mc{T}$ permutes $s$ and $\bar{s}$, $H^2(p6m\times SO(3)\times \Z_2^T, \Z_{4}\oplus\Z_{4})$ or $H^2(p6m\times SO(3)\times \Z_2^T, \Z_{4}\oplus\Z_{4})$ is isomorphic to $H^2(p6m\times SO(3), \Z_{4})$ or $H^2(p4m\times SO(3), \Z_4)$, respectively, with the $\Z_4$ corresponding to the diagonal $(0,0),(1,1),(2,2),(3,3)$ anyons. As discussed at the beginning of the appendix,  $H^2(G, \Z_{4})$ can all be obtained from the $\Z$ cohomology or $\Z_2$ cohomology of the symmetry groups, and we list the $\Z$ cohomolgy of $p6m$ or $p4m$ in Appendix \ref{subapp: Z4 TO}. It turns out that all symmetry fractionalization classes lead to anomaly-free states.

\section{Anomaly indicators} \label{app: anomaly indicators}

In this appendix, we first write down the anomaly indicators for $\Z_2^T$, $\Z_2^T\times \Z_2^T$, $\Z_2\times \Z_2$ and $SO(3)\times \Z_2^T$ symmetries, where $\Z_2^T$ denotes an anti-unitary order 2 symmetry group. These anomaly indicators are all derived in Ref.~\cite{Ye2022} (also see Ref.~\cite{Wang2017anomaly,walker2019} for the $\Z_2^T$ symmetry). As explained in Ref.~\cite{Ye2022} in detail (see Sec. VI therein), the anomaly indicators of many other groups, including $p6\times SO(3)$, $p4\times SO(3)$, $p6m\times SO(3)\times\Z_2^T$, $p4m\times SO(3)\times\Z_2^T$, $p6m\times\Z_2^T$ and $p4m\times\Z_2^T$, can be obtained by restricting these groups to some of their $\Z_2^T$, $\Z_2^T\times\Z_2^T$ and $\Z_2\times\Z_2$ subgroups. So we can use the known anomaly indicators in Ref. \cite{Ye2022} to write down all anomaly indicators for all symmetry groups appearing in this paper, and identify the anomaly accordingly. These anomaly indicators are also recorded in this appendix. Using the expressions of the anomaly of each lattice homotopy class for these symmetries in Ref. \cite{Ye2021a}, which are written in terms of group cohomology, we can further obtain the values of the anomaly indicators for each lattice homotopy class of these symmetries.

\begin{itemize}

\item {$\Z_2^T$}

The anomalies for the group $\Z_2^T$ in (2+1)-d are classified by $\Z_2\oplus \Z_2$, and the two $\Z_2$ factors correspond to the ``in-cohomology'' and ``beyond-cohomology'' piece of the anomaly, respectively. The anomaly indicator for the beyond-cohomology piece is given by
\beq
\mc{I}_0 = \frac{1}{D} \sum_a d_a^2 \theta_a\,.
\eeq
This is also related to the chiral central charge $c_-$ by the formula $\mc{I}_0 = \exp\left(2\pi i\frac{c_-}{8}\right)$. The anomaly indicator for the in-cohomology piece is given by
\beq\label{eq:indicator_z2T}
\mc{I}_1(\mc{T})= \frac{1}{D}\sum_{\substack{a\\\,^{\mc{T}}a = a}}{d_a}\theta_a\times\eta_a(\mc{T}, \mc{T})\,
\eeq
where $\mc{T}$ is the generator of the $\Z_2^T$ symmetry.

\item {$\Z_2^T\times \Z_2^T$}

The anomalies for the group $\Z_2^T\times \Z_2^T$ in (2+1)-d are classified by $(\Z_2)^4$. Suppose the two anti-unitary generators of $\Z_2^T\times \Z_2^T$ are $\mc{T}_1$ and $\mc{T}_2$. The four anomaly indicators can be given by $\mc{I}_0$, $\mc{I}_1(\mc{T}_1)$, $\mc{I}_1(\mc{T}_2)$ and
\beq\label{eq:indicator_with_z2Tz2T}
\begin{aligned}
\mathcal{I}_2&\left(\mc{T}_1, \mc{T}_2\right) = 
\frac{1}{D^3}\sum_{\substack{a,b,c,x,y,u,v\\ \mu_x\nu_x\mu_y\nu_y\tilde{\mu_x}\tilde{\nu_x}\tilde{\mu_y}\tilde{\nu_y}\rho\sigma\tau\alpha\beta\gamma\delta\\
^{\mc{T}_1}a \times ^{\mc{T}_2}c \times c \rightarrow a\\
^{\mc{T}_1}c \times c \times b \rightarrow ^{\mc{T}_2}b}}
{d_c d_v}\frac{\theta_v}{\theta_a\theta_b}
\left(R_u^{^{\mc{T}_1}c, ^{\mc{T}_2}c}\right)_{\rho\sigma}
\\
&\times\left(F_{v}^{a,^{\mc{T}_1\mc{T}_2}c, ^{\mc{T}_2}y}\right)^*_{(^{\mc{T}_1}x,\tilde \mu_x,\alpha)(b,\tilde\mu_y,\tau)}
\left(F_{^{\mc{T}_2}y}^{^{\mc{T}_2}c,^{\mc{T}_1}c,y}\right)^*_{(u,\rho,\beta)(^{\mc{T}_2}b,\mu_y,\tilde\nu_y)}\\
&\times 
\left(F_{x}^{^{\mc{T}_1}x, ^{\mc{T}_1}c,^{\mc{T}_2}c}\right)^*_{(^{\mc{T}_1}a, \tilde v_x,\mu_x)(u,\sigma,\gamma)}
\left(F_v^{^{\mc{T}_1}x, u, y}\right)^*_{(x, \gamma, \delta)(^{\mc{T}_2}y, \beta,\alpha)}
\left(F_v^{x, c, b}\right)^*_{(a, \nu_x,\tau)(y, \nu_y, \delta)}\\
&\times 
U_{\mc{T}_1}^{-1}(^{\mc{T}_1}a,^{\mc{T}_2}c;x)_{\mu_x\tilde\mu_x} 
U_{\mc{T}_1}^{-1}(x,c;a)_{\nu_x\tilde\nu_x}U_{\mc{T}_2}^{-1}(^{\mc{T}_1}c,y;^{\mc{T}_2}b)^*_{\mu_y\tilde\mu_y}U_{\mc{T}_2}^{-1}(c, b;y)^*_{\nu_y\tilde\nu_y}
\times \eta_a(\mc{T}_1, \mc{T}_1)\eta_b(\mc{T}_2, \mc{T}_2)\frac{\eta_c(\mc{T}_2, \mc{T}_1)}{\eta_c(\mc{T}_1, \mc{T}_2)}
\end{aligned}
\eeq

\item {$\Z_2\times \Z_2$}

The anomalies for the group $\Z_2\times \Z_2$ in (2+1)-d are classified by $(\Z_2)^2$. Suppose the two generators of $\Z_2\times \Z_2$ are $C_1$ and $C_2$. The two anomaly indicators can be given by $\mc{I}_3(C_1, C_2)$ and $\mc{I}_3(C_2, C_1)$, where
\beq\label{eq:indicator_with_z2z2}
\begin{aligned}
\mathcal{I}_3\left(C_1, C_2\right) = \frac{1}{D^2}\sum_{\substack{a,b,x,u\\ \mu\nu\tilde{\mu}\tilde\nu\rho\sigma\alpha\\
^{C_1}a = a\\
a \times b \times ^{C_1}b \rightarrow ^{C_2}a}}
&{d_b}\frac{\theta_x}{\theta_a}
\left(R_u^{b, ^{C_1}b}\right)_{\rho\sigma}  \left(F_{{^{C_2}}a}^{a,b,{^{C_1}}b}\right)^*_{(x,\tilde{\mu},\tilde{\nu})(u,\sigma,\alpha)}
\left(F_{{^{C_2}}a}^{a,{^{C_1}}b,b}\right)_{({^{C_1}}x,\mu,\nu)(u,\rho,\alpha)}\\
&\times U_{C_1}^{-1}(a,b;x)_{\tilde{\mu}\mu} U_{C_1}^{-1}(x,^{C_1}b;^{C_2}a)_{\tilde{\nu}\nu}\times \frac{1}{\eta_b(C_1, C_1)}\frac{\eta_a(C_2, C_1)}{\eta_a(C_1, C_2)}
\end{aligned}
\eeq

\item {$SO(3)\times\Z_2^T$}

The anomalies for the group $SO(3)\times\Z_2^T\equiv SO(3)\times \Z_2^T$ in (2+1)-d are classified by $(\Z_2)^4$. Suppose that the generator of $\Z_2^T$ is $\mc{T}$ and let $U_\pi$ be a $\pi$ rotation in $SO(3)$. The four anomaly indicators can be given by $\mc{I}_0$, $\mc{I}(\mc{T})$, $\mc{I}(\mc{T}U_\pi)$, and 
\beq
\mc{I}_4 = \frac{1}{D}\sum_a d_a^2 \theta_a e^{i2\pi q_a}\,
\eeq
where $q_a\in \{0, \frac{1}{2}\}$ denotes whether anyon $a$ carries linear representation ($q_a=0$) or spinor representation ($q_a=\frac{1}{2}$) under $SO(3)$ symmetry.

From these known anomaly indicators, we can construct the anomaly indicators of the symmetry groups appearing in this paper by restricting to  subgroups. We need to find a complete list of subgroups such that all nontrivial elements are nonzero after restricting to at least one such subgroup. If the condition is satisfied, we indeed find a complete set of anomaly indicators.

\item {$p6\times SO(3)$}

The anomalies for the group $p6\times SO(3)$ in (2+1)-d are classified by $(\Z_2)^2$. The two anomaly indicators can be given by 
\beq
\mathsf{I}_1 = \mc{I}_3(C_2U_\pi, C_2 U'_\pi)\,,\quad 
\mathsf{I}_2 = \mc{I}_3(T_1C_2U_\pi, T_1C_2 U'_\pi)\,. 
\eeq
The values of these anomaly indicators in each lattice homotopy class with $p6\times SO(3)$ symmetry are given in Table \ref{tab: anomaly indicators p6 x SO(3)}.

\item {$p4\times SO(3)$}

The anomalies for the group $p4\times SO(3)$ in (2+1)-d are classified by $(\Z_2)^3$. The three anomaly indicators can be given by
\beq
\mathsf{I}_1 = \mc{I}_3(C_2U_\pi, C_2 U'_\pi)\,,\quad
\mathsf{I}_2 = \mc{I}_3(T_1T_2C_2U_\pi, T_1T_2C_2 U'_\pi)\,,\quad    
\mathsf{I}_3 = \mc{I}_3(T_1C_2U_\pi, T_1C_2 U'_\pi)\,.
\eeq
The values of these anomaly indicators in each lattice homotopy class with $p6\times SO(3)$ symmetry are given in Table \ref{tab: anomaly indicators p4 x SO(3)}.

\item {$p6m\times SO(3)\times\Z_2^T$ and $p6m\times \Z_2^T$}

The anomalies for the group $p6m\times SO(3)\times\Z_2^T$ in (2+1)-d are classified by $(\Z_2)^{22}$. The complete list of anomaly indicators can be given by
\begin{equation}
\begin{aligned}
    &\mathsf{I}_0 = \mathcal{I}_0\\ \\
    &\mathsf{I}_1 = \mathcal{I}_1(\mc{T})\quad 
    \mathsf{I}_2 = \mathcal{I}_1(M)\quad 
    \mathsf{I}_3 = \mathcal{I}_1(C_2\mc{T})\quad 
    \mathsf{I}_4 = \mathcal{I}_1(C_2M)\quad \\
    &\mathsf{I}_5 = \mathcal{I}_2(\mc{T}, C_2\mc{T})\quad 
    \mathsf{I}_6 = \mathcal{I}_2(\mc{T}, M)\quad 
    \mathsf{I}_7 = \mathcal{I}_2(C_2\mc{T}, M)\quad 
    \mathsf{I}_8 = \mathcal{I}_2(C_2\mc{T}, C_2M)\quad 
    \mathsf{I}_9 = \mathcal{I}_2(M, C_2M)\\ 
    &\mathsf{I}_{10} = \mathcal{I}_1(T_1T_2C_2\mc{T})\quad
    \mathsf{I}_{11} = \mathcal{I}_2(M, T_1T_2C_2\mc{T})\quad
    \mathsf{I}_{12} = \mathcal{I}_1(\mc{T}, T_1T_2C_2\mc{T})\quad
    \mathsf{I}_{13} = \mathcal{I}_1(M, T_1T_2C_2M)\\ \\
    &\mathsf{I}_{14} = \mathcal{I}_1(\mc{T}U_\pi)\quad
    \mathsf{I}_{15} = \mathcal{I}_1(MU_\pi)\quad
    \mathsf{I}_{16} = \mathcal{I}_1(C_2\mc{T}U_\pi)\quad
    \mathsf{I}_{17} = \mathcal{I}_1(C_2MU_\pi)\\
    &\mathsf{I}_{18} = \mathcal{I}_1(T_1T_2C_2\mc{T}U_\pi)\quad
    \mathsf{I}_{19} = \mathcal{I}_4\quad
    \mathsf{I}_{20} = \mathcal{I}_3(C_2U_\pi, U'_\pi)\quad
    \mathsf{I}_{21} = \mathcal{I}_3(M\mc{T}U_\pi, U'_\pi)\quad
\end{aligned}
\end{equation}

The values of these anomaly indicators in each lattice homotopy class with $p6m\times SO(3)\times\Z_2^T$ symmetry are given in Table \ref{tab: anomaly indicators p6m x O(3)T}.

The anomalies for the group $p6m\times\Z_2^T$ in (2+1)-d are classified by $(\Z_2)^{14}$. The complete set of anomaly indicators for $p6m\times \Z_2^T$ can be obtained by simply ignoring all anomaly indicators involving $U_\pi$, \ie this set consists of $\mathsf{I}_{0}$ to $\mathsf{I}_{13}$. The values of these anomaly indicators in each lattice homotopy class with $p6m\times\Z_2^T$ symmetry are also given by Table \ref{tab: anomaly indicators p6m x O(3)T} (after removing $\mathsf{I}_{14}$ to $\mathsf{I}_{21}$).

\begin{table}[!htbp]
\centering
\renewcommand{\arraystretch}{1.3}
\begin{tabular}{c|c|c|c|c}
\toprule[2pt]
 & 0 & a & c & a+c \\ \hline
$\mathsf{I}_3$ & 1 & $-1$ & 1 & $-1$\\
$\mathsf{I}_5$ & 1 & $-1$ & $1$ & $-1$ \\
$\mathsf{I}_{10}$ & 1 & $1$ & $-1$ & $-1$ \\
$\mathsf{I}_{12}$ & 1 & $1$ & $-1$ & $-1$ \\
$\mathsf{I}_{20}$ & 1 & $-1$ & $1$ & $-1$ \\
\bottomrule[2pt]
\end{tabular}
\caption{Values of the anomaly indicators for the 4 lattice homotopy classes with symmetry group $p6m\times SO(3)\times\Z_2^T$. The anomaly indicators not listed in the table are all 1 for all lattice homotopy classes.}
\label{tab: anomaly indicators p6m x O(3)T}
\end{table}

\item {$p4m\times SO(3)\times\Z_2^T$ and $p4m\times \Z_2^T$}

The anomalies for the group $p4m\times SO(3)\times\Z_2^T$ in (2+1)-d are classified by $(\Z_2)^{31}$. The complete list of anomaly indicators can be given by
\begin{equation}
\begin{aligned}
    &\mathsf{I}_0 = \mathcal{I}_0\\ 
    \\ 
    &\mathsf{I}_1 = \mathcal{I}_1(\mc{T})\quad 
    \mathsf{I}_2 = \mathcal{I}_1(M)\quad 
    \mathsf{I}_3 = \mathcal{I}_1(C_2\mc{T})\quad 
    \mathsf{I}_4 = \mathcal{I}_1(C_4M)\\
    &\mathsf{I}_5 = \mathcal{I}_2(\mc{T}, C_2\mc{T})\quad 
    \mathsf{I}_6 = \mathcal{I}_2(\mc{T}, M)\quad 
    \mathsf{I}_7 = \mathcal{I}_2(\mc{T}, C_4M)\quad 
    \mathsf{I}_8 = \mathcal{I}_2(C_2\mc{T}, M)\quad
    \mathsf{I}_9 = \mathcal{I}_2(C_2\mc{T}, C_4M)\\
    &\mathsf{I}_{10} = \mathcal{I}_1(T_1M)\quad
    \mathsf{I}_{11} = \mathcal{I}_1(T_1C_2\mc{T})\quad
    \mathsf{I}_{12} = \mathcal{I}_1(T_1T_2C_2\mc{T})\\
    &    \mathsf{I}_{13} = \mathcal{I}_2(\mc{T}, T_1M)\quad
    \mathsf{I}_{14} = \mathcal{I}_2(\mc{T}, T_1C_2\mc{T})\quad
    \mathsf{I}_{15} = \mathcal{I}_2(\mc{T}, T_1T_2C_2\mc{T})\\
    &\mathsf{I}_{16} = \mathcal{I}_2(T_2C_2\mc{T}, M)\quad
    \mathsf{I}_{17} = \mathcal{I}_2(M, T_2C_2M)\quad
    \mathsf{I}_{18} = \mathcal{I}_2(T_1T_2^{-1}C_2\mc{T}, C_4M) \quad
    \mathsf{I}_{19} = \mathcal{I}_2(T_1T_2C_2\mc{T},T_1M) \\
    \\   
    &\mathsf{I}_{20} = \mathcal{I}_4\quad
    \mathsf{I}_{21} = \mathcal{I}_1(\mc{T}U_\pi)\quad
    \mathsf{I}_{22} = \mathcal{I}_1(MU_\pi)\quad
    \mathsf{I}_{23} = \mathcal{I}_1(C_2\mc{T}U_\pi)\quad
    \mathsf{I}_{24} = \mathcal{I}_1(C_4MU_\pi)\\
    &\mathsf{I}_{25} = \mathcal{I}_1(T_1MU_\pi)\quad
    \mathsf{I}_{26} = \mathcal{I}_1(T_1C_2\mc{T}U_\pi)\quad
    \mathsf{I}_{27} = \mathcal{I}_1(T_1T_2C_2\mc{T}U_\pi)\\
    &\mathsf{I}_{28} = \mathcal{I}_3(C_4M\mc{T}U_\pi, U'_\pi)\quad
    \mathsf{I}_{29} = \mathcal{I}_3(M\mc{T}U_\pi, U'_\pi)\quad
    \mathsf{I}_{30} = \mathcal{I}_3(T_1M\mc{T}U_\pi, U'_\pi)\quad
\end{aligned}
\end{equation}

The values of these anomaly indicators in each lattice homotopy class with $p4m\times SO(3)\times\Z_2^T$ symmetry are given in Table \ref{tab: anomaly indicators p4m x O(3)T}.

The anomalies for the group $p4m\times\Z_2^T$ in (2+1)-d are classified by $(\Z_2)^{20}$. The complete set of anomaly indicators for $p4m\times \Z_2^T$ can be obtained by simply ignoring all anomaly indicators involving $U_\pi$, \ie this set consists of $\mathsf{I}_{0}$ to $\mathsf{I}_{19}$. The values of these anomaly indicators in each lattice homotopy class with $p4m\times\Z_2^T$ symmetry can be obtained from Table \ref{tab: anomaly indicators p4m x O(3)T} (after removing $\mathsf{I}_{20}$ to $\mathsf{I}_{30}$).

\begin{table}[!htbp]
\centering
\renewcommand{\arraystretch}{1.3}
\begin{tabular}{c|c|c|c|c|c|c|c|c}
\toprule[2pt]
 & 0 & a & b & c & a+b & a+c & b+c & a+b+c \\ \hline
$\mathsf{I}_3$ & $1$ & $-1$ & 1 & $1$ & $-1$ & $-1$ & 1 & $-1$\\
$\mathsf{I}_5$ & 1 & $-1$ & $1$ & $1$ & $-1$ & $-1$ & $1$ & $-1$\\
$\mathsf{I}_{11}$ & 1 & $1$ & $1$ & $-1$ & 1 & $-1$ & $-1$ & $-1$\\
$\mathsf{I}_{12}$ & 1 & $1$ & $-1$ & $1$ & $-1$ & $1$ & $-1$ & $-1$\\
$\mathsf{I}_{14}$ & 1 & $1$ & $1$ & $-1$ & 1 & $-1$ & $-1$ & $-1$\\
$\mathsf{I}_{15}$ & 1 & $1$ & $-1$ & $1$ & $-1$ & $1$ & $-1$ & $-1$\\
\bottomrule[2pt]
\end{tabular}
\caption{Values of the anomaly indicators for the 8 lattice homotopy classes with symmetry group $p4m\times SO(3)\times\Z_2^T$. The anomaly indicators not listed in the table are all 1 for all lattice homotopy classes.}
\label{tab: anomaly indicators p4m x O(3)T}
\end{table}

\end{itemize}

\section{Symmetry fractionalization classes of the ``beyond-parton" $\Z_2$ topological quantum spin liquids} \label{app: beyond parton}

In Sec. \ref{sec: Zn}, we have found 117 different $p4m\times\Z_2^T$ symmetry-enriched $\Z_2$ topological quantum spin liquids in lattice homotopy class a, where 64 of them were identified using the parton-mean-field approach \cite{Yang2015}, while the other 53 are beyond the usual parton mean field. It turns out that there is no anyon permutaion for any of these 117 states. In this appendix, we present the details of the symmetry fractionalization class of each of these 53 ``beyond-parton" states, summarized in Table \ref{tab:SF beyond parton}. 

According to Table \ref{tab:SF Z2 TO}, without anyon permutation the symmetry fractionalization classes are classified by $(\Z_2)^{20}$, which can be viewed as 10 different quantum numbers for each of $e$ and $m$. These quantum numbers are recorded in Table \ref{tab:SF beyond parton} for each of the 53 states. Their physical meanings are clear. For example, the column for $(C_2)^2$ being 1 ($-1$) for an anyon means this anyon carries trivial (nontrivial) projective quantum number under $C_2$, which roughly speaking says that $C_2^2=-1$ for this anyon. Similarly, $T_1\mc{T}T_1^{-1}\mc{T}^{-1}$ being $1$ ($-1$) for an anyon means that the translation $T_1$ and time reversal $\mc{T}$ commute (anti-commute) for this anyon.

From Table \ref{tab:SF beyond parton} we can see that in all these 53 states, the anyon $e$ is a Kramers doublet under the time reversal symmetry, while the anyon $m$ is a Kramers singlet. Furthermore, for the anyon $m$ there is always some nontrivial symmetry fractionalization simultaneously involving the lattice and time reversal symmetries. For example, translation and time reversal may not commute on $m$. In contrast, in all 64 states identified using the parton-mean-field approach in Ref. \cite{Yang2015}, $m$ experiences no symmetry fractionalization that involves both lattice and time reversal symmetries. Moreover, we remark that for all 117 states, the $C_2\equiv C_4^2$ symmetry fractionalizes on the $m$ anyon, \ie effectively $C_2^2=-1$ for $m$. Usually, the interpretation of this phenomenon is that there is a background $e$ anyon at each square lattice site (the $C_4$ center), and the mutual braiding statistics between $e$ and $m$ yields $C_2^2=-1$. However, for 16 of the 53 ``beyond-parton" states, $(T_1C_2)^2=(T_2C_2)^2=-1$ for $m$, which seems to suggest that there are also background $e$ anyons at the 2-fold rotation centers of $T_1C_2$ and $T_2C_2$, although microscopically there is no spin at those positions. So the analysis based on anomaly matching suggests that the simple picture where the fractionalization of rotational symmetries purely comes from background anyons is actually incomplete.

\begin{table}[!htbp]
\centering
\renewcommand{\arraystretch}{1.3}
\resizebox{\columnwidth}{!}{%
\begin{tabular}{|c|c|c|c|c|c|c|c|c|c|c|c|c|c|c|c|c|c|c|c|}
\toprule[2pt]

\multicolumn{10}{|c|}{$e$} & \multicolumn{10}{|c|}{$m$} \\\hline
$(C_2)^2$ & $(T_1T_2C_2)^2$ & $(T_1C_2)^2$ & $M^2$ & $(T_1M)^2$ & $(C_4M)^2$ & $T_1\mc{T}T_1^{-1}\mc{T}^{-1}$ & $C_4\mc{T}C_4^{-1}\mc{T}^{-1}$ & $M\mc{T}M^{-1}\mc{T}^{-1}$ & $\mc{T}^2$ & $(C_2)^2$ & $(T_1T_2C_2)^2$ & $(T_1C_2)^2$ & $M^2$ & $(T_1M)^2$ & $(C_4M)^2$ & $T_1\mc{T}T_1^{-1}\mc{T}^{-1}$ & $C_4\mc{T}C_4^{-1}\mc{T}^{-1}$ & $M\mc{T}M^{-1}\mc{T}^{-1}$ & $\mc{T}^2$ \\\hline

 1 & 1 & 1 & 1 & 1 & 1 & 1 & 1 & 1 & -1 & -1 & 1 & -1 & 1 & 1 & 1 & -1 & 1 & 1 & 1 \\\hline
 1 & 1 & 1 & 1 & 1 & -1 & 1 & -1 & 1 & -1 & -1 & 1 & -1 & 1 & 1 & 1 & -1 & 1 & 1 & 1 \\\hline
 1 & 1 & -1 & 1 & -1 & 1 & 1 & 1 & 1 & -1 & -1 & 1 & -1 & 1 & 1 & 1 & -1 & 1 & 1 & 1 \\\hline
 1 & 1 & -1 & 1 & -1 & -1 & 1 & -1 & 1 & -1 & -1 & 1 & -1 & 1 & 1 & 1 & -1 & 1 & 1 & 1 \\\hline
 1 & 1 & 1 & 1 & 1 & 1 & 1 & 1 & 1 & -1 & -1 & 1 & -1 & 1 & 1 & 1 & -1 & 1 & -1 & 1 \\\hline
 1 & 1 & -1 & 1 & -1 & 1 & 1 & 1 & 1 & -1 & -1 & 1 & -1 & 1 & 1 & 1 & -1 & 1 & -1 & 1 \\\hline
 1 & 1 & 1 & 1 & 1 & 1 & 1 & 1 & 1 & -1 & -1 & 1 & -1 & 1 & 1 & 1 & -1 & -1 & 1 & 1 \\\hline
 1 & 1 & -1 & 1 & -1 & 1 & 1 & 1 & 1 & -1 & -1 & 1 & -1 & 1 & 1 & 1 & -1 & -1 & 1 & 1 \\\hline
 1 & 1 & 1 & 1 & 1 & 1 & 1 & 1 & 1 & -1 & -1 & 1 & -1 & 1 & 1 & 1 & -1 & -1 & -1 & 1 \\\hline
 1 & 1 & 1 & 1 & 1 & -1 & 1 & -1 & 1 & -1 & -1 & 1 & -1 & 1 & 1 & 1 & -1 & -1 & -1 & 1 \\\hline
 1 & 1 & -1 & 1 & -1 & 1 & 1 & 1 & 1 & -1 & -1 & 1 & -1 & 1 & 1 & 1 & -1 & -1 & -1 & 1 \\\hline
 1 & 1 & -1 & 1 & -1 & -1 & 1 & -1 & 1 & -1 & -1 & 1 & -1 & 1 & 1 & 1 & -1 & -1 & -1 & 1 \\\hline
 1 & -1 & 1 & 1 & 1 & 1 & 1 & 1 & 1 & -1 & -1 & 1 & -1 & 1 & 1 & 1 & -1 & -1 & -1 & 1 \\\hline
 1 & -1 & 1 & 1 & 1 & -1 & 1 & -1 & 1 & -1 & -1 & 1 & -1 & 1 & 1 & 1 & -1 & -1 & -1 & 1 \\\hline
 1 & -1 & -1 & 1 & -1 & 1 & 1 & 1 & 1 & -1 & -1 & 1 & -1 & 1 & 1 & 1 & -1 & -1 & -1 & 1 \\\hline
 1 & -1 & -1 & 1 & -1 & -1 & 1 & -1 & 1 & -1 & -1 & 1 & -1 & 1 & 1 & 1 & -1 & -1 & -1 & 1 \\\hline
 1 & 1 & 1 & 1 & 1 & 1 & -1 & 1 & 1 & -1 & -1 & 1 & 1 & 1 & -1 & 1 & 1 & 1 & -1 & 1 \\\hline
 1 & 1 & 1 & 1 & 1 & 1 & -1 & 1 & 1 & -1 & -1 & 1 & 1 & 1 & -1 & 1 & 1 & -1 & -1 & 1 \\\hline
 1 & 1 & 1 & 1 & 1 & -1 & -1 & -1 & 1 & -1 & -1 & 1 & 1 & 1 & -1 & 1 & 1 & -1 & -1 & 1 \\\hline
 1 & -1 & 1 & 1 & 1 & 1 & -1 & 1 & 1 & -1 & -1 & 1 & 1 & 1 & -1 & 1 & 1 & -1 & -1 & 1 \\\hline
 1 & -1 & 1 & 1 & 1 & -1 & -1 & -1 & 1 & -1 & -1 & 1 & 1 & 1 & -1 & 1 & 1 & -1 & -1 & 1 \\\hline
 1 & 1 & 1 & 1 & 1 & 1 & -1 & 1 & 1 & -1 & -1 & 1 & 1 & 1 & -1 & 1 & -1 & 1 & 1 & 1 \\\hline
 1 & 1 & 1 & 1 & 1 & -1 & -1 & -1 & 1 & -1 & -1 & 1 & 1 & 1 & -1 & 1 & -1 & 1 & 1 & 1 \\\hline
 1 & -1 & 1 & 1 & 1 & 1 & -1 & 1 & 1 & -1 & -1 & 1 & 1 & 1 & -1 & 1 & -1 & 1 & 1 & 1 \\\hline
 1 & -1 & 1 & 1 & 1 & -1 & -1 & -1 & 1 & -1 & -1 & 1 & 1 & 1 & -1 & 1 & -1 & 1 & 1 & 1 \\\hline
 1 & 1 & 1 & 1 & 1 & 1 & -1 & 1 & 1 & -1 & -1 & 1 & 1 & 1 & -1 & 1 & -1 & -1 & 1 & 1 \\\hline
 1 & 1 & 1 & 1 & 1 & 1 & 1 & 1 & 1 & -1 & -1 & 1 & 1 & 1 & 1 & 1 & 1 & 1 & -1 & 1 \\\hline
 1 & 1 & -1 & 1 & -1 & 1 & 1 & 1 & 1 & -1 & -1 & 1 & 1 & 1 & 1 & 1 & 1 & 1 & -1 & 1 \\\hline
 1 & 1 & 1 & 1 & 1 & 1 & 1 & 1 & 1 & -1 & -1 & 1 & 1 & 1 & 1 & 1 & 1 & -1 & 1 & 1 \\\hline
 1 & 1 & 1 & 1 & 1 & 1 & -1 & 1 & 1 & -1 & -1 & 1 & 1 & 1 & 1 & 1 & 1 & -1 & 1 & 1 \\\hline
 1 & 1 & 1 & -1 & -1 & 1 & 1 & -1 & -1 & -1 & -1 & 1 & 1 & 1 & 1 & 1 & 1 & -1 & 1 & 1 \\\hline
 1 & 1 & 1 & -1 & -1 & 1 & -1 & -1 & -1 & -1 & -1 & 1 & 1 & 1 & 1 & 1 & 1 & -1 & 1 & 1 \\\hline
 1 & 1 & 1 & 1 & -1 & 1 & 1 & 1 & 1 & -1 & -1 & 1 & 1 & 1 & 1 & 1 & 1 & -1 & 1 & 1 \\\hline
 1 & 1 & 1 & 1 & -1 & 1 & -1 & 1 & 1 & -1 & -1 & 1 & 1 & 1 & 1 & 1 & 1 & -1 & 1 & 1 \\\hline
 1 & 1 & 1 & -1 & 1 & 1 & 1 & -1 & -1 & -1 & -1 & 1 & 1 & 1 & 1 & 1 & 1 & -1 & 1 & 1 \\\hline
 1 & 1 & 1 & -1 & 1 & 1 & -1 & -1 & -1 & -1 & -1 & 1 & 1 & 1 & 1 & 1 & 1 & -1 & 1 & 1 \\\hline
 1 & 1 & -1 & 1 & -1 & 1 & 1 & 1 & 1 & -1 & -1 & 1 & 1 & 1 & 1 & 1 & 1 & -1 & 1 & 1 \\\hline
 1 & 1 & -1 & 1 & -1 & 1 & -1 & 1 & 1 & -1 & -1 & 1 & 1 & 1 & 1 & 1 & 1 & -1 & 1 & 1 \\\hline
 1 & 1 & -1 & -1 & 1 & 1 & 1 & -1 & -1 & -1 & -1 & 1 & 1 & 1 & 1 & 1 & 1 & -1 & 1 & 1 \\\hline
 1 & 1 & -1 & -1 & 1 & 1 & -1 & -1 & -1 & -1 & -1 & 1 & 1 & 1 & 1 & 1 & 1 & -1 & 1 & 1 \\\hline
 1 & 1 & -1 & 1 & 1 & 1 & 1 & 1 & 1 & -1 & -1 & 1 & 1 & 1 & 1 & 1 & 1 & -1 & 1 & 1 \\\hline
 1 & 1 & -1 & 1 & 1 & 1 & -1 & 1 & 1 & -1 & -1 & 1 & 1 & 1 & 1 & 1 & 1 & -1 & 1 & 1 \\\hline
 1 & 1 & -1 & -1 & -1 & 1 & 1 & -1 & -1 & -1 & -1 & 1 & 1 & 1 & 1 & 1 & 1 & -1 & 1 & 1 \\\hline
 1 & 1 & -1 & -1 & -1 & 1 & -1 & -1 & -1 & -1 & -1 & 1 & 1 & 1 & 1 & 1 & 1 & -1 & 1 & 1 \\\hline
 1 & 1 & 1 & 1 & 1 & 1 & 1 & 1 & 1 & -1 & -1 & 1 & 1 & 1 & 1 & 1 & 1 & -1 & -1 & 1 \\\hline
 1 & 1 & 1 & 1 & 1 & -1 & 1 & -1 & 1 & -1 & -1 & 1 & 1 & 1 & 1 & 1 & 1 & -1 & -1 & 1 \\\hline
 1 & 1 & -1 & 1 & -1 & 1 & 1 & 1 & 1 & -1 & -1 & 1 & 1 & 1 & 1 & 1 & 1 & -1 & -1 & 1 \\\hline
 1 & 1 & -1 & 1 & -1 & -1 & 1 & -1 & 1 & -1 & -1 & 1 & 1 & 1 & 1 & 1 & 1 & -1 & -1 & 1 \\\hline
 1 & 1 & 1 & 1 & 1 & 1 & -1 & 1 & 1 & -1 & -1 & 1 & 1 & 1 & 1 & 1 & -1 & 1 & -1 & 1 \\\hline
 1 & 1 & 1 & 1 & 1 & 1 & -1 & 1 & 1 & -1 & -1 & 1 & 1 & 1 & 1 & 1 & -1 & -1 & -1 & 1 \\\hline
 1 & 1 & 1 & 1 & 1 & -1 & -1 & -1 & 1 & -1 & -1 & 1 & 1 & 1 & 1 & 1 & -1 & -1 & -1 & 1 \\\hline
 1 & -1 & 1 & 1 & 1 & 1 & -1 & 1 & 1 & -1 & -1 & 1 & 1 & 1 & 1 & 1 & -1 & -1 & -1 & 1 \\\hline
 1 & -1 & 1 & 1 & 1 & -1 & -1 & -1 & 1 & -1 & -1 & 1 & 1 & 1 & 1 & 1 & -1 & -1 & -1 & 1 \\\hline
\bottomrule[2pt]	
\end{tabular}
}
\caption{The symmetry fractionalization class of the 53 ``beyond-parton" $p4m\times\Z_2^T$ symmetry-enriched $\Z_2$ topological quantum spin liquids in lattice homotopy class a.}
\label{tab:SF beyond parton}
\end{table}

\end{appendices}

\clearpage
\twocolumngrid

\bibliography{ref.bib}

\end{document}